\documentclass[reprint, amsmath,amssymb, aps, superscriptaddress,prl,longbibliography]{revtex4-2}
\usepackage{bm}
\usepackage[normalem]{ulem}

\usepackage{amsthm}
\usepackage{mathtools}
\usepackage{dsfont}
\usepackage{physics}
\usepackage{graphicx}
\usepackage{adjustbox}
\usepackage{placeins}
\usepackage[T1]{fontenc}
\usepackage{lipsum}
\usepackage{csquotes}
\usepackage{stmaryrd}
\usepackage[compat=0.6]{yquant}

\usepackage{tikz}

\usepackage[pdftex, pdftitle={Article}, pdfauthor={Author}]{hyperref} 
\usepackage{hyperref}
\newcommand{\Harvard}{Department of Physics, Harvard University, Cambridge, Massachusetts 02138, USA}
\newcommand{\Caltech}{Walter Burke Institute for Theoretical Physics and Department of Physics, California Institute of Technology, Pasadena, CA 91125, USA}
\newcommand{\Chicago}{Pritzker School of Molecular Engineering, University of Chicago, Chicago, IL 60637, USA}

\usetikzlibrary{decorations.markings}
\tikzset{->-/.style={decoration={markings, mark=at position .5 with {\arrow{>}}},postaction={decorate}}}
\tikzset{-<-/.style={decoration={markings, mark=at position .5 with {\arrow{<}}},postaction={decorate}}}
\usetikzlibrary{decorations.pathmorphing, decorations.pathreplacing, decorations.shapes}

\begin{document}
\title{Protocols for Creating Anyons and Defects via Gauging}

\author{Anasuya Lyons}
\email{anasuya\_lyons@g.harvard.edu}
\affiliation{\Harvard}
\author{Chiu Fan Bowen Lo}
\affiliation{\Harvard}
\author{Nathanan Tantivasadakarn}
\affiliation{\Caltech}
\author{Ashvin Vishwanath}
\affiliation{\Harvard}
\author{Ruben Verresen}
\affiliation{\Chicago}
\affiliation{\Harvard}

\date{\today}

\begin{abstract}
Creating and manipulating anyons and symmetry defects in topological phases, especially those with a non-Abelian character, constitutes a primitive for topological quantum computation. We provide a physical protocol for implementing the ribbon operators of non-Abelian anyons and symmetry defects. We utilize dualities, in particular the Kramers-Wannier or gauging map, which have previously been used to construct topologically ordered ground states by relating them to simpler states. In this work, ribbon operators are implemented by applying a gauging procedure to a lower-dimensional region of such states. This protocol uses sequential unitary circuits or, in certain cases, constant-depth adaptive circuits. We showcase this for anyons and defects in the $\mathbb{Z}_3$ toric code and $S_3$ quantum double. The general applicability of our method is demonstrated by deriving unitary expressions for ribbon operators of various (twisted) quantum doubles.
\end{abstract}

\maketitle
\emph{Introduction:} Duality transformations are able to map complicated theories to simpler ones. Case in point is the Kramers-Wannier transformation \cite{kramers-wannier,Wegner:1971app, RevModPhys.51.659, Trebst_2007, Tupitsyn:2008ah, Cobanera_2011, Haegeman_2015, PhysRevB.94.235157, Yoshida_2017,williamson2017symmetryenrichedtopologicalordertensor,kubica2018ungaugingquantumerrorcorrectingcodes,Shirley_2019,Tantivasadakarn_2020,Tantivasadakarn_20201,radicevic2020systematicconstructionsfractontheories,Dolev_2022,Rayhaun_2023,okuda2024anomalyinflowcssfractonic,Seiberg24a,Seiberg24b}, which maps the topologically-ordered toric code \cite{kitaevFaulttolerantQuantumComputation2003} to a trivial paramagnet. Its group-based generalizations similarly relate product states to any quantum double $D(G)$ \cite{Brell_2015, Haegeman_2015}. These represent deconfined phases of $G$ gauge theories \cite{kitaevFaulttolerantQuantumComputation2003} whose anyonic quasiparticles exhibit generalized exchange statistics \cite{leinaas-myrheim, wilczek1982, goldin1985, Frohlich1990, wen1991}, implying topological degeneracy.
Recently, the duality maps for solvable groups $G$ have been rewritten as finite-depth measurement-based circuits, allowing for efficient preparation of a variety of interesting topological phases \cite{tantivasadakarnLongrangeEntanglementMeasuring2022a, tantivasadakarnShortestRouteNonAbelian2022, verresenEfficientlyPreparingSchr2022, bravyiAdaptiveConstantdepthCircuits2022,tantivasadakarnHierarchyTopologicalOrder2022,Li_2023,Sukeno_2024,li2023measuringtopologicalfieldtheories}. For instance, such a measurement-based scheme has been used to create $D_4$ topological order in the lab \cite{iqbalCreationNonAbelianTopological2023}.

\begin{figure}
    \centering
    \includegraphics[width=0.98\linewidth]{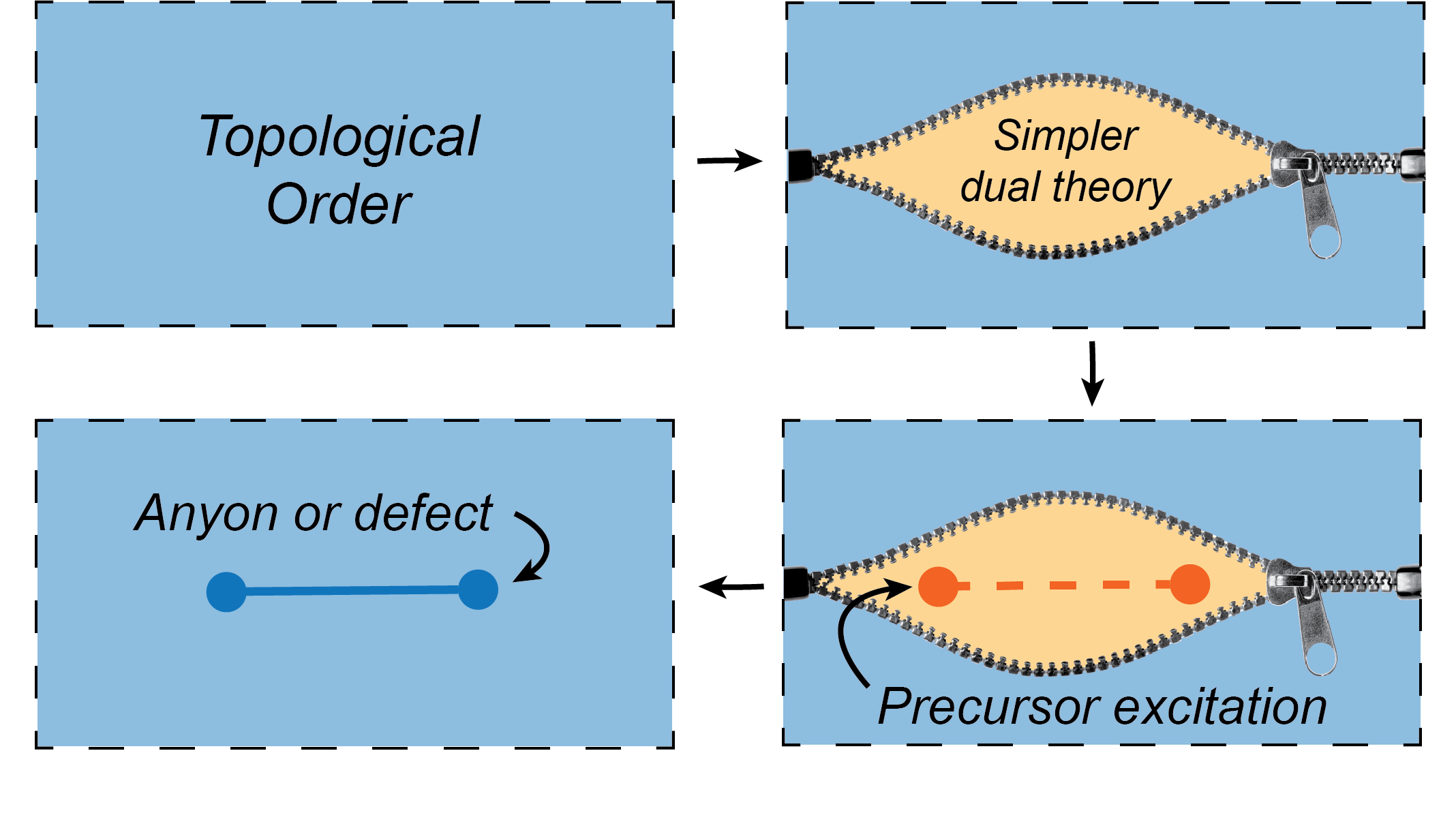}
    \caption{\textbf{Taking advantage of dualities:} To create a given anyon or defect in a topological order, we apply a duality mapping in a finite region, ``unzipping'' to the simpler dual theory. Then, we create a ``precursor'' excitation, typically with a simple finite-depth circuit. Finally, we ``zip'' back up to the original theory, leaving behind the desired excitation.}
    \label{fig:intro}
\end{figure}

Here, we use dualities to construct ribbon operators \cite{kitaevFaulttolerantQuantumComputation2003, bombinFamilyNonAbelianKitaev2008, chenRibbonOperatorsGeneralized2022} for topological defects \cite{barkeshli2010,bombinTopologicalOrderTwist2010,youProjectiveNonAbelianStatistics2012,barkeshliTwistDefectsProjective2013,youSyntheticNonAbelianStatistics2013,barkeshliTheoryDefectsAbelian2013,Barkeshliclassification13,barkeshliSymmetryFractionalizationDefects2019} and anyons. The main conceptual result is illustrated in Fig. \ref{fig:intro}: we first map a one-dimensional region of a topological order to a simpler theory, then create a precursor to a given topological excitation. By re-applying the duality map we return to the original topological order, which now contains the desired excitation. We note that Ref.~\onlinecite{Li_2023} also used the interplay between gauging and anyon operators to diagnose certain symmetry-enriched topological orders (SETs). 

We first exemplify our method for the $\mathbb{Z}_3$ toric code, then generalize to any quantum double $D(G)$. Finally, we discuss the applicability of our protocols for understanding \emph{twisted} quantum double models \cite{Dijkgraaf:1989pz, Roche:1990hs, propitiusTopologicalInteractionsBroken1995, huTwistedQuantumDouble2013}. In contrast to the existing literature on quantum double ribbons \cite{kitaevFaulttolerantQuantumComputation2003,bombinFamilyNonAbelianKitaev2008, chenRibbonOperatorsGeneralized2022}, our approach is both conceptually and algebraically simple, relying solely on the properties of the $G$-gauging map \cite{Brell_2015, tantivasadakarnLongrangeEntanglementMeasuring2022a}. This simplicity underscores its practicality, as we can easily obtain efficient circuits for non-Abelian excitations ideal for near-term quantum devices. Indeed, in a companion paper using a trapped-ion quantum processor \cite{z3-paper}, we implement the $\mathbb{Z}_3$ toric code symmetry defect circuits introduced in the present work. Adaptive, efficient circuits for the creation of anyons in quantum doubles of solvable groups have been proposed before \cite{bravyiAdaptiveConstantdepthCircuits2022}; our approach pinpoints the gauging and ungauging layers inherent in the ribbon structure as the steps able to be made finite-depth. This conceptual insight (i) yields intuitive unitary ribbon operators for \emph{any} group-based non-Abelian anyon, (ii) gives a systematic construction for a wide range of \emph{symmetry defect} ribbons, and (iii) opens up the possibility of constructing efficient ribbon operators for topological orders outside of the Kitaev quantum double.

\emph{$\mathbb{Z}_3$ toric code symmetry defects:} We first discuss how to create symmetry defects in the $\mathbb{Z}_3$ toric code \cite{kitaevFaulttolerantQuantumComputation2003}. The computational qutrit basis is labeled $\ket{n}$, with $n = 0, 1, 2$. Define the qutrit clock and shift operators:
\begin{equation}
 \begin{aligned}
 \mathcal{Z}\ket{n} = \omega^n \ket{n},  &\quad \mathcal{X}\ket{n} = \ket{n+1}  \\
 \mathcal{Z}^2  = \mathcal{Z}^\dagger, \quad \mathcal{X}^2 = &\mathcal{X}^\dagger, 
 \quad \mathcal{ZX} = \omega \mathcal{XZ},
 \end{aligned}
 \end{equation}
with $\omega = e^{i \frac{2\pi}{3}}$. The Hamiltonian is given by $H_{\mathbb{Z}_3} = -\sum_v A^{(v)} - \sum_p B^{(p)} + \mathrm{h.c.}$, where \footnote{We are making a choice of lattice orientation where horizontal edges point right and vertical edges point upwards. This convention is constant throughout.}:
\begin{equation}
\raisebox{-17pt}{
\begin{tikzpicture}
\node at (2.5,0){
    \begin{tikzpicture}
    \node at (-1.1,0) {$A^{(v)} =$};
    \draw[-,opacity=0.2] (-0.6,0) -- (0.6,0);
    \draw[-,opacity=0.2] (0,-0.6) -- (0,0.6);
    \node at (0.3,0) {$\mathcal X$};
    \node at (0,0.3) {$\mathcal X$};
    \node at (-0.3,0) {$\mathcal X^\dagger$};
    \node at (0,-0.3) {$\mathcal X^\dagger$};
    \end{tikzpicture}
};

\node at (6.2,0){
    \begin{tikzpicture}
    \node at (-0.8,0) {$B_p =$};
    \draw[-,opacity=0.2] (-0.2,0.3) -- (0.9,0.3);
    \draw[-,opacity=0.2] (-0.2,-0.3) -- (0.9,-0.3);
    \draw[-,opacity=0.2] (0,-0.55) -- (0,0.55);
    \draw[-,opacity=0.2] (0.7,-0.55) -- (0.7,0.55);
    \node at (0,0) {$\mathcal Z^\dagger$};
    \node at (0.42,0.3) {$\mathcal Z^\dagger$};
    \node at (0.35,-0.3) {$\mathcal Z$};
    \node at (0.7,0) {$\mathcal Z$};
    \end{tikzpicture}
};
\end{tikzpicture}
}.
\end{equation}
A note on notation: throughout, any sub- or superscripts in parentheses denote a physical index. Violations of the vertex stabilizers are known as $e$ ($\omega$ eigenvalue) and $e^2$ ($\omega^*$ eigenvalue) anyons , while plaquette violations are $m$ and $m^2$ respectively.

The $\mathbb{Z}_3$ toric code has a charge conjugation symmetry permuting $e \leftrightarrow e^2$ and $m \leftrightarrow m^2$, not present in its $\mathbb{Z}_2$ counterpart \cite{barkeshliTheoryDefectsAbelian2013,youSyntheticNonAbelianStatistics2013, barkeshliSymmetryFractionalizationDefects2019}. A $\mathbb{Z}_3$ anyon encircling a charge-conjugation defect is transformed to its conjugate particle. Thus, these defects are non-Abelian, as they can absorb and emit Abelian particles \cite{bombinTopologicalOrderTwist2010, youProjectiveNonAbelianStatistics2012}. The charge-conjugation symmetry is realized onsite by the charge conjugation operator $\mathcal{C}$: 
\begin{equation}
 \mathcal{C}\ket{n} = \ket{-n \! \! \! \mod 3} \; \; \Rightarrow \; \; \mathcal{CXC} = \mathcal{X}^\dagger, \; \; \mathcal{CZC} = \mathcal{Z}^\dagger
\end{equation}
A closed charge conjugation boundary is created by acting $\prod \mathcal{C}$ in a finite region of the lattice. Since inside this region, $\mathcal CA^{(v)} \mathcal C = A^{(v)\dagger}$ and $\mathcal C B^{(p)} \mathcal C = B^{(p)\dagger}$, the Hamiltonian only changes near the \emph{boundary} of this region:
\begin{equation}
\raisebox{-23pt}{
    \begin{tikzpicture}
        \node at (0,0){
            \begin{tikzpicture}[scale=0.75]
            \draw[-,opacity=0.2] (0,0) -- (2,0);
            \draw[-,opacity=0.2] (0,0.75) -- (2,0.75);
            \draw[-,opacity=0.2] (0,-0.75) -- (2,-0.75);
            \draw[-,opacity=0.2] (0,-0.75) -- (0,0.75);
            \draw[-,opacity=0.2] (1,-0.75) -- (1,0.75);
            \draw[-,opacity=0.2] (2,-0.75) -- (2,0.75);
        
            \node at (0.5, 0) {$\mathcal C$};
            \node at (1.5, 0) {$\mathcal C$};
            \node at (0.5, -0.75) {$\mathcal C$};
            \node at (1.5, -0.75) {$\mathcal C$};
            \node at (0, -0.375) {$\mathcal C$};
            \node at (1, -0.375) {$\mathcal C$};
            \node at (2, -0.375) {$\mathcal C$};
        
            \node[color=black!60] at (1,0) {\small $v$};
            \node[color=black!60] at (1.5,0.375) {\small $p$};
        \end{tikzpicture}
        };
        \node at (2.5, 0){
            \begin{tikzpicture}
            \node at (-1.2,0) {$\tilde{A}^{(v)} =$};
            \draw[-,opacity=0.2] (-0.6,0) -- (0.6,0);
            \draw[-,opacity=0.2] (0,-0.6) -- (0,0.6);
            \node at (0.3,0) {$\mathcal X^\dagger$};
            \node at (0,0.3) {$\mathcal X$};
            \node at (-0.3,0) {$\mathcal X$};
            \node at (0,-0.3) {$\mathcal X$};
            \end{tikzpicture}
        };
        \node at (5,0){
            \begin{tikzpicture}
            \node at (-1,0) {$\tilde{B}^{(p)} =$};
            \draw[-,opacity=0.2] (-0.2,0.3) -- (0.9,0.3);
            \draw[-,opacity=0.2] (-0.2,-0.3) -- (0.9,-0.3);
            \draw[-,opacity=0.2] (0,-0.55) -- (0,0.55);
            \draw[-,opacity=0.2] (0.7,-0.55) -- (0.7,0.55);
            \node at (0,0) {$\mathcal Z^\dagger$};
            \node at (0.42,0.3) {$\mathcal Z^\dagger$};
            \node at (0.35,-0.3) {$\mathcal Z^\dagger$};
            \node at (0.7,0) {$\mathcal Z$};
            \end{tikzpicture}
        };
    \end{tikzpicture}
    }
\label{eq:bulk-cc}
\end{equation}

\begin{figure}
    \centering
    \includegraphics[width=1\linewidth]{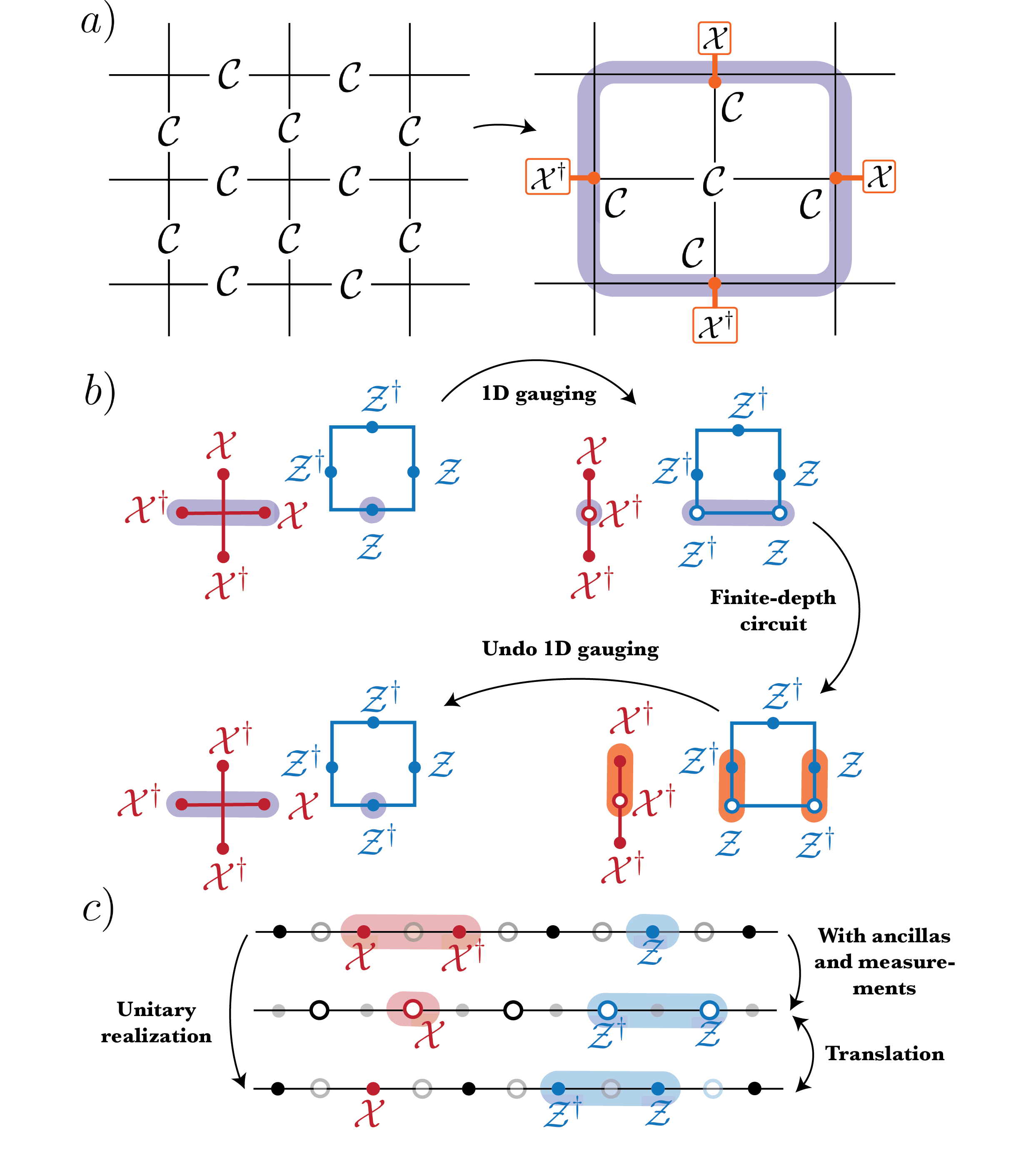}
    \caption{\textbf{Creating a charge conjugation defect in $\mathbb{Z}_3$ toric code:} a) A charge conjugation symmetry membrane reduces to a boundary circuit of controlled-$\mathcal{X}$ gates by using variables of the pre-gauged theory (See Eq.~\eqref{eq:defect-circuit}). b) To create charge conjugation defects: Map back to the necessary vertices using the 1D gauging map (affected degrees of freedom are highlighted in purple), then apply the derived boundary action (support highlighted in orange). Finally, map back to edges. 
    c) The action of the measurement-based and unitary realizations of $\mathsf{KW}_{\mathbb{Z}_3}^{\text{1D}}$ differ by half a lattice translation. Here, solid circles mark edges, open circles mark vertices.}
    \label{fig:1d-ungauging}
\end{figure}

Any anyon string that crosses the defect line must transform to its conjugate, or violate these modified stabilizers. To create \emph{point} charge-conjugation defects, we must terminate the defect line. This is not possible using a bulk symmetry membrane---we need a circuit localized to the defect line itself. We will show that such a circuit exists, whose form we will derive and unpack:
\begin{equation}
\raisebox{-35pt}{
\begin{tikzpicture}
    \node at (-3, 0){$U^\gamma_{c.c.} = \mathsf{KW}^{\text{1D};\gamma}_{\mathbb{Z}_3} \left ( \prod\limits_{v \in \gamma} \text{C}\mathcal{X}_{(v \rightarrow e_v)}^{\eta(e_v)} \right) \mathsf{KW}^{\text{1D};\gamma \dagger}_{\mathbb{Z}_3}$};
    \node at (-2.4, -1.25){
        \begin{tikzpicture}
            \node at (0, 0){
                \begin{tikzpicture}
                    \draw[-,opacity=0.2] (-0.5,0) -- (0.5,0);
                    \draw[-,opacity=0.2] (0,-0.3) -- (0, 0.3);
                    \node at (0.3,0) {$\mathcal{X}^\dagger$};
                    \node at (-0.3,0) {$\mathcal X$};
                \end{tikzpicture}
            };
            \draw[->] (0.7,0) -- node[above,midway] {\footnotesize $\mathsf{KW}^{\text{1D}}_{\mathbb{Z}_3}$} (1.5,0);
            \node at (2, 0){
                \begin{tikzpicture}
                    \draw[-,opacity=0.2] (-0.4,0) -- (0.4,0);
                    \draw[-,opacity=0.2] (0,-0.3) -- (0, 0.3);
                    \node at (0,0) {$\mathcal X$};
                \end{tikzpicture}
            };
            \node at (3.6, 0){
                \begin{tikzpicture}
                    \draw[-,opacity=0.2] (-0.5,0) -- (0.5,0);
                    \draw[-,opacity=0.2] (-0.3,-0.3) -- (-0.3, 0.3);
                    \draw[-,opacity=0.2] (0.3,-0.3) -- (0.3, 0.3);
                    \node at (0,0) {$\mathcal Z$};
                \end{tikzpicture}
            };
            \draw[->] (4.3,0) -- node[above,midway] {\footnotesize $\mathsf{KW}^{\text{1D}}_{\mathbb{Z}_3}$} (5.1,0);
            \node at (5.8, 0){
                \begin{tikzpicture}
                    \draw[-,opacity=0.2] (-0.5,0) -- (0.5,0);
                    \draw[-,opacity=0.2] (-0.25,-0.3) -- (-0.25, 0.3);
                    \draw[-,opacity=0.2] (0.25,-0.3) -- (0.25, 0.3);
                    \node at (-0.25,0) {$\mathcal{Z}^\dagger$};
                    \node at (0.25,0) {$\mathcal{Z}$};
                \end{tikzpicture}
            };
        \end{tikzpicture}
    };
    \node at (0.25, -0.1) {;};
    \node at (0.85, 0){
        \begin{tikzpicture}[scale=1.25]
            \draw[-,opacity=0.2] (-0.25,0) -- (0.25,0);
            \draw[-,opacity=0.2] (0,0) -- (0, 0.5);
            \node[color=black!60] at (0,0) {\footnotesize $v$};
            \node[color=black!60] at (0,0.3) {\footnotesize $e_v$};
        \end{tikzpicture}
    };
\end{tikzpicture}
}
\label{eq:cc-defect-circuit}
\end{equation}
The action of $U^\gamma_{c.c.}$ is represented in Figs.~\ref{fig:1d-ungauging}a-b. 
$\mathsf{KW}_{\mathbb{Z}_3}^{1D; \gamma \dagger}$ is the 1D (un)gauging map along an open path $\gamma$. This map may be more familiar as the Kramers-Wannier duality of the 1D Ising model: it maps the ferromagnetic phase (which we can think of as living on the edges along $\gamma$) to the paramagnetic phase (which lives on the \emph{vertices} along $\gamma$) and vice versa. This duality is a result of gauging the $\prod X$ symmetry of the Ising paramagnet. The definition of the mapping given in the second line of Eq.~\eqref{eq:cc-defect-circuit} is the $\mathbb{Z}_3$ analogue. We can roughly think of our topological state, restricted to a line, as being the ferromagnet--- we need this 1D mapping to access its dual paramagnet, where the action of the symmetry is simpler. Note that we write $\mathsf{KW}_{\mathbb{Z}_3}^{1D}$ rather than an explicit circuit as both measurement-based \cite{tantivasadakarnLongrangeEntanglementMeasuring2022a} and unitary \cite{Seiberg24a,Chen24} realizations of the same mapping exist, and simply differ by half a lattice translation--- see Fig. \ref{fig:1d-ungauging}c for the action of the 1D gauging map in both cases \footnote{The measurement-based circuit involves explicitly introducing new degrees of freedom on the vertices to map from the ferromagnetic phase to the paramagnetic, while the unitary circuit maps between wavefunctions on the same degrees of freedom.}. The unitary implementation of $\mathsf{KW}_{\mathbb{Z}_3}^{1D}$ is a sequential circuit of qutrit CNOT gates: $\prod_{i \in \gamma} C\mathcal{X}_{i \rightarrow i+1}$.

The middle portion of $U^\gamma_{c.c.}$ acts between the uncovered vertex degrees of freedom and their neighboring edges perpendicular to $\gamma$, with $\text{C}\mathcal{X}\ket{n, m} = \ket{n, m+n~(\text{mod } 3)}$. Here, $\eta(e_v) = \pm 1$ is some sign structure depending on the lattice, to be specified later. This part of the circuit implements the charge conjugation symmetry on the hybrid state created by the first application of $\mathsf{KW}_{\mathbb{Z}_3}^{1D}$.

To derive and explain Eq. \eqref{eq:cc-defect-circuit}, note that the $\mathbb{Z}_3$ paramagnet $\prod_{v} \left( \ket{0} + \ket{1} + \ket{2} \right)$ has the global $\mathbb{Z}_2$ symmetry $\prod \mathcal{C}$, along with its $\mathbb{Z}_3$ symmetry $\prod \mathcal{X}$. Upon gauging the latter using the \emph{2D} gauging map (defined in Eq.~\eqref{eq:z3-gauging-props} below) the $\mathbb{Z}_3$ paramagnet is mapped to the $\mathbb{Z}_3$ toric code. Charge conjugation acts trivially in a finite region of the $\mathbb{Z}_3$ paramagnet---therefore, all the key properties of the charge conjugation symmetry membrane on the $\mathbb{Z}_3$ toric code are captured by the 2D gauging map itself. By `pushing' the membrane from the $\mathbb{Z}_3$ toric code through the $\mathbb{Z}_3$ gauging map back to the paramagnet, we can simplify the bulk operator to some action purely at the membrane boundary.

More precisely, we define $\mathsf{KW}^{2D}_{\mathbb{Z}_3}$ as the map that takes vertices to edges in the following way:
\begin{equation}
\raisebox{-17pt}{
\begin{tikzpicture}
\node at (0,0){
    \begin{tikzpicture}
    \draw[-,opacity=0.2] (-0.6,0) -- (0.6,0);
    \draw[-,opacity=0.2] (0,-0.6) -- (0,0.6);
    \node at (0,0) {$\mathcal X$};
    \end{tikzpicture}
};
\draw[->] (0.85,0) -- node[above,midway] {\footnotesize $\mathsf{KW}^{2D}_{\mathbb{Z}_3}$} (1.65,0);
\node at (2.5,0){
    \begin{tikzpicture}
    \draw[-,opacity=0.2] (-0.6,0) -- (0.6,0);
    \draw[-,opacity=0.2] (0,-0.6) -- (0,0.6);
    \node at (0.3,0) {$\mathcal X$};
    \node at (0,0.3) {$\mathcal X$};
    \node at (-0.3,0) {$\mathcal X^\dagger$};
    \node at (0,-0.3) {$\mathcal X^\dagger$};
    \end{tikzpicture}
};
\node at (4.5,0){
    \begin{tikzpicture}
    \draw[-,opacity=0.2] (-0.2,0.3) -- (0.2,0.3);
    \draw[-,opacity=0.2] (-0.2,-0.3) -- (0.2,-0.3);
    \draw[-,opacity=0.2] (0,-0.55) -- (0,0.55);
    \node at (0,0.3) {$\mathcal Z^\dagger$};
    \node at (0,-0.3) {$\mathcal Z$};
    \end{tikzpicture}
};
\draw[->] (0.85+4.1,0) -- node[above,midway] {\footnotesize $\mathsf{KW}^{2D}_{\mathbb{Z}_3}$} (1.65+4.1,0);
\node at (6.2,0){
    \begin{tikzpicture}
    \draw[-,opacity=0.2] (-0.2,0.3) -- (0.2,0.3);
    \draw[-,opacity=0.2] (-0.2,-0.3) -- (0.2,-0.3);
    \draw[-,opacity=0.2] (0,-0.55) -- (0,0.55);
    \node at (0,0) {$\mathcal Z$};
    \end{tikzpicture}
};
\end{tikzpicture}
}
\label{eq:z3-gauging-props}
\end{equation}
The above fully specifies how $\mathsf{KW}^{2D}_{\mathbb{Z}_3}$ transforms $\mathcal{C}$, as
\begin{equation}
    \mathcal{C} = \ket{0}\bra{0} + \mathcal{X}\ket{1}\bra{1} + \mathcal{X}^\dagger\ket{2}\bra{2}  =\sum_{n=0}^2 \mathcal X^n T_{n}
\label{eq:cc-decomp}
\end{equation}
where $T_n = \ket{n}\bra{n} =\frac{1}{3}(1+ \omega^{-n}\mathcal Z +\omega^n\mathcal Z^\dagger)$.
As a single vertex $\mathcal{Z}$ is not symmetric, we leave it behind when pushing $\mathcal{C}^{(\text{vertex})}$ through the gauging map. Similarly, we leave behind the $\mathcal X$ operators involved in $\mathcal C^{(\text{edge})}$:
\begin{equation}
\raisebox{-35pt}{
\begin{tikzpicture}
\node at (0, 0) {
    \begin{tikzpicture}
        \node at (-1, 0) { \footnotesize $\mathsf{KW}^{2D}_{\mathbb{Z}_3} ~\cdot$};
        \node at (0,0){
            \begin{tikzpicture}[scale=0.7]
            \draw[-,opacity=0.2] (-0.6,0) -- (0.6,0);
            \draw[-,opacity=0.2] (0,-0.6) -- (0,0.6);
            \node at (0,0) {$\mathcal C$};
            \end{tikzpicture}
        };
        \node at (1, 0) {$= \sum\limits_{n=0}^2 $}; 
        \node at (2.2,0){
            \begin{tikzpicture}[scale=0.7]
            \draw[-,opacity=0.2] (-0.6,0) -- (0.6,0);
            \draw[-,opacity=0.2] (0,-0.6) -- (0,0.6);
            \node at (0, 0) {$(A^{(v)})^n$};
            \end{tikzpicture}
        };
        \node at (3.4, 0) {\footnotesize $\cdot ~ \mathsf{KW}^{2D}_{\mathbb{Z}_3} ~\cdot$};
        \node at (4.5, 0) {
            \begin{tikzpicture}[scale=0.7]
            \draw[-,opacity=0.2] (-0.6,0) -- (0.6,0);
            \draw[-,opacity=0.2] (0,-0.6) -- (0,0.6);
            \node at (0,0) {$T_n$};
            \end{tikzpicture}
        };
    \end{tikzpicture}
}; 
\node at (0, -1.35) {
    \begin{tikzpicture}
        \node at (0,-1.5){
            \begin{tikzpicture}
            \draw[-,opacity=0.2] (-0.2,0.3) -- (0.2,0.3);
            \draw[-,opacity=0.2] (-0.2,-0.3) -- (0.2,-0.3);
            \draw[-,opacity=0.2] (0,-0.55) -- (0,0.55);
            \node at (0,0) {$\mathcal C$};
            \end{tikzpicture}
        };
        \node at (1.2, -1.65) {\footnotesize $\cdot~ \mathsf{KW}^{2D}_{\mathbb{Z}_3} = \sum\limits_{n, n'}$};
        \node at (2.9,-1.5){
            \begin{tikzpicture}
            \draw[-,opacity=0.2] (-0.2,0.3) -- (0.2,0.3);
            \draw[-,opacity=0.2] (-0.2,-0.3) -- (0.2,-0.3);
            \draw[-,opacity=0.2] (0,-0.55) -- (0,0.55);
            \node at (0.3, 0) {$\mathcal{X}^{n-n'}$};
            \end{tikzpicture}
        };
        \node at (4.2, -1.5) {\footnotesize $\cdot~ \mathsf{KW}^{2D}_{\mathbb{Z}_3} ~\cdot$};
        \node at (5.1, -1.5) {
            \begin{tikzpicture}
                \draw[-,opacity=0.2] (-0.2,0.3) -- (0.2,0.3);
                \draw[-,opacity=0.2] (-0.2,-0.3) -- (0.2,-0.3);
                \draw[-,opacity=0.2] (0,-0.55) -- (0,0.55);
                \node at (0, -0.3) {$T_n$};
                \node at (0, 0.3) {$T_{n'}$};
            \end{tikzpicture}
        };
    \end{tikzpicture}
};
\end{tikzpicture}
}
\label{eq:cc-pushthrough}
\end{equation}
Thus, applying $\mathcal{C}$ to a pre-gauged vertex $v$ is equivalent to applying $A^{(v)}$ after the gauging map, controlled on $v$. Acting $\mathcal{C}$ on an edge is equivalent to a product of $\text{C}\mathcal{X}$, controlled on the neighboring vertices. 

Our first key observation is that Eq.~\eqref{eq:cc-pushthrough} allows us to write down a purely one-dimensional operator which creates a charge-conjugation boundary (see Fig.~\ref{fig:1d-ungauging}a):
\begin{equation}
\prod_{v \in \gamma} \text{C}\mathcal{X}_{(v \rightarrow e_v)}^{\eta(e_v)}
\label{eq:defect-circuit}
\end{equation}
where $\gamma$ is a closed path, $e_v$ is defined as in Eq.~\eqref{eq:cc-defect-circuit}, and $\eta(e) = +1 (-1)$ if $e_v$ points south, west (north, east). 

\footnotetext[10]{For a 2D model, a 1-form symmetry lives on deformable 1D submanifolds; the logical operators of the toric code are an example.}

We can create point defects by acting with a finite segment of this string operator. Starting from the toric code, however, we naively only have access to the edges, while Eq.~\eqref{eq:defect-circuit} requires we use vertices as control qutrits. Our second key observation is that we can access the necessary vertices by simply \emph{ungauging} the toric code along a line. We can interpret this as gauging a 1-form symmetry \cite{Roumpedakis_2023, Note10}: along the line, the 1-form symmetry looks like a 0-form symmetry of a 1D system. This means we can use $\mathsf{KW}_{\mathbb{Z}_3}^{\text{1D}}$ to gauge it, allowing simultaneous access to vertex and edge degrees of freedom.


We have arrived at a 1D protocol to create charge-conjugation defects: (1) Access the vertices needed to act as controls by applying the 1D gauging map along the desired path. (2) Apply $C\mathcal{X}$ between the (ungauged) vertices along the line and the (gauged) edges just outside. (3) Undo the 1D gauging, mapping back to edge degrees of freedom everywhere. Combining all steps into a single unitary yields Eq.~\eqref{eq:cc-defect-circuit}. Fig.~\ref{fig:1d-ungauging}b shows that this circuit gives exactly the same boundary stabilizers as the bulk charge conjugation membrane (Eq.~\eqref{eq:bulk-cc}). As we are acting on an open line, there will be non-symmetric inputs to the gauging map at the endpoints; the resultant non-local stabilizers store the internal state of the non-Abelian defect pair (see \cite{supplemental} for details). 

\emph{$D(S_3)$ quantum double:} The quantum double $D(S_3)$ is obtained by gauging the $\prod \mathcal C$ symmetry of the $\mathbb{Z}_3$ toric code \cite{verresenEfficientlyPreparingSchr2022}---upon gauging, the charge-conjugation defects become deconfined non-Abelian anyons. We immediately arrive at ribbon operators for these anyons by sandwiching the defect circuit with $\mathsf{KW}^{\text{1D}}_{\mathbb{Z}_2}$. Other anyons in $D(S_3)$ originate from anyons of the $\mathbb{Z}_3$ toric code itself and so are created by gauging-map decorated Abelian strings (see Table \ref{tab:summary}).

We can implement the 1D gauging map via a sequential unitary circuit \cite{Chen24}, or a finite-depth adaptive circuit (for solvable symmetry groups) \cite{tantivasadakarnLongrangeEntanglementMeasuring2022a,tantivasadakarnHierarchyTopologicalOrder2022,bravyiAdaptiveConstantdepthCircuits2022}. The latter choice yields adaptive, efficient ribbons for non-Abelian excitations. Such circuits have been written down before for quantum doubles with solvable gauge group \cite{bravyiAdaptiveConstantdepthCircuits2022}---we emphasize, however, that our scheme applies more generally, as we will illustrate below. Moreover, it gives manifestly unitary ribbon operators for general non-Abelian anyons, which was left as an open question in Ref.~\onlinecite{bravyiAdaptiveConstantdepthCircuits2022}.

\usetikzlibrary { decorations.pathmorphing, decorations.pathreplacing, decorations.shapes, }
\begin{table*}
    \centering
    \begin{tabular}{cccc}
        \hline 
        \hline
        \textbf{Topological Order}& \textbf{Duality Map} & \textbf{Precursor Phase} & \textbf{Ribbons}\\
        \hline 
         a) $D(S_3)$ & $\mathsf{KW}_{\mathbb{Z}_2}$ & $D(\mathbb{Z}_3)$ & \begin{tikzpicture}[baseline=0.2cm]
            \node at (0, 0) {$[2]:$};
            \node at (1.25, 0) {
                \begin{tikzpicture}[scale=0.5]
                    \filldraw[-, color=black, fill = blue!10] (0, 0) -- (0, 0.5) -- (3, 0.5) -- (3, 0) -- (0, 0);
                    \node[node font=\tiny] at (1.5, 0.25) {Un-gauging};
                    \draw[-, opacity=0.2] (0.5, -1) -- (0.5, 0);
                    \draw[-, opacity=0.2] (1.5, -1) -- (1.5, 0);
                    \draw[-, opacity=0.2] (2.5, -1) -- (2.5, 0);
                    \filldraw[-, color=black, fill = blue!10] (0, -1) -- (0, -1.5) -- (3, -1.5) -- (3, -1) -- (0, -1);
                    \node[node font=\tiny] at (1.5, -1.25) {Re-gauging};

                    \draw[decorate, decoration={snake, segment length = 2pt, amplitude=1pt}, red] (0.1, -0.5) -- (2.9, -0.5);
                    \filldraw[color=black, fill = red] (0.5, -0.5) circle (3pt);
                    \filldraw[color=black, fill = red] (1.5, -0.5) circle (3pt);
                    \filldraw[color=black, fill = red] (2.5, -0.5) circle (3pt);

                    \node[red] at (3.1, -0.5) {$e$};
                \end{tikzpicture}
            };
            \node at (2.7, 0) {$C_3 :$};
            \node at (3.95, 0) {
                \begin{tikzpicture}[scale=0.5]
                    \filldraw[-, color=black, fill = blue!10] (0, 0) -- (0, 0.5) -- (3, 0.5) -- (3, 0) -- (0, 0);
                    \node[node font=\tiny] at (1.5, 0.25) {Un-gauging};
                    \draw[-, opacity=0.2] (0.5, -1) -- (0.5, 0);
                    \draw[-, opacity=0.2] (1.5, -1) -- (1.5, 0);
                    \draw[-, opacity=0.2] (2.5, -1) -- (2.5, 0);
                    \filldraw[-, color=black, fill = blue!10] (0, -1) -- (0, -1.5) -- (3, -1.5) -- (3, -1) -- (0, -1);

                    \draw[decorate, decoration={snake, segment length = 2pt, amplitude=1pt}, blue] (0.1, -0.5) -- (2.9, -0.5);
                    \filldraw[color=black, fill = blue] (0.5, -0.5) circle (3pt);
                    \filldraw[color=black, fill = blue] (1.5, -0.5) circle (3pt);
                    \filldraw[color=black, fill = blue] (2.5, -0.5) circle (3pt);
                    \node[node font=\tiny] at (1.5, -1.25) {Re-gauging};

                    \node[blue] at (3.2, -0.5) {$m$};
                \end{tikzpicture}
            };
         \end{tikzpicture}\\
         b) $D(D_{4})$ & $\mathsf{KW}_{\mathbb{Z}_2^3}$ & Type III $\mathbb{Z}_2^3$ SPT \cite{Yoshida_2016} & 
         \begin{tikzpicture}[baseline=-0.1cm]
            \node at (-1.1, 0) {$m_b: $};
            \node at (0, 0){
                \begin{tikzpicture}[scale=0.5]
                    \draw[-, opacity=0.2] (0, 1) -- (0, 0) -- (0.5*1.73, -0.5) -- (1.73, 0) -- (1.5*1.73, -0.5) -- (1.5*1.73, 0.5) -- (1.73, 1) -- (0.5*1.73, 0.5) -- (0, 1);
                    \draw[-, opacity=0.2] (0.5*1.73, -0.5) -- (0.5*1.73, 0.5);
                    \draw[-, opacity=0.2] (1.73, 0) -- (1.73, 1);
                    \draw[-, blue, very thick] (0, 0) -- (0.5*1.73, 0.5) -- (1.73, 0) -- (1.5*1.73, 0.5);

                    \draw[decorate, decoration={snake, segment length = 2pt, amplitude=1pt}] (0, 0.5) -- (0.5*1.73, 0);
                    \draw[decorate, decoration={snake, segment length = 2pt, amplitude=1pt}] (1.73, 0.5) -- (1.5*1.73, 0);
                    \draw[decorate, decoration={snake, segment length = 2pt, amplitude=1pt}] (0, 0.5) -- (1.5*1.73, 0);
                    \filldraw[color = black, fill = green, thick] (0, 0.5) circle (3.5pt);
                    \filldraw[color = black, fill = green, thick] (1.73, 0.5) circle (3.5pt);
                    \filldraw[color = black, fill = magenta, thick] (0.5*1.73, 0) circle (3.5pt);
                    \filldraw[color = black, fill = magenta, thick] (1.5*1.73, 0) circle (3.5pt);
                \end{tikzpicture}
             };
            \node at (1.4, 0) {$m_g: $};
            \node at (2.5, 0){
                \begin{tikzpicture}[scale=0.5]
                    \draw[-, opacity=0.2] (0, 1) -- (0, 0) -- (0.5*1.73, -0.5) -- (1.73, 0) -- (1.5*1.73, -0.5) -- (1.5*1.73, 0.5) -- (1.73, 1) -- (0.5*1.73, 0.5) -- (0, 1);
                    \draw[-, opacity=0.2] (0.5*1.73, -0.5) -- (0.5*1.73, 0.5);
                    \draw[-, opacity=0.2] (1.73, 0) -- (1.73, 1);
                    \draw[-, green, very thick] (0, 0) -- (0.5*1.73, 0.5) -- (1.73, 0) -- (1.5*1.73, 0.5);

                    \draw[decorate, decoration={snake, segment length = 2pt, amplitude=1pt}] (0, 0.5) -- (0.5*1.73, 0);
                    \draw[decorate, decoration={snake, segment length = 2pt, amplitude=1pt}] (1.73, 0.5) -- (1.5*1.73, 0);
                    \draw[decorate, decoration={snake, segment length = 2pt, amplitude=1pt}] (0, 0.5) -- (1.5*1.73, 0);
                    \filldraw[color = black, fill = magenta, thick] (0, 0.5) circle (3.5pt);
                    \filldraw[color = black, fill = magenta, thick] (1.73, 0.5) circle (3.5pt);
                    \filldraw[color = black, fill = blue, thick] (0.5*1.73, 0) circle (3.5pt);
                    \filldraw[color = black, fill = blue, thick] (1.5*1.73, 0) circle (3.5pt);
                \end{tikzpicture}
             };
             \node at (3.75, 0.25) {\tiny $X$:};
             \node at (3.75, -0.25) {\tiny $CZ$:};

             \draw[-, green, very thick] (4.1, 0.3) -- (4.6, 0.3);
             \draw[-, blue, very thick] (4.1, 0.2) -- (4.6, 0.2);

            \draw[decorate, decoration={snake, segment length = 2pt, amplitude=1pt}] (4.1, -0.15) -- (4.6, -0.15);
            \filldraw[color = black, fill = magenta, thick] (4.1, -0.15) circle (1.5pt);
            \filldraw[color = black, fill = blue, thick] (4.6, -0.15) circle (1.5pt);

            \draw[decorate, decoration={snake, segment length = 2pt, amplitude=1pt}] (4.1, -0.35) -- (4.6, -0.35);
            \filldraw[color = black, fill = magenta, thick] (4.1, -0.35) circle (1.5pt);
            \filldraw[color = black, fill = green, thick] (4.6, -0.35) circle (1.5pt);
             
         \end{tikzpicture}
         \\
         c) Quaternions ($D(Q_8)$) & $\mathsf{KW}_{\mathbb{Z}_2^3}$ & Type I + III  $\mathbb{Z}_2^3$ SPT \cite{propitiusDiscreteGaugeTheories1996} & 
         \begin{tikzpicture}[baseline=-0.1cm]
            \node at (-1.55, 0) {$m_g: $};
            \node at (0, 0){
                \begin{tikzpicture}[scale=0.5]
                    \draw[-, opacity=0.2] (0, 1) -- (0, 0) -- (0.5*1.73, -0.5) -- (1.73, 0) -- (1.5*1.73, -0.5) -- (2*1.73, 0) -- (2.5*1.73, -0.5) -- (2.5*1.73, 0.5) -- (2*1.73, 1) -- (1.5*1.73, 0.5) -- (1.73, 1) -- (0.5*1.73, 0.5) -- (0, 1);
                    \draw[-, opacity=0.2] (0.5*1.73, -0.5) -- (1.73, -1) -- (1.5*1.73, -0.5) -- (2*1.73, -1) -- (2.5*1.73, -0.5);
                    \draw[-, opacity=0.2] (0, 1) -- (0.5*1.73, 1.5) -- (1.73, 1) -- (1.5*1.73, 1.5) -- (2*1.73, 1);
                    \draw[-, opacity=0.2] (0.5*1.73, -0.5) -- (0.5*1.73, 1.5);
                    \draw[-, opacity=0.2] (1.73, -1) -- (1.73, 1);
                    \draw[-, opacity=0.2] (1.5*1.73, -0.5) -- (1.5*1.73, 1.5);
                    \draw[-, opacity=0.2] (2*1.73, -1) -- (2*1.73, 1);
                    
                    \draw[-, green, very thick] (0, 0) -- (0.5*1.73, 0.5) -- (1.73, 0) -- (1.5*1.73, 0.5) -- (2*1.73, 0) -- (2.5*1.73, 0.5);
                    \draw[-, green, very thick] (0.5*1.73, 0.5) -- (0.5*1.73, 1.5);
                    \draw[-, green, very thick] (1.73, 0) -- (1.73, -1);
                    \draw[-, green, very thick] (1.5*1.73, 0.5) -- (1.5*1.73, 1.5);
                    \draw[-, green, very thick] (2*1.73, 0) -- (2*1.73, -1);

                    \draw[decorate, decoration={snake, segment length = 2.5pt, amplitude=0.5pt}] (0, 0.5) arc[start angle=180, end angle=290, radius=0.6cm];
                    
                    \draw[decorate, decoration={snake, segment length = 2.5pt, amplitude=0.5pt}] (0, 0.5) arc[start angle=180, end angle=330, x radius=1.38cm, y radius=0.95cm];
                    
                    \draw[decorate, decoration={snake, segment length = 2.5pt, amplitude=0.5pt}] (0, 0.5) arc[start angle=180, end angle=330, x radius=2.3cm, y radius=1.2cm];

                    \draw[decorate, decoration={snake, segment length = 2.5pt, amplitude=0.5pt}] (1.73, 0.5) arc[start angle=180, end angle=290, radius=0.6cm];
                    
                    \draw[decorate, decoration={snake, segment length = 2.5pt, amplitude=0.5pt}] (1.73, 0.5) arc[start angle=180, end angle=330, x radius=1.38cm, y radius=0.95cm];
                    
                    \draw[decorate, decoration={snake, segment length = 2.5pt, amplitude=0.5pt}] (2*1.73, 0.5) arc[start angle=180, end angle=290, radius=0.6cm];

                    \filldraw[color=green, fill=green] (0.5*1.73, 1) circle (3.5pt); 
                    \filldraw[color=green, fill=green] (1.5*1.73, 1) circle (3.5pt); 
                    \filldraw[color=green, fill = green] (1.73, -0.5) circle (3.5pt); 
                    \filldraw[color=green, fill=green] (2*1.73, -0.5) circle (3.5pt);
                    \filldraw[color=black, fill=green, thick] (0.25*1.73, 0.25) circle (3.5pt); 
                    \filldraw[color=green, fill=black, thick] (0.75*1.73, 0.25) circle (3.5pt); 
                    \filldraw[color=black, fill = green, thick] (1.25*1.73, 0.25) circle (3.5pt); 
                    \filldraw[color=green, fill=black, thick] (1.75*1.73, 0.25) circle (3.5pt); 
                    \filldraw[color=black, fill = green, thick] (2.25*1.73, 0.25) circle (3.5pt);

                    \filldraw[color = black, fill = magenta, thick] (0, 0.5) circle (3pt);
                    \filldraw[color = black, fill = magenta, thick] (1.73, 0.5) circle (3pt);
                    \filldraw[color = black, fill = magenta, thick] (2*1.73, 0.5) circle (3pt);

                    \filldraw[color = black, fill = blue, thick] (0.5*1.73, 0) circle (3pt);
                    \filldraw[color = black, fill = blue, thick] (1.5*1.73, 0) circle (3pt);
                    \filldraw[color = black, fill = blue, thick] (2.5*1.73, 0) circle (3pt);
                    
                \end{tikzpicture}
             };
            \node at (1.8, 0) {$\bar{m}_g: $};
            \node at (3.35, 0){
                \begin{tikzpicture}[scale=0.5]
                    \draw[-, opacity=0.2] (0, 1) -- (0, 0) -- (0.5*1.73, -0.5) -- (1.73, 0) -- (1.5*1.73, -0.5) -- (2*1.73, 0) -- (2.5*1.73, -0.5) -- (2.5*1.73, 0.5) -- (2*1.73, 1) -- (1.5*1.73, 0.5) -- (1.73, 1) -- (0.5*1.73, 0.5) -- (0, 1);
                    \draw[-, opacity=0.2] (0.5*1.73, -0.5) -- (1.73, -1) -- (1.5*1.73, -0.5) -- (2*1.73, -1) -- (2.5*1.73, -0.5);
                    \draw[-, opacity=0.2] (0, 1) -- (0.5*1.73, 1.5) -- (1.73, 1) -- (1.5*1.73, 1.5) -- (2*1.73, 1);
                    \draw[-, opacity=0.2] (0.5*1.73, -0.5) -- (0.5*1.73, 1.5);
                    \draw[-, opacity=0.2] (1.73, -1) -- (1.73, 1);
                    \draw[-, opacity=0.2] (1.5*1.73, -0.5) -- (1.5*1.73, 1.5);
                    \draw[-, opacity=0.2] (2*1.73, -1) -- (2*1.73, 1);
                    
                    \draw[-, green, very thick] (0, 0) -- (0.5*1.73, 0.5) -- (1.73, 0) -- (1.5*1.73, 0.5) -- (2*1.73, 0) -- (2.5*1.73, 0.5);
                    \draw[-, green, very thick] (0.5*1.73, 0.5) -- (0.5*1.73, 1.5);
                    \draw[-, green, very thick] (1.73, 0) -- (1.73, -1);
                    \draw[-, green, very thick] (1.5*1.73, 0.5) -- (1.5*1.73, 1.5);
                    \draw[-, green, very thick] (2*1.73, 0) -- (2*1.73, -1);

                    \draw[decorate, decoration={snake, segment length = 2.5pt, amplitude=0.5pt}] (0, 0.5) arc[start angle=180, end angle=290, radius=0.6cm];
                    
                    \draw[decorate, decoration={snake, segment length = 2.5pt, amplitude=0.5pt}] (0, 0.5) arc[start angle=180, end angle=330, x radius=1.38cm, y radius=0.95cm];

                    \draw[decorate, decoration={snake, segment length = 2.5pt, amplitude=0.5pt}] (0, 0.5) arc[start angle=180, end angle=330, x radius=2.3cm, y radius=1.2cm];

                    \draw[decorate, decoration={snake, segment length = 2.5pt, amplitude=0.5pt}] (1.73, 0.5) arc[start angle=180, end angle=290, radius=0.6cm];

                    \draw[decorate, decoration={snake, segment length = 2.5pt, amplitude=0.5pt}] (1.73, 0.5) arc[start angle=180, end angle=330, x radius=1.38cm, y radius=0.95cm];

                    \draw[decorate, decoration={snake, segment length = 2.5pt, amplitude=0.5pt}] (2*1.73, 0.5) arc[start angle=180, end angle=290, radius=0.6cm];

                     \filldraw[color=green, fill=green] (0.5*1.73, 1) circle (3.5pt); 
                    \filldraw[color=green, fill=green] (1.5*1.73, 1) circle (3.5pt); 
                    \filldraw[color=green, fill = white, thick] (1.73, -0.5) circle (3.5pt); 
                    \filldraw[color=green, fill=white, thick] (2*1.73, -0.5) circle (3.5pt);
                    \filldraw[color=black, fill=green, thick] (0.25*1.73, 0.25) circle (3.5pt); 
                    \filldraw[color=green, fill=black, thick] (0.75*1.73, 0.25) circle (3.5pt); 
                    \filldraw[color=black, fill = green, thick] (1.25*1.73, 0.25) circle (3.5pt); 
                    \filldraw[color=green, fill=black, thick] (1.75*1.73, 0.25) circle (3.5pt); 
                    \filldraw[color=black, fill = green, thick] (2.25*1.73, 0.25) circle (3.5pt);

                    \filldraw[color = black, fill = magenta, thick] (0, 0.5) circle (3pt);
                    \filldraw[color = black, fill = magenta, thick] (1.73, 0.5) circle (3pt);
                    \filldraw[color = black, fill = magenta, thick] (2*1.73, 0.5) circle (3pt);

                    \filldraw[color = black, fill = blue, thick] (0.5*1.73, 0) circle (3pt);
                    \filldraw[color = black, fill = blue, thick] (1.5*1.73, 0) circle (3pt);
                    \filldraw[color = black, fill = blue, thick] (2.5*1.73, 0) circle (3pt);
                    
                \end{tikzpicture}
             };
             \node at (5, 0) {
                 \begin{tikzpicture}
                    \node at (0.15, 0) {\tiny $R$:};
                    \filldraw[color=green, fill=green] (0.4, 0) circle (2pt); 
                    \node at (0, -0.25) {\tiny $ZR$:};
                    \filldraw[color=green, fill=white, thick] (0.4, -0.25) circle (2pt); 
                    \node at (0, -0.55) {\tiny $XR$:};
                    \filldraw[color=black, fill=green, thick] (0.4, -0.55) circle (2pt); 
                    \node at (0, -0.85) {\tiny $RX$:};
                    \filldraw[color=green, fill=black, thick] (0.4, -0.85) circle (2pt); 
                \end{tikzpicture}
             };
             
             \end{tikzpicture}\\
         \hline 
         \hline
    \end{tabular}
    \caption{\textbf{Examples:} We list exemplar topological orders whose ribbon operators are derived using the symmetry-membrane approach; further details can be found in \cite{supplemental}. a) All $D(S_3)$ anyons can be mapped back to anyons and defects of the $\mathbb{Z}_3$ toric code, as the two theories are related by gauging charge conjugation symmetry. $D(S_3)$ ribbons are equivalent to $\mathbb{Z}_3$ ribbons decorated with $\mathsf{KW}^{\text{1D}}_{\mathbb{Z}_2}$.
    b,c) Both $D(D_4)$ and $D(Q_8)$ are isomorphic to \emph{twisted} quantum doubles \cite{Dijkgraaf:1989pz, Roche:1990hs, propitiusTopologicalInteractionsBroken1995, huTwistedQuantumDouble2013}. We illustrate flux anyon ribbons corresponding to different $\mathbb{Z}_2$ symmetries: the legend indicates the constituent gates ($R = e^{i \frac{\pi}{4} Z}$).}
    \label{tab:summary}
\end{table*}

\begin{figure}
    \centering
    \includegraphics[width=1\linewidth]{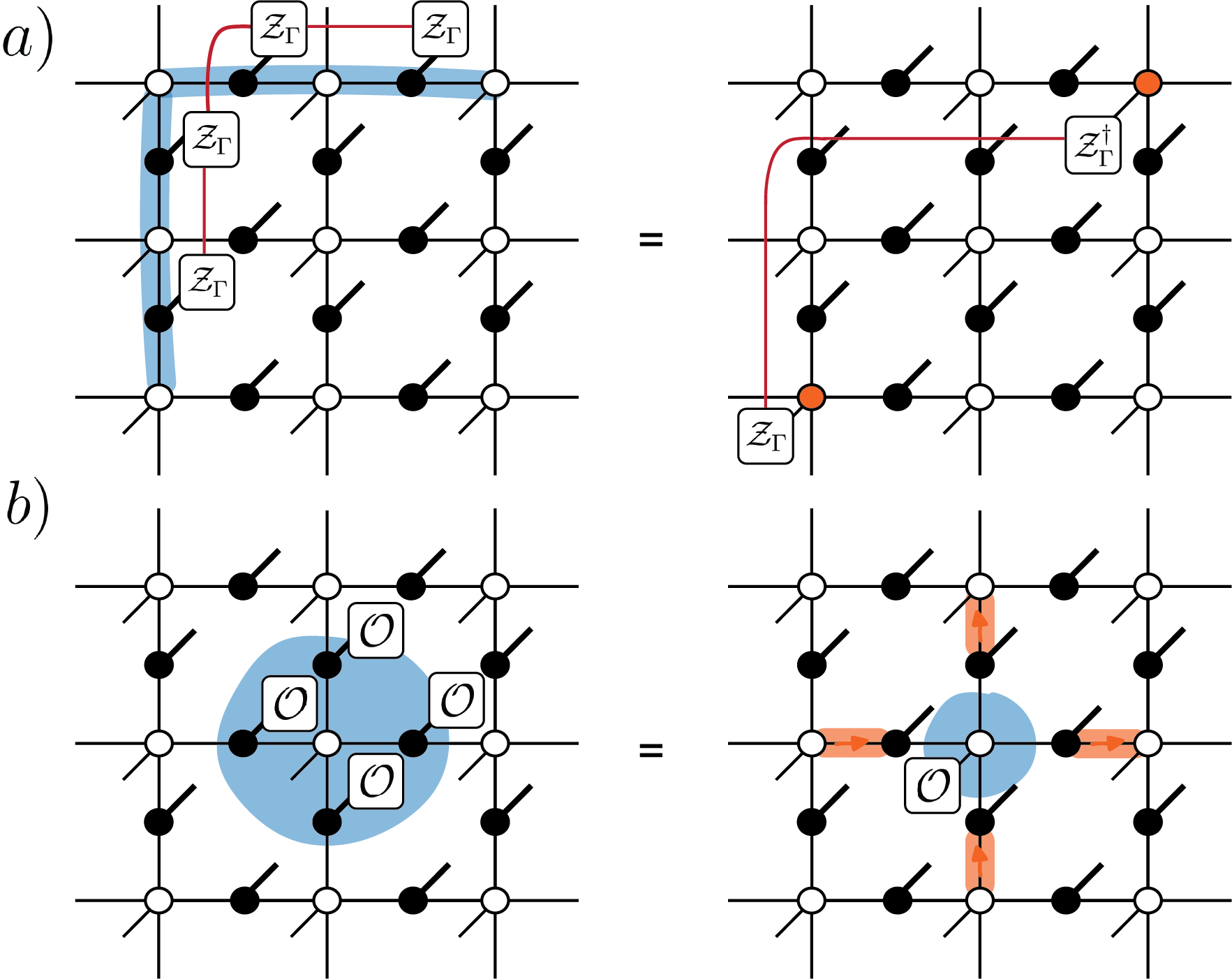}
    \caption{\textbf{Anyons and defects from the $G$-gauging map:} a) A charge anyon ribbon maps to a pair of charges on top of the paramagnet. b) The automorphism symmetry membrane on edges pushes through the gauging map the vertices, now decorated with a boundary circuit (orange lines). If the initial state is a paramagnet, the right-hand membrane disappears, leaving only the boundary circuit.}
    \label{fig:cluster-state}
\end{figure}

\emph{General $G$-gauging map:} Our protocol to construct ribbons for the $\mathbb{Z}_3$ toric code and $D(S_3)$ can be generalized to general quantum double $D(G)$ by using the finite-group $G$ gauging map $\mathsf{KW}_G$ \cite{Brell_2015, tantivasadakarnHierarchyTopologicalOrder2022,fechisin2024noninvertiblesymmetryprotectedtopologicalorder}. First, we define group-based Pauli operators for finite group $G$: 
\begin{equation}
\begin{aligned}
L_g\ket{h} = \ket{gh}, &\qquad R_g\ket{h} = \ket{h\bar g}\\
\mathcal{Z}_{\Gamma_{ij}}\ket{h} = \Gamma_{ij}(h)\ket{h}, &\qquad T_{g}\ket{h} =  \delta_{g,h}\ket{h}
\end{aligned}
\label{eq:group-valued-operators}
\end{equation}
where $g, h \in G$, $\Gamma \in \text{Rep}(G)$, and $i, j$ are indices in the virtual representation space. $\mathsf{KW}_G$ gauges the $\prod R_g$ symmetry of the $G$-paramagnet $\otimes_v \big( \sum_{g\in G} \ket{g}_v \big)$ to yield $D(G)$. It is defined by the following transformation rules:
\begin{equation}
\raisebox{-17pt}{
\begin{tikzpicture}
\node at (0,0){
    \begin{tikzpicture}[scale=0.7]
    \draw[-,opacity=0.2] (-0.6,0) -- (0.6,0);
    \draw[-,opacity=0.2] (0,-0.6) -- (0,0.6);
    \node at (0,0) {$L_g$};
    \end{tikzpicture}
};
\draw[->] (0.6,0) -- node[above,midway] {\tiny $\mathsf{KW}_{G}$} (1.1,0);
\node at (1.75, 0) {$A_g = $};
\node at (2.9,0){
    \begin{tikzpicture}[scale=0.8]
    \draw[-,opacity=0.2] (-0.6,0) -- (0.6,0);
    \draw[-,opacity=0.2] (0,-0.6) -- (0,0.6);
    \node at (0.5,0) {$L_g$};
    \node at (0,0.5) {$L_g$};
    \node at (-0.5,0) {$R_g$};
    \node at (0.05,-0.55) {$R_g$};
    
    \end{tikzpicture}
};
\node at (4.25, -0.15) {$\sum\limits_k$};
\node at (4.9,0){
    \begin{tikzpicture}
    \draw[-,opacity=0.2] (-0.2,0.3) -- (0.2,0.3);
    \draw[-,opacity=0.2] (-0.2,-0.3) -- (0.2,-0.3);
    \draw[-,opacity=0.2] (0,-0.55) -- (0,0.55);
    \node at (0,0.3) {$\mathcal{Z}^\dagger_{\Gamma^{kj}}$};
    \node at (0,-0.3) {$\mathcal{Z}_{\Gamma^{ik}}$};
    \end{tikzpicture}
};
\draw[->] (5.35,0) -- node[above,midway] {\tiny $\mathsf{KW}_{G}$} (5.85,0);
\node at (6.3,0){
    \begin{tikzpicture}
    \draw[-,opacity=0.2] (-0.2,0.3) -- (0.2,0.3);
    \draw[-,opacity=0.2] (-0.2,-0.3) -- (0.2,-0.3);
    \draw[-,opacity=0.2] (0,-0.55) -- (0,0.55);
    \node at (0,0) {$\mathcal{Z}_{\Gamma_{ij}}$};
    \end{tikzpicture}
};
\end{tikzpicture}
}
\label{eq:G-gauging-props}
\end{equation}

To create generic anyons and defects in $D(G)$, we first identify the corresponding ``pre-cursor'' excitations in the $G$-paramagnet. Charge anyons originate as charges of the $G$ paramagnet, created by $\mathcal Z^{(v')\dagger}_{\Gamma} \cdot \mathcal Z^{(v')}_{\Gamma}$ for some vertices $v,v'$---here, $(-~\cdot~-)$ indicates matrix multiplication in the virtual index space. After gauging, these charges live at the ends of open $\text{Rep}(G)$ 1-form symmetry operators (see Fig. \ref{fig:cluster-state}a). 

Flux anyons and symmetry defects are both related to automorphisms of the underlying group $G$ \footnote{Alternatively, fluxes can be traced back to charges on top of a $\text{Rep}(G)$ paramagnet. See \cite{supplemental} for more information, including a derivation of the usual flux ribbon operators using the properties of the $\text{Rep}(G)$ gauging map.}. When a flux $g$ encircles some region, it transforms any flux $h$ inside to $g h \bar{g}$. Conjugations by $g$ generate $\text{Inn}(G)$, the `inner' automorphism group. Extrinsic symmetry defects correspond to `outer' automorphisms in $\text{Out}(G) = \text{Aut}(G)/\text{Inn}(G)$, which cannot be expressed as conjugation by any $g \in G$. For example, $\text{Out}(\mathbb{Z}_3) = \{1, \mathcal{C}\} \simeq \mathbb{Z}_2$. 

Given an automorphism $\mathcal{O} \in \text{Aut}(G)$, $\mathsf{KW}_G$ maps $\prod_v \mathcal O^{(v)}$ into $\prod_e \mathcal O^{(e)}$---therefore, an automorphism symmetry membrane is a sensible operator to push through $\mathsf{KW}_G$. $\mathcal{O}$ can be decomposed in terms of Eq.~\eqref{eq:group-valued-operators}:
\begin{equation}
    \mathcal{O} = \sum_{g\in G} L_{\Omega(g)} T_g  = \sum_{g\in G} R_{\Omega(\bar{g})} T_{g},
\label{eq:outer-decomp}
\end{equation}
where $\Omega(g) \equiv \mathcal{O}(g)\bar{g}$. Combining Eqs. \eqref{eq:G-gauging-props}, \eqref{eq:outer-decomp} yields:
\begin{equation}
\raisebox{-35pt}{
\begin{tikzpicture}
\node at (0, 0) {
    \begin{tikzpicture}
        \node at (-1, 0) { \footnotesize $\mathsf{KW}_{G} ~\cdot$};
        \node at (0,0){
            \begin{tikzpicture}[scale=0.7]
                \draw[-,opacity=0.2] (-0.6,0) -- (0.6,0);
                \draw[-,opacity=0.2] (0,-0.6) -- (0,0.6);
                \node at (0,0) {$\mathcal O$};
            \end{tikzpicture}
        };
        \node at (0.9, -0.15) {$ = \sum\limits_g$};
        \node at (1.9, 0) {
            \begin{tikzpicture}[scale=0.7]
                \draw[-,opacity=0.2] (-0.6,0) -- (0.6,0);
                \draw[-,opacity=0.2] (0,-0.6) -- (0,0.6);
                \node at (0, 0) {$A_{\Omega(g)}$};
            \end{tikzpicture}
        };
        
        \node at (3, 0) {\footnotesize $\cdot ~ \mathsf{KW}_{G} ~\cdot$};
        \node at (4.2, 0) {
            \begin{tikzpicture}[scale=0.7]
            \draw[-,opacity=0.2] (-0.6,0) -- (0.6,0);
            \draw[-,opacity=0.2] (0,-0.6) -- (0,0.6);
            \node at (0,0) {$T_g$};
            \end{tikzpicture}
        };
    \end{tikzpicture}
}; 
\node at (0, -1.35) {
    \begin{tikzpicture}
        \node at (0,-1.5){
            \begin{tikzpicture}
            \draw[-,opacity=0.2] (-0.2,0.3) -- (0.2,0.3);
            \draw[-,opacity=0.2] (-0.2,-0.3) -- (0.2,-0.3);
            \draw[-,opacity=0.2] (0,-0.55) -- (0,0.55);
            \node at (0,0) {$\mathcal O$};
            \end{tikzpicture}
        };
        \node at (1.1, -1.65) {\footnotesize $\cdot~ \mathsf{KW}_{G} = \sum\limits_{g, h}$};
        \node at (2.9,-1.5){
            \begin{tikzpicture}
            \draw[-,opacity=0.2] (-0.2,0.3) -- (0.2,0.3);
            \draw[-,opacity=0.2] (-0.2,-0.3) -- (0.2,-0.3);
            \draw[-,opacity=0.2] (0,-0.55) -- (0,0.55);
            \node at (0.2, 0) {$R_{\Omega(\bar{g})} L_{\Omega(h)}$};
            \end{tikzpicture}
        };
        \node at (4.4, -1.5) {\footnotesize $\cdot~ \mathsf{KW}_{G} ~\cdot$};
        \node at (5.3, -1.5) {
            \begin{tikzpicture}
                \draw[-,opacity=0.2] (-0.2,0.3) -- (0.2,0.3);
                \draw[-,opacity=0.2] (-0.2,-0.3) -- (0.2,-0.3);
                \draw[-,opacity=0.2] (0,-0.55) -- (0,0.55);
                \node at (0, -0.3) {$T_h$};
                \node at (0, 0.3) {$T_{g}$};
            \end{tikzpicture}
        };
    \end{tikzpicture}
};
\end{tikzpicture}
}
\label{eq:outer-identities}
\end{equation}

Thus, a finite membrane of $\mathcal{O}^{(e)}$ is decorated by a boundary circuit of control gates when pushed through $\mathsf{KW}_G$ (see Fig. \ref{fig:cluster-state}b). An open defect or flux ribbon is constructed by truncating this circuit to an open line. The pre-gauged degrees of freedom acting as controls can be accessed by applying $\mathsf{KW}^{\text{1D}}_G$ along path $\gamma$, realized unitarily by $\prod_{i \in \gamma} CL^{(i \rightarrow i+1)}$, where $CL \ket{h, g} = \ket{h, hg}$.

\emph{Twisted quantum doubles:} So far, we have leveraged the duality between the trivial $G$-paramagnet and $D(G)$. Nontrivial symmetry-protected topological (SPT) phases are dual to \emph{twisted} quantum doubles \cite{Dijkgraaf:1989pz, Roche:1990hs, propitiusTopologicalInteractionsBroken1995, huTwistedQuantumDouble2013}---using this duality, we can construct ribbon operators for a broader set of topological phases. In particular, we obtain simple circuits for doubled semion order \cite{Freedman04, Levin_2005, Levin_2012}, $D(D_4)$, and $D(Q_8)$ ribbons \footnote{While the latter two examples are conventional quantum doubles for the non-Abelian groups $D_4$ and $Q_8$, they are each isomorphic to a \emph{twisted} quantum double of $\mathbb{Z}_2^3$.}---see Table.~\ref{tab:summary}. We provide detailed derivations for all three examples in the Supplementary Material \cite{supplemental}.

Here, we briefly discuss $D(D_4)$, which has recently been experimentally realized \cite{iqbalCreationNonAbelianTopological2023}. $D(D_4)$ is isomorphic to a $\mathbb{Z}_2^3$ twisted quantum double, obtained from gauging the type-III $\mathbb{Z}_2^3$ SPT \cite{propitiusTopologicalInteractionsBroken1995,Yoshida_2016}. This SPT can be written as $\prod_\triangle CCZ \ket{+}$ on the triangular lattice. The non-Abelian fluxes of $D(D_4)$ correspond to finite membranes of the $\prod X$ sublattice symmetries of the SPT---such membranes fractionalize on the SPT degrees of freedom to a product of $CZ$ gates spanning different sublattices. When we push these gates through the gauging map, they will become non-local operators, which can be written as linear-depth circuits (see Table.~\ref{tab:summary}), making the corresponding flux anyons non-Abelian. 

\emph{Discussion:} 
We have presented a method for deriving anyon and (outer automorphism) defect ribbons for $D(G)$. We obtain ribbon circuits spanning both vertices and edges by pushing precursor operators through the gauging map; applying a further 1D gauging map ensures the circuits are supported only on gauged degrees of freedom. Our protocol yields unitary expressions for anyons of generic $D(G)$---additionally, we have introduced completely new ribbons for symmetry defects. In the case where only Abelian or solvable symmetry groups are considered, our method gives a simple way to construct \emph{finite-depth} ribbons for non-Abelian excitations, by replacing all gauging steps with their efficient, measurement-based realizations. This is a powerful practical tool for constructing ribbon circuits for near-term quantum devices. 

Non-Abelian excitations provide a possible route to universal, topologically-protected quantum computing \cite{kitaevAnyonsExactlySolved2006, nayakNonAbelianAnyonsTopological2008}---as non-Abelian phases and excitations become more accessible to quantum platforms \cite{Iqbal:2023shx,andersen2023nonabelianbraidinggraphvertices,iqbalCreationNonAbelianTopological2023,Xu_2024,minev2024realizingstringnetcondensationfibonacci,z3-paper}, it is necessary to find concrete circuit realizations for the creation of any desired topological excitation. Scalability is also a key consideration given finite coherence times on near-term devices; while \emph{unitary} non-Abelian ribbon operators cannot be finite-depth \cite{Shi_2019, bravyiAdaptiveConstantdepthCircuits2022}, we provide a framework for constructing efficient \emph{non-unitary} ribbons thereby extending the work of Ref.~\onlinecite{bravyiAdaptiveConstantdepthCircuits2022}. Recently, finite-depth unitary circuits forgoing strict spatial locality have been proposed to prepare topologically ordered states \cite{sheffer2024preparingtopologicalstatesfinite}---such methods may be useful in our framework as well to develop efficient unitary ribbons. An interesting direction would be to evaluate and improve the practicality of our circuits for fault-tolerant computing.

We emphasize that our ideas are not restricted to conventional quantum doubles. Twisted quantum double ribbons can be derived by using a nontrivial SPT as input to the gauging map: symmetry fractionalization on the input SPT must also be considered. Using the higher-dimensional gauging map, we can generalize to topological orders in $d\geq2$. More exotic symmetry defects like the electromagnetic duality can also be accommodated in our framework, given a lattice realization of the symmetry.

Recent progress has been made on understanding string-net states \cite{Levin_2005} as gauge theories \cite{kawagoe2024levinwengaugetheoryentanglement,zhao2024noninvertiblegaugesymmetry21d}; constructing a string-net gauging map could allow us to generalize our results to generic string-net states. We note that unitary, linear-depth non-Abelian string operators for string-net states have been introduced in Refs.~\onlinecite{liu2022} and \onlinecite{ minev2024realizingstringnetcondensationfibonacci}---it would be interesting to compare with the ribbons obtained in our framework.

\begin{acknowledgments}
\textbf{Acknowledgements.} A.L. and C.F.B.L acknowledge support from the National Science Foundation Graduate Research Fellowship Program (NSF GRFP).
This work is in part supported by the DARPA MeasQuIT program.
 N.T. is supported by the Walter Burke Institute
for Theoretical Physics at Caltech.
A.V. and R.V. are supported by the Simons Collaboration on Ultra-Quantum Matter, which is a grant
from the Simons Foundation (618615, A.V.).
We thank the authors of Ref.~\onlinecite{Ren24} for informing us of their concurrent work, which discusses efficiently pairing up anyons using adaptive circuits.

\end{acknowledgments}

\bibliography{literature.bib}

\onecolumngrid
\clearpage
\begin{center}
\textbf{\large Supplementary Material for:\\
``Protocols for Creating Anyons and Defects via Gauging''}
\end{center}

\begin{center}
Anasuya Lyons\textsuperscript{1},
Chiu Fan Bowen Lo\textsuperscript{1},
Nathanan Tantivasadakarn\textsuperscript{2},
Ashvin Vishwanath\textsuperscript{1}, and
Ruben Verresen\textsuperscript{3,1}
\end{center}

\begin{center}
    \textsuperscript{1} \textit{Department of Physics, Harvard University, Cambridge, Massachusetts 02138, USA}\\
    \textsuperscript{2} \textit{Walter Burke Institute for Theoretical Physics and Department of \\Physics, California Institute of Technology, Pasadena, CA 91125, USA}\\
    \textsuperscript{3} \textit{Pritzker School of Molecular Engineering, University of Chicago, Chicago, IL 60637, USA}
\end{center}

\setcounter{equation}{0}
\setcounter{figure}{0}
\setcounter{table}{0}
\setcounter{secnumdepth}{2}
\makeatletter
\renewcommand{\theequation}{S\arabic{equation}}
\renewcommand{\thefigure}{S\arabic{figure}}

\tableofcontents

\section{General 2D \emph{G}-Gauging Map}
In this section, we detail the properties of the two-dimensional gauging map for a general finite group $G$ \cite{Brell_2015, tantivasadakarnHierarchyTopologicalOrder2022}. We begin by defining the group-valued Pauli operators, generalizations of the usual qubit Pauli operators to any finite group $G$. 
\begin{equation}
    \begin{aligned}
        R_g &= \sum_{h \in G} \ket{h \bar{g}}\bra{h} \\
        L_g &= \sum_{h \in G} \ket{gh}\bra{h}\\
        \mathcal{Z}_{\Gamma_{ij}} &= \sum_{g \in G} \Gamma(g)_{ij} \ket{g}\bra{g}\\
        T_h = \ket{h}\bra{h} &= \sum_{\Gamma \in \text{Rep}(G)} \frac{d_\Gamma}{\abs{G}} \sum_{i,j = 0}^{d_\Gamma} \bar{\Gamma}_{ij}(h) \mathcal{Z}_{\Gamma_{ij}}, \quad \bar{\Gamma}_{ij}(h) \equiv [\Gamma_{ij}(h)]^* 
    \end{aligned}
\end{equation}

Note that group multiplication by left or right takes the place of a Pauli ``shift'' operation, while the clock operation $\mathcal{Z}$ is extended to general higher dimensional irreducible representations. We will also need group-valued controlled-multiplication. We need to specify whether we are using the right or left action of the group, giving rise to two distinct controlled gates: 
\begin{equation}
    \begin{aligned}
        CL_{i \rightarrow j} \ket{g}_i \ket{h}_j &= \ket{g}_i\ket{gh}_j\\
        CR_{i \rightarrow j} \ket{g}_i \ket{h}_j  &= \ket{g}_i \ket{h\bar{g}}_j\\
    \end{aligned}
\end{equation}

We consider a square lattice with degrees of freedom on both edges $E$ and vertices $V$, with edge orientations as specified in Fig. \ref{fig:g-pepo} (see \cite{Brell_2015} for a generic definition of $\mathsf{KW}_G$ that works for any bipartite graph). We initialize the edges in the identity group element $\ket{I}$ (the +1 $Z$ eigenstate, analogue of $\ket{0}$ in the qubit computational basis) and vertices in the trivial irrep $\ket{[+]}$ (the +1 $X$ eigenstate, analogue of qubit $\ket{+}$). The $G$-gauging map taking vertex degrees of freedom to edges is defined in the following way: 
\begin{equation} \mathsf{KW}_G = \bigotimes_{v \in V} \bra{[+]}_{v} \left (\prod_{\langle e,v\rangle}^{\#} C(L/R)_{v \rightarrow e} \right) \bigotimes_{e\in E}\ket{I}_{e}
\end{equation}
A few things need to be clarified--- Firstly, the choice of left or right controlled-multiplication is determined by the orientation of the edge $e$. If $e$ points out of $v$, we apply $CR$, and if $e$ points into $v$, we apply $CL$. Secondly, the order \# in which the controlled gates are applied around each even site must be specified; this is because, for a non-Abelian group, the controlled gates will not generally commute on their targets. 

\subsection{PEPO representation}

\begin{figure}
    \centering
    \includegraphics[width=0.75\linewidth]{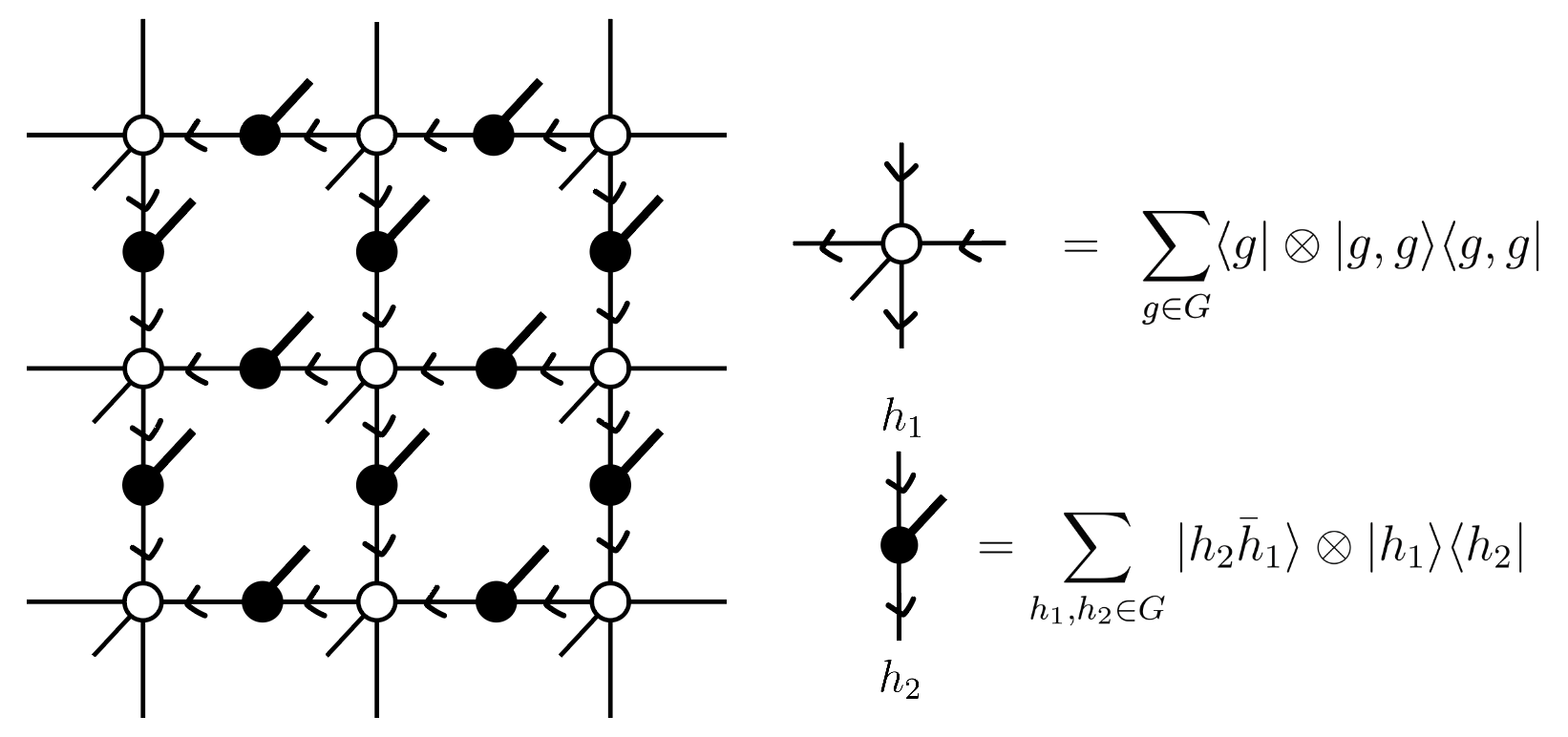}
    \caption{$\mathsf{KW}_G$ \textbf{projective entangled-pair operators:} The tensors making up the PEPO expression for $\mathsf{KW}_G$ are pictured. We also specify the orientations we will use going forwards.}
    \label{fig:g-pepo}
\end{figure}

We define the projective entangled-pair operator (PEPO) expression for $\mathsf{KW}_G$ as defined above. Here, we do not actually need to specify the order of controlled-gates applied to the edges, since the only two incident edges on each vertex are of opposite orientations, and so the control gates automatically commute. The tensors are also pictured in Fig. \ref{fig:g-pepo}.
    
The tensors defined in Fig. \ref{fig:g-pepo} have various symmetries:
\begin{equation}
\raisebox{-60pt}{
\begin{tikzpicture}
    \node at (0, 0) {
        \begin{tikzpicture}
            \node at (0.25, 0) {
            \begin{tikzpicture}[scale=0.75]
                \draw[thick, -<-, opacity=0.2] (-1.1, 0) -- (0, 0);
    	    \draw[thick, -<-, opacity=0.2] (0, 0) -- (1.1, 0);
    	    \draw[thick, -<-, opacity=0.2] (0, 0) -- (0, 1.1);
    	    \draw[thick, ->-, opacity=0.2] (0, 0) -- (0, -1.1);
    	    \draw[thick] (0, 0) -- (-0.5, -0.75);
                \filldraw[color=black, fill=white, thick] (0, 0) circle (4pt);
            \end{tikzpicture}
        };
        \node at (1.75, 0) {$=$};
        \node at (3.5, 0) {
            \begin{tikzpicture}[scale=0.75]
                \draw[thick, -<-, opacity=0.2] (-1, 0) -- (0, 0);
    	    \node [fill=white,inner sep=1pt] at (-1.1,0) {$L_g$};
    	    \draw[thick, -<-, opacity=0.2] (0, 0) -- (1, 0);
    	    \node [fill=white,inner sep=1pt] at (1.1,0) {$L^\dagger_g$};
    	    \draw[thick, -<-, opacity=0.2] (0, 0) -- (0, 1);
    	    \node [fill=white,inner sep=1pt] at (0,1.1) {$L^\dagger_g$};
    	    \draw[thick, ->-, opacity=0.2] (0, 0) -- (0, -1);
    	    \node [fill=white,inner sep=1pt] at (0,-1.1) {$L_g$};
    	    \draw[thick] (0, 0) -- (-0.5, -0.75);
                \node [fill=white,inner sep=1pt] at (-0.6, -0.85) {$L_g$};
                \filldraw[color=black, fill=white, thick] (0, 0) circle (4pt);
            \end{tikzpicture}
        };
        \node at (5.25, 0) {$=$};
        \node at (7, 0) {
            \begin{tikzpicture}[scale=0.75]
                \draw[thick, -<-, opacity=0.2] (-1, 0) -- (0, 0);
    	    \node [fill=white,inner sep=1pt] at (-1.1,0) {$R_g$};
    	    \draw[thick, -<-, opacity=0.2] (0, 0) -- (1, 0);
    	    \node [fill=white,inner sep=1pt] at (1.1,0) {$R^\dagger_g$};
    	    \draw[thick, -<-, opacity=0.2] (0, 0) -- (0, 1);
    	    \node [fill=white,inner sep=1pt] at (0,1.1) {$R^\dagger_g$};
    	    \draw[thick, ->-, opacity=0.2] (0, 0) -- (0, -1);
    	    \node [fill=white,inner sep=1pt] at (0,-1.1) {$R_g$};
    	    \draw[thick] (0, 0) -- (-0.5, -0.75);
                \node [fill=white,inner sep=1pt] at (-0.6, -0.85) {$R_g$};
                \filldraw[color=black, fill=white, thick] (0, 0) circle (4pt);
            \end{tikzpicture}
        };
        \end{tikzpicture}
    };
    \node at (0, -2.5) {
        \begin{tikzpicture}
            \node at (0.25, -2.5) {
            \begin{tikzpicture}[scale=0.75]
                \draw[thick, -<-, opacity=0.2] (-1, 0) -- (0, 0);
    	    \draw[thick, -<-, opacity=0.2] (0, 0) -- (1, 0);
    	    \draw[thick, -<-, opacity=0.2] (0, 0) -- (0, 1);
    	    \draw[thick, ->-, opacity=0.2] (0, 0) -- (0, -1);
                \draw[thick] (0, 0) -- (-0.5, -0.75);
                \node [fill=white,inner sep=1pt] at (-0.6, -0.85) {$\mathcal{Z}_{\Gamma_{ij}}$};
                \filldraw[color=black, fill=white, thick] (0, 0) circle (4pt);
            \end{tikzpicture}
        };
        \node at (8, -2.5) {$=$};
        \node at (9.5, -2.75) {
            \begin{tikzpicture}[scale=0.75]
                \draw[thick, -<-, opacity=0.2] (-1, 0) -- (0, 0);
    	    \draw[thick, -<-, opacity=0.2] (0, 0) -- (1, 0);
    	    \draw[thick, -<-, opacity=0.2] (0, 0) -- (0, 1);
    	    \draw[thick, ->-, opacity=0.2] (0, 0) -- (0, -1);
                \node [fill=white,inner sep=1pt] at (0,-1.1) {$\mathcal{Z}_{\Gamma_{ij}}$};
    	    \draw[thick] (0, 0) -- (-0.5, -0.75);
                \filldraw[color=black, fill=white, thick] (0, 0) circle (4pt);
            \end{tikzpicture}
        };
        \node at (11, -2.5) {$=$};
        \node at (12.5, -2.5) {
            \begin{tikzpicture}[scale=0.75]
                \draw[thick, -<-, opacity=0.2] (-1, 0) -- (0, 0);
    	    \node [fill=white,inner sep=1pt] at (-1.1,0) {$\mathcal{Z}_{\Gamma_{ij}}$};
    	    \draw[thick, -<-, opacity=0.2] (0, 0) -- (1, 0);
    	    \draw[thick, -<-, opacity=0.2] (0, 0) -- (0, 1);
    	    \draw[thick, ->-, opacity=0.2] (0, 0) -- (0, -1);
    	    \draw[thick] (0, 0) -- (-0.5, -0.75);
                \filldraw[color=black, fill=white, thick] (0, 0) circle (4pt);
            \end{tikzpicture}
        };
        \node at (1.75, -2.5) {$=$};
        \node at (3.5, -2.5) {
            \begin{tikzpicture}[scale=0.75]
                \draw[thick, -<-, opacity=0.2] (-1, 0) -- (0, 0);
    	    \draw[thick, -<-, opacity=0.2] (0, 0) -- (1, 0);
                \node [fill=white,inner sep=1pt] at (1.1,0) {$\mathcal{Z}_{\Gamma_{ij}}$};
    	    \draw[thick, -<-, opacity=0.2] (0, 0) -- (0, 1);
    	    \draw[thick, ->-, opacity=0.2] (0, 0) -- (0, -1);
    	    \draw[thick] (0, 0) -- (-0.5, -0.75);
                \filldraw[color=black, fill=white, thick] (0, 0) circle (4pt);
            \end{tikzpicture}
        };
        \node at (5.25, -2.5) {$=$};
        \node at (6.75, -2.5) {
            \begin{tikzpicture}[scale=0.75]
                \draw[thick, -<-, opacity=0.2] (-1, 0) -- (0, 0);
    	    \draw[thick, -<-, opacity=0.2] (0, 0) -- (1, 0);
    	    \draw[thick, -<-, opacity=0.2] (0, 0) -- (0, 1);
                \node [fill=white,inner sep=1pt] at (0,1.1) {$\mathcal{Z}_{\Gamma_{ij}}$};
    	    \draw[thick, ->-, opacity=0.2] (0, 0) -- (0, -1);
    	    \draw[thick] (0, 0) -- (-0.5, -0.75);
                \filldraw[color=black, fill=white, thick] (0, 0) circle (4pt);
            \end{tikzpicture}
        };
        \end{tikzpicture}
    };
\end{tikzpicture}
}
\label{eq:app-vertex-tensor-identities}
\end{equation}

\begin{equation}
    \raisebox{-50pt}{
    \begin{tikzpicture}
        \node at (0, 0.15) {
            \begin{tikzpicture}[scale=0.75]
                \draw[thick, ->-, opacity=0.2] (0, 1) -- (0, 0);
                \draw[thick, ->-, opacity=0.2] (0, 0) -- (0, -1);
                \draw[thick] (0, 0) -- (0.5, 0.75);
                \filldraw[black] (0, 0) circle (4pt);
            \end{tikzpicture}
        };
        \node at (0.75, 0.15) {$=$};
        \node at (1.75, 0) {
            \begin{tikzpicture}[scale=0.75]
                \draw[thick, ->-, opacity=0.2] (0, 1) -- (0, 0);
                \node [fill=white,inner sep=1pt] at (0,-1.1) {$L^\dagger_g$};
                \draw[thick, ->-, opacity=0.2] (0, 0) -- (0, -1);
                \draw[thick] (0, 0) -- (0.5, 0.75);
                \node [fill=white,inner sep=1pt] at (0.6,0.85) {$L_g$};
                \filldraw[black] (0, 0) circle (4pt);
            \end{tikzpicture}
        };
        \node at (2.5, 0.15) {$=$};
        \node at (3.5, 0.25) {
            \begin{tikzpicture}[scale=0.75]
                \draw[thick, ->-, opacity=0.2] (0, 1) -- (0, 0);
                \draw[thick, ->-, opacity=0.2] (0, 0) -- (0, -1);
                \node [fill=white,inner sep=1pt] at (0,1.1) {$L_g$};
                \draw[thick] (0, 0) -- (0.5, 0.75);
                \node [fill=white,inner sep=1pt] at (0.6,0.85) {$R_g$};
                \filldraw[black] (0, 0) circle (4pt);
            \end{tikzpicture}
        };
        \node at (4.15, 0.15) {$=$};
        \node at (4.75, 0) {\large $\sum\limits_{h}$};
        \node at (0.75+5, 0.15) {
            \begin{tikzpicture}[scale=0.75]
                \draw[thick, ->-, opacity=0.2] (0, 1) -- (0, 0);
                \draw[thick, ->-, opacity=0.2] (0, 0) -- (0, -1);
                \node [fill=white,inner sep=1pt] at (0,1.1) {$R_g T_h$};
                \draw[thick] (0, 0) -- (0.5, 0.75);
                \node [fill=white,inner sep=1pt] at (0.85, 0.55) {$R^\dagger_{hg\bar{h}}$};
                \filldraw[black] (0, 0) circle (4pt);
            \end{tikzpicture}
        };
        \node at (7, 0.15) {$=$};
        \node at (7.6, 0) {\large $\sum\limits_{h}$};
        \node at (8.5, 0) {
            \begin{tikzpicture}[scale=0.75]
                \draw[thick, ->-, opacity=0.2] (0, 1) -- (0, 0);
                \node [fill=white,inner sep=1pt] at (0,-1.1) {$R^\dagger_g T_h$};
                \draw[thick, ->-, opacity=0.2] (0, 0) -- (0, -1);
                \draw[thick] (0, 0) -- (0.5, 0.75);
                \node [fill=white,inner sep=1pt] at (0.6,0.85) {$L^\dagger_{hg\bar{h}}$};
                \filldraw[black] (0, 0) circle (4pt);
            \end{tikzpicture}
        };

        \node at (3, -2.25) {
            \begin{tikzpicture}[scale=0.75]
                \draw[thick, ->-, opacity=0.2] (0, 1) -- (0, 0);
                \draw[thick, ->-, opacity=0.2] (0, 0) -- (0, -1);
                \draw[thick] (0, 0) -- (0.5, 0.75);
                \node [fill=white,inner sep=1pt] at (0.9,0.65) {$\mathcal{Z}_{\Gamma_{ij}}$};
                \filldraw[black] (0, 0) circle (4pt);
            \end{tikzpicture}
        };
        \node at (4, -2.25) {$=$};
        \node at (5, -2.25) {\large $\sum\limits_k$};
        \node at (5.75, -2.25) {
            \begin{tikzpicture}[scale=0.75]
                \draw[thick, ->-, opacity=0.2] (0, 1) -- (0, 0);
                \node [fill=white,inner sep=1pt] at (0,1.1) {$\mathcal{Z}^\dagger_{\Gamma_{kj}}$};
                \draw[thick, ->-, opacity=0.2] (0, 0) -- (0, -1);
                \node [fill=white,inner sep=1pt] at (0,-1.1) {$\mathcal{Z}_{\Gamma_{ik}}$};
                \draw[thick] (0, 0) -- (0.5, 0.75);
                \filldraw[black] (0, 0) circle (4pt);
            \end{tikzpicture}
        };
        
    \end{tikzpicture}
    }
\label{eq:app-edge-tensor-identities}
\end{equation}
These identities can be combined to derive the defining properties of the $\mathsf{KW}_G$ map:
\begin{equation}
\raisebox{-23pt}{
\begin{tikzpicture}
\node at (0,0){
    \begin{tikzpicture}
    \draw[-,opacity=0.2] (-0.6,0) -- (0.6,0);
    \draw[-,opacity=0.2] (0,-0.6) -- (0,0.6);
    \node at (0,0) {$L_g$};
    \end{tikzpicture}
};
\draw[->] (0.85,0) -- node[above,midway] {\footnotesize $\mathsf{KW}_{G}$} (1.65,0);
\node at (2.7,0){
    \begin{tikzpicture}
    \draw[-,opacity=0.2] (-0.6,0) -- (0.6,0);
    \draw[-,opacity=0.2] (0,-0.6) -- (0,0.6);
    \node at (0.5,0) {$L_g$};
    \node at (0,0.5) {$L_g$};
    \node at (-0.5,0) {$R_g$};
    \node at (0.05,-0.55) {$R_g$};
    \end{tikzpicture}
};
\node at (4.3, -0.2) {\large $\sum\limits_k$};
\node at (5,0){
    \begin{tikzpicture}
    \draw[-,opacity=0.2] (-0.2,0.3) -- (0.2,0.3);
    \draw[-,opacity=0.2] (-0.2,-0.3) -- (0.2,-0.3);
    \draw[-,opacity=0.2] (0,-0.55) -- (0,0.55);
    \node at (0,0.3) {$\mathcal{Z}^\dagger_{\Gamma^{kj}}$};
    \node at (0,-0.3) {$\mathcal{Z}_{\Gamma^{ik}}$};
    \end{tikzpicture}
};
\draw[->] (0.85+4.1+0.5,0) -- node[above,midway] {\footnotesize $\mathsf{KW}_{G}$} (1.65+4.1+0.5,0);
\node at (6.2+0.5,0){
    \begin{tikzpicture}
    \draw[-,opacity=0.2] (-0.2,0.3) -- (0.2,0.3);
    \draw[-,opacity=0.2] (-0.2,-0.3) -- (0.2,-0.3);
    \draw[-,opacity=0.2] (0,-0.55) -- (0,0.55);
    \node at (0,0) {$\mathcal{Z}_{\Gamma_{ij}}$};
    \end{tikzpicture}
};
\end{tikzpicture}
}
\label{eq:app-G-gauging-props}
\end{equation}
The decomposition of $\mathcal{O} \in Aut(G)$ in terms of $L_g$, $R_g$, and $T_g$ operators as discussed in the main text allows the derivation of ``push-through'' relations for both inner and outer automorphisms from the vertex degrees of freedom to the edges and vice versa. 
\begin{equation}
    \begin{aligned}
        \mathcal{O} &= \sum_{g\in G} L_{\mathcal{O}(g)\bar{g}} T_g \equiv \sum_{g\in G} \mathcal{O}^L_g T_g \\
    \mathcal{O} &= \sum_{g\in G} R_{\mathcal{O}(\bar{g})g} T_{g} \equiv \sum_{g\in G} \mathcal{O}^R_{g} T_{g}
    \end{aligned}
\label{eq:out-decomp}
\end{equation}
Using Eqs. \eqref{eq:app-G-gauging-props}, \eqref{eq:out-decomp}, we find:
\begin{equation}
    \raisebox{-60pt}{
    \begin{tikzpicture}
        \node at (-0.2,0){
            \begin{tikzpicture}[scale=1.25]
                \draw[-,opacity=0.2] (-0.6,0) -- (0.6,0);
                \draw[-,opacity=0.2] (0,-0.6) -- (0,0.6);
                \node at (0,0) {$\mathcal{O}$};
            \end{tikzpicture}
        };
        \draw[->] (0.7,0) -- node[above,midway] {\footnotesize $\mathsf{KW}_{G}$} (1.5,0);
        \node at (5,0){
            \begin{tikzpicture}[scale=1.25]
                \node at (-1.4, 0) {$\sum\limits_{g \in G}$};
                \draw[-,opacity=0.2] (-0.6,0) -- (0.6,0);
                \draw[-,opacity=0.2] (0,-0.6) -- (0,0.6);
                \node at (0.7,0) {$L_{\mathcal{O}(g)\bar{g}}$};
                \node at (0,0.6) {$L_{\mathcal{O}(g)\bar{g}}$};
                \node at (0,0) {$T_g$};
                \node at (-0.6,0) {$R_{\mathcal{O}(g)\bar{g}}$};
                \node at (0.05,-0.6) {$R_{\mathcal{O}(g)\bar{g}}$};

                \node at (1.5, 0) {$=$};
                \node at (2.75, 0) {
                    \begin{tikzpicture}
                        \draw[-,opacity=0.2] (-0.6,0) -- (0.6,0);
                        \draw[-,opacity=0.2] (0,-0.6) -- (0,0.6);
                        \draw[-,opacity=0.5, thick] (-0.4,0) -- (0.4,0);
                        \draw[-,opacity=0.5, thick] (0,-0.4) -- (0,0.4);
                        \node at (0.7,0) {\footnotesize $C\mathcal{O}^L$};
                        \node at (0,0.6) {\footnotesize $C\mathcal{O}^L$};
                        \node at (-0.6,0) {\footnotesize $C\mathcal{O}^R$};
                        \node at (0.05,-0.6) {\footnotesize $C\mathcal{O}^R$};

                        \filldraw[color=black] (0, 0) circle (2pt);
                        
                    \end{tikzpicture}
                };
            \end{tikzpicture}
        };
        \node at (-0.4, -2){
            \begin{tikzpicture}[scale=1.25]
                \draw[-,opacity=0.2] (-0.2,0.3) -- (0.2,0.3);
                \draw[-,opacity=0.2] (-0.2,-0.3) -- (0.2,-0.3);
                \draw[-,opacity=0.2] (0,-0.55) -- (0,0.55);
                \node at (0,0) {$\mathcal{O}$};
            \end{tikzpicture}
        };
        \draw[->] (0.3,-2) -- node[above,midway] {\footnotesize $\mathsf{KW}_{G}$} (1.1,-2);
        \node at (3.8,-2.25) {$\sum\limits_{h_1, h_2 \in G} \ket{\mathcal{O}(h_2) \mathcal{O}(\bar{h}_1)} \otimes \ket{h_1} \bra{h_2} ~=~$};
        \node at (7.25, -2.25) {
            \begin{tikzpicture}[scale=1.5]
                \draw[-,opacity=0.2] (-0.2,0.5) -- (0.2,0.5);
                \draw[-,opacity=0.5, thick] (0,0.5) -- (0,0.3);
                \draw[-,opacity=0.2] (-0.2,-0.5) -- (0.2,-0.5);
                \draw[-,opacity=0.5, thick] (0,-0.5) -- (0,-0.3);
                \draw[-,opacity=0.2] (0,-0.65) -- (0,0.65);
                \node at (0,0.25) {\footnotesize $C\mathcal{O}^R$};
                \node at (0,-0.25) {\footnotesize $C\mathcal{O}^L$};
                \filldraw[color=black] (0, 0.5) circle (2pt);
                \filldraw[color=black] (0, -0.5) circle (2pt);
            \end{tikzpicture}
        };
    \end{tikzpicture}
    }
\end{equation}

Here, $C\mathcal{O}^L$ and $C\mathcal{O}^R$ are defined:
\begin{equation}
    \begin{aligned}
        C\mathcal{O}^L_{ij} &= \sum_{g \in G} \mathcal{O}^{L;(j)}_g T^{(i)}_g \\
        C\mathcal{O}^R_{ij} &= \sum_{g \in G} \mathcal{O}^{R;(j)}_{\bar{g}} T^{(i)}_{g} \\
    \end{aligned}
\end{equation}

These relations also follow from the symmetries of the tensors themselves under automorphisms $\mathcal{O}$: 
\begin{equation}
\raisebox{-35pt}{
    \begin{tikzpicture}
        \node at (0, 0){
            \begin{tikzpicture}[scale=0.75]
                \draw[thick, -<-, opacity=0.2] (-1, 0) -- (0, 0);
    	    
    	    \draw[thick, -<-, opacity=0.2] (0, 0) -- (1, 0);
    	    
    	    \draw[thick, -<-, opacity=0.2] (0, 0) -- (0, 1);
    	    
    	    \draw[thick, ->-, opacity=0.2] (0, 0) -- (0, -1);
    	    
    	    \draw[thick] (0, 0) -- (-0.5, -0.75);
                \node [fill=white,inner sep=1pt] at (-0.6, -0.85) {$\mathcal O$};
                \filldraw[color=black, fill=white, thick] (0, 0) circle (4pt);
            \end{tikzpicture}
        };
        \node at (1.25, 0) {$=$};
        \node at (2.5, 0){
            \begin{tikzpicture}[scale=0.75]
                \draw[thick, -<-, opacity=0.2] (-1, 0) -- (0, 0);
    	    \node [fill=white,inner sep=1pt] at (-1.1,0) {$\mathcal{O}^\dagger$};
    	    \draw[thick, -<-, opacity=0.2] (0, 0) -- (1, 0);
    	    \node [fill=white,inner sep=1pt] at (1.1,0) {$\mathcal{O}$};
    	    \draw[thick, -<-, opacity=0.2] (0, 0) -- (0, 1);
    	    \node [fill=white,inner sep=1pt] at (0,1.1) {$\mathcal{O}$};
    	    \draw[thick, ->-, opacity=0.2] (0, 0) -- (0, -1);
    	    \node [fill=white,inner sep=1pt] at (0,-1.1) {$\mathcal{O}^\dagger$};
    	    \draw[thick] (0, 0) -- (-0.5, -0.75);
                
                \filldraw[color=black, fill=white, thick] (0, 0) circle (4pt);
            \end{tikzpicture}
        };
        \node at (3.75, 0) {$,$};
        \node at (5, 0){
            \begin{tikzpicture}[scale=0.75]
                \draw[thick, ->-, opacity=0.2] (0, 1) -- (0, 0);
                \node [fill=white,inner sep=2pt] at (0,1.1) {$\mathcal{O}^\dagger$};
                \draw[thick, ->-, opacity=0.2] (0, 0) -- (0, -1);
                \draw[thick] (0, 0) -- (0.5, 0.75);
                \filldraw[black] (0, 0) circle (4pt);
            \end{tikzpicture}
        };
        \node at (5.6, 0) {$=$};
        \node at (6.5, 0){
            \begin{tikzpicture}[scale=0.75]
                \draw[thick, ->-, opacity=0.2] (0, 1) -- (0, 0);
                \filldraw[black] (0, 1) circle (2pt);

                \draw[thick, ->-, opacity=0.2] (0, 0) -- (0, -1);
                \draw[thick] (0, 0) -- (0.5, 0.75);
                \draw[thick, dashed] (0, 1) -- (0.5, 0.75);
                \node[draw = black, fill = white, inner sep = 2pt] at (0.75, 0.8) {$\mathcal{O}^R$};
                \filldraw[black] (0, 0) circle (4pt);
            \end{tikzpicture}
        };
        \node at (6.75, 0) {$,$};
        \node at (8, 0){
            \begin{tikzpicture}[scale=0.75]
                \draw[thick, ->-, opacity=0.2] (0, 1) -- (0, 0);
                \node [fill=white,inner sep=2pt] at (0,-1.1) {$\mathcal{O}$};
                \draw[thick, ->-, opacity=0.2] (0, 0) -- (0, -1);
                \draw[thick] (0, 0) -- (0.5, 0.75);
                \filldraw[black] (0, 0) circle (4pt);
            \end{tikzpicture}
        };
        \node at (8.5, 0) {$=$};
        \node at (9.5, 0){
            \begin{tikzpicture}[scale=0.75]
                \draw[thick, ->-, opacity=0.2] (0, 1) -- (0, 0);
                \filldraw[black] (0, -1) circle (2pt);
                \draw[thick, ->-, opacity=0.2] (0, 0) -- (0, -1);
                \draw[thick] (0, 0) -- (0.5, 0.75);
                \draw[thick, dashed] (0, -1) -- (0.65, 0.75);
                \node[draw = black, fill = white, inner sep = 2pt] at (0.65, 0.8) {$\mathcal{O}^L$};
                \filldraw[black] (0, 0) circle (4pt);
            \end{tikzpicture}
        };
        \node at (9.75, 0) {$,$};
        \node at (11, 0){
            \begin{tikzpicture}[scale=0.75]
                \draw[thick, ->-, opacity=0.2] (0, 1) -- (0, 0);
                \node [fill=white,inner sep=2pt] at (0,1.1) {$\mathcal{O}^\dagger$};
                \draw[thick, ->-, opacity=0.2] (0, 0) -- (0, -1);
                \node [fill=white,inner sep=2pt] at (0,-1.1) {$\mathcal{O}$};
                \draw[thick] (0, 0) -- (0.5, 0.75);
                \filldraw[black] (0, 0) circle (4pt);
            \end{tikzpicture}
        };
        \node at (11.5, 0) {$=$};
        \node at (12.25, 0){
            \begin{tikzpicture}[scale=0.75]
                \draw[thick, ->-, opacity=0.2] (0, 1) -- (0, 0);
                \draw[thick, ->-, opacity=0.2] (0, 0) -- (0, -1);
                \draw[thick] (0, 0) -- (0.5, 0.75);
                \node[fill = white, inner sep = 2pt] at (0.75, 0.8) {$\mathcal{O}$};
                \filldraw[black] (0, 0) circle (4pt);
            \end{tikzpicture}
        };
    \end{tikzpicture}
}
\end{equation}
The dotted lines indicate the control degrees of freedom for $C\mathcal{O}^{L/R}$.

\subsection{Symmetries}

We utilized local symmetries of the $\mathsf{KW}_G$ PEPO tensors to derive push-through relations for local operators in the above. Now we consider global (or higher-form) symmetries of the $\mathsf{KW}_G$ map. $\mathsf{KW}_G$ has two distinct $G$ global symmetries originating from the left and right actions of $G$: 
\begin{equation}
    \begin{aligned}
    \mathsf{KW}_G &\cdot \prod_{v}R^{(v)}_{g} = \mathsf{KW}_G\\
    \prod_{e}C^{(e)}_g &\cdot \mathsf{KW}_G \cdot \prod_{v}L^{(v)}_{g} = \mathsf{KW}_G
    \end{aligned}
\end{equation}
where $v$ runs over all vertices and $e$ runs over all edges. These symmetry operators form a representation of $G$, inherited from $L_g$ and $R_g$:
\begin{equation}
    L_g L_h = L_{gh}, \qquad R_g R_h = R_{gh}
\end{equation}
We note that the two $G$ symmetries can be combined to yield an $Inn(G)$ symmetry: 
\begin{equation}
    \prod_{e}C^{(e)}_g \cdot \mathsf{KW}_G \cdot \prod_{v}C^{(v)}_{g} = \mathsf{KW}_G
\end{equation}
There is also a $Rep(G)$ 1-form symmetry acting only on the edges:
\begin{equation}
\begin{aligned}
    \mathrm{Tr}\left(\prod_{e\in \gamma}Z^{(b)}_{\Gamma}\right) &\cdot \mathsf{KW}_G = \mathsf{KW}_G \\
    \mathrm{Tr}\left (Z^{(e)}_{\Gamma} \right ) &= \sum_{i=1}^{d_\Gamma} Z^{(e)}_{\Gamma^{ii}}
\end{aligned}
\end{equation}
where $\gamma$ is any closed path living on the edge degrees of freedom. See \cite{fechisin2024noninvertiblesymmetryprotectedtopologicalorder} for a proof that these operators have the same fusion properties as $Rep(G)$:
\begin{equation}
    \Gamma_1 \times \Gamma_2 = \sum_{\Gamma} N^{\Gamma}_{\Gamma_1, \Gamma_2} \Gamma
\end{equation}
where $N^{\Gamma}_{\Gamma_1, \Gamma_2}$ gives the number of ways $\Gamma_1, \Gamma_2$ can fuse to $\Gamma$. 
We also have an $Out(G)$ global symmetry, which acts on both edges and vertices: 
\begin{equation}
    \prod_e \mathcal{O}^{(e)} \cdot \mathsf{KW}_G \cdot \prod_v \mathcal{O}^{(v)}
\end{equation}
The group multiplication law of $Out(G)$ can be thought of as function composition; the operators $\mathcal{O}$ naturally obey the correct composition laws.

\subsection{Charge ribbons from \it{Rep(G)} 1-form}

We demonstrate that the $Rep(G)$ 1-form symmetry of $\mathsf{KW}_G$ gives exactly the form of the usual charge ribbons in the quantum double $D(G)$. Using the definition given by Bombin \cite{bombinFamilyNonAbelianKitaev2008}, a pure charge ribbon has the following form: 
\begin{equation}
    F_{\gamma}^{\Gamma; ij} = \frac{d_\Gamma}{\abs{G}} \sum_{g \in G} \bar{\Gamma}_{ij}(g) \sum_{\substack{\{h_k\} \text{ s.t. }\\ \prod_k h_k = g}} \prod_{e_k \in \gamma} T^{e_k}_{h_k}
\end{equation}
The operator $F_\gamma^{\Gamma; ij}$ creates charges corresponding to irrep $\Gamma$ at the endpoints of the path $\gamma$, with internal states specified by indices $i, j$. Using the decomposition of $T_{h_i}$ in terms of $\mathcal{Z}$ operators, we find: 
\begin{equation}
    \begin{aligned}
        F_{\gamma}^{\Gamma; ij} &= \frac{d_\Gamma}{\abs{G}} \sum_{g \in G} \bar{\Gamma}_{ij}(g) \sum_{\substack{\{h_k\} \text{ s.t. }\\ \prod_k h_k = g}} \prod_{e_k \in \gamma} \left (\sum_{\Lambda} \frac{d_\Lambda}{\abs{G}} \sum_{l, m = 1}^{d_\Lambda} \bar{\Lambda}_{lm}(h_k) \mathcal{Z}^{(e_k)}_{\Lambda_{lm}} \right)\\
        &= \frac{d_\Gamma}{\abs{G}} \sum_{g \in G} \sum_{\Lambda} \frac{d_\Lambda}{\abs{G}} \sum_{l, m = 1}^{d_\Lambda} \bar{\Gamma}_{ij}(g)   \bar{\Lambda}_{lm}(g) \left (\prod_{e_k \in \gamma} \mathcal{Z}^{(e_k)}_{\Lambda}\right )_{lm}\\
        &= \frac{d_\Gamma}{\abs{G}} \sum_{\Lambda} \frac{d_\Lambda}{\abs{G}} \sum_{l, m = 1}^{d_\Lambda}\left (\sum_{g \in G} \bar{\Gamma}_{ij}(g)   \bar{\Lambda}_{lm}(g) \right )\left (\prod_{e_k \in \gamma} \mathcal{Z}^{(e_k)}_{\Lambda}\right )_{lm}\\
        &= \frac{d_\Gamma}{\abs{G}} \sum_{\Lambda} \sum_{l, m = 1}^{d_\Lambda} \delta_{\Gamma, \bar{\Lambda}} \delta_{il} \delta_{jm}\left (\prod_{e_k \in \gamma} \mathcal{Z}^{(e_k)}_{\Lambda}\right )_{lm}\\
        &= \frac{d_\Gamma}{\abs{G}} \left (\prod_{e_k \in \Gamma} \mathcal{Z}^{(e_k)}_{\gamma}\right )_{ij}\\
    \end{aligned}
\end{equation}
where we have used the Schur orthogonality relations going from line 3 to line 4. The final expression is simply an open path version of the closed-loop $Rep(G)$ symmetry defined in the previous section, where we have enforced the internal indices of the $\mathcal{Z}$ operators to be $i, j$ at the endpoints. If we want, we can construct a closed loop ribbon which forces $i=j$, and sum over all $i$, giving exactly the form of the $Rep(G)$ higher form symmetry. So we see that charges of $D(G)$ live at the ends of open $Rep(G)$ 1-form symmetry operators. 

\subsection{1D \emph{G}-gauging map}
The unitary sequential circuit realization of the Kramers-Wannier map is given by:
\begin{equation}
    \prod_{i \in \gamma} CL_{i \rightarrow i+1}
\label{eq:1d-g-kw}
\end{equation}
This transforms operators along the 1D path $\gamma$ as follows:
\begin{equation}
    \begin{aligned}
        R^{(i)}_g L^{(i+1)}_g &\rightarrow R^{(i)}_g \\
        T^{(i+1)}_g &\rightarrow \sum_h T^{(i)}_h T^{(i+1)}_{hg}
    \end{aligned}
\end{equation}
This is analogous to the way $\mathsf{KW}^{\text{1D}}_{\mathbb{Z}_2}$ and $\mathsf{KW}^{\text{1D}}_{\mathbb{Z}_3}$ act, as discussed in the main text. By applying Eq. \eqref{eq:1d-g-kw} to our desired ribbon path $\gamma$, we can access the needed degrees of freedom for a generic $D(G)$ anyon or defect ribbon.

\section{\emph{Rep(G)} gauging map}
For Abelian groups, there is a natural isomorphism between charges and fluxes; each one-dimensional irrep can be consistently mapped onto a group element. In this case, $G \simeq Rep(G)$. Consequently, one can construct two distinct gauging maps, related by a product of Hadamard gates on the edge degrees of freedom. For example, we can construct a $\mathbb{Z}_2$ gauging map from $CZ$ gates rather than $CX$. 

For non-Abelian groups, there is no such isomorphism between $G$ and $Rep(G)$ as $Rep(G)$ will have the structure of a fusion category rather than a group. However, we can still construct a distinct gauging map for generic finite group $G$, with a different set of symmetries. In fact, it will possess a 1-form symmetry whose endpoints are fluxes of $D(G)$, rather than charges. We leave the full characterization of the symmetries of this dual gauging map to future work, but one might assume that there should be some $Rep(G)$ global symmetry whose finite membranes on the dual paramagnet correspond to closed pure charge ribbons. 

In the following, we construct this dual gauging map, which we will call $\mathsf{KW}_{Rep(G)}$, and use it to derive the exact form of the quantum double flux ribbons using its PEPO representation. We note that the Choi state for this map was first introduced by Brell \cite{Brell_2015}, who noted that partially projecting the vertices of this generalized cluster state yielded the $D(G)$ ground state. 

\subsection{PEPO representation}

\begin{figure}
    \centering
    \includegraphics[width=0.75\linewidth]{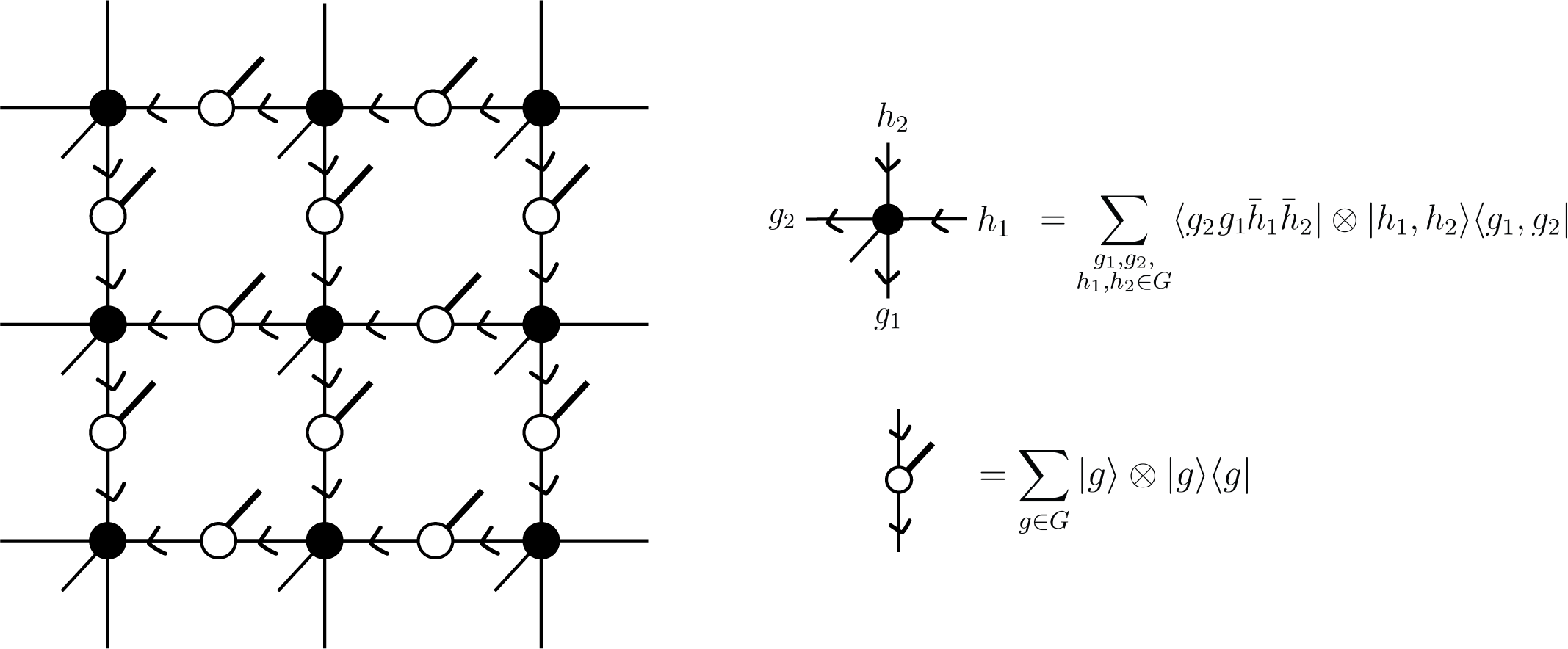}
    \caption{\textbf{Rep(G) gauging map PEPO tensors:} The tensors making up the PEPO expression for $\mathsf{KW}_{Rep(G)}$ are pictured. The structure of the vertex and edge tensors have been reversed relative to the $\mathsf{KW}_G$ tensors.}
    \label{fig:rep-g-pepo}
\end{figure}

We consider the same 2D square lattice as for $\mathsf{KW}_G$. Now, however, we initialize the edges in the trivial irreducible representation $\ket{[+]}$ and vertices in the identity group element $\ket{I}$. The $Rep(G)$-gauging map between the even and odd sublattices is defined in the following way: 
\begin{equation} 
\mathsf{KW}_{Rep(G)} = \otimes_{e\in E} \bra{[+]}_{e}\left(\prod_{\langle e,v\rangle}^{\#} C(L/R)_{e \rightarrow v} \right )  \otimes_{v\in V}\ket{I}_{v}
\label{eq:rep-g-gauging-map}
\end{equation}
Again, the choice of left or right controlled-multiplication is determined by the orientation of the edge $e$---however, the correspondence is reverse. If $e$ points out of $v$, choose \emph{left} multiplication. If the $e$ points into $v$, choose \emph{right} multiplication. See Fig. \ref{fig:rep-g-pepo} for the PEPO tensors.

We will not enumerate all symmetries of the PEPO tensors here, but we do highlight some key identities that we will use to derive the structure of the $G$-one form symmetry. 

\begin{equation}
\raisebox{-35pt}{
    \begin{tikzpicture}
        \node at (0.25, 0) {
            \begin{tikzpicture}[scale=0.75]
                \draw[thick, -<-, opacity=0.2] (-1.1, 0) -- (0, 0);
    	    \draw[thick, -<-, opacity=0.2] (0, 0) -- (1.1, 0);
    	    \draw[thick, -<-, opacity=0.2] (0, 0) -- (0, 1.1);
    	    \draw[thick, ->-, opacity=0.2] (0, 0) -- (0, -1.1);
    	    \draw[thick] (0, 0) -- (-0.5, -0.75);
                \filldraw[color=black, fill=black, thick] (0, 0) circle (4pt);
            \end{tikzpicture}
        };
        \node at (1.75, 0) {$=$};
        \node at (3.5, 0) {
            \begin{tikzpicture}[scale=0.75]
                \draw[thick, -<-, opacity=0.2] (-1, 0) -- (0, 0);
    	    \node [fill=white,inner sep=1pt] at (-1.1,0) {$L_g$};
    	    \draw[thick, -<-, opacity=0.2] (0, 0) -- (1, 0);
    	    \draw[thick, -<-, opacity=0.2] (0, 0) -- (0, 1);
    	    \draw[thick, ->-, opacity=0.2] (0, 0) -- (0, -1);
    	    \draw[thick] (0, 0) -- (-0.5, -0.75);
                \node [fill=white,inner sep=1pt] at (-0.6, -0.85) {$L_g$};
                \filldraw[color=black, fill=black, thick] (0, 0) circle (4pt);
            \end{tikzpicture}
        };
        \node at (5, 0) {$=$};
        \node at (6.25, 0) {
            \begin{tikzpicture}[scale=0.75]
                \draw[thick, -<-, opacity=0.2] (-1, 0) -- (0, 0);
    	    \draw[thick, -<-, opacity=0.2] (0, 0) -- (1, 0);
    	    \draw[thick, -<-, opacity=0.2] (0, 0) -- (0, 1);
    	    \draw[thick, ->-, opacity=0.2] (0, 0) -- (0, -1);
                \node [fill=white,inner sep=1pt] at (0, 1.1) {$L_g$};
    	    \draw[thick] (0, 0) -- (-0.5, -0.75);
                \node [fill=white,inner sep=1pt] at (-0.6, -0.85) {$R_g$};
                \filldraw[color=black, fill=black, thick] (0, 0) circle (4pt);
            \end{tikzpicture}
        };
        \node at (8.25, 0) {$=~ \sum_{h \in G}$};
        \node at (10.25, 0) {
            \begin{tikzpicture}[scale=0.75]
                \draw[thick, -<-, opacity=0.2] (-1, 0) -- (0, 0);
                \node [fill=white,inner sep=1pt] at (-1.1,0) {$T_h$};
    	    \draw[thick, -<-, opacity=0.2] (0, 0) -- (1, 0);
    	    \draw[thick, -<-, opacity=0.2] (0, 0) -- (0, 1);
    	    \draw[thick, ->-, opacity=0.2] (0, 0) -- (0, -1);
                \node [fill=white,inner sep=1pt] at (0,-1.2) {$L^\dagger_{\bar{h}gh}$};
    	    \draw[thick] (0, 0) -- (-0.5, -0.75);
                \node [fill=white,inner sep=1pt] at (-0.75, -0.6) {$L^\dagger_g$};
                \filldraw[color=black, fill=black, thick] (0, 0) circle (4pt);
            \end{tikzpicture}
        };
    \end{tikzpicture}
}
\label{eq:rep-g-vertex-tensor-identities}
\end{equation}

\begin{equation}
    \raisebox{-35pt}{
    \begin{tikzpicture}
        \node at (0, 0.15) {
            \begin{tikzpicture}[scale=0.75]
                \draw[thick, ->-, opacity=0.2] (0, 1) -- (0, 0);
                \draw[thick, ->-, opacity=0.2] (0, 0) -- (0, -1);
                \draw[thick] (0, 0) -- (0.5, 0.75);
                \filldraw[black, fill=white] (0, 0) circle (4pt);
            \end{tikzpicture}
        };
        \node at (0.75, 0.15) {$=$};
        \node at (1.75, 0) {
            \begin{tikzpicture}[scale=0.75]
                \draw[thick, ->-, opacity=0.2] (0, 1) -- (0, 0);
                \node [fill=white,inner sep=1pt] at (0,1.1) {$L_g$};
                \draw[thick, ->-, opacity=0.2] (0, 0) -- (0, -1);
                \node [fill=white,inner sep=1pt] at (0,-1.1) {$L^\dagger_g$};
                \draw[thick] (0, 0) -- (0.5, 0.75);
                \node [fill=white,inner sep=1pt] at (0.6,0.85) {$L_g$};
                \filldraw[black, fill=white] (0, 0) circle (4pt);
            \end{tikzpicture}
        };
        \node at (2.5, 0.15) {$=$};
        \node at (3.5, 0.25) {
            \begin{tikzpicture}[scale=0.75]
                \draw[thick, ->-, opacity=0.2] (0, 1) -- (0, 0);
                \draw[thick, ->-, opacity=0.2] (0, 0) -- (0, -1);
                \node [fill=white,inner sep=1pt] at (0,1.1) {$R_g$};
                \node [fill=white,inner sep=1pt] at (0,-1.1) {$R^\dagger_g$};
                \draw[thick] (0, 0) -- (0.5, 0.75);
                \node [fill=white,inner sep=1pt] at (0.6,0.85) {$R_g$};
                \filldraw[black, fill=white] (0, 0) circle (4pt);
            \end{tikzpicture}
        };
        \node at (4.15, 0.15) {$=$};
        \node at (5, 0.15) {
            \begin{tikzpicture}[scale=0.75]
                \draw[thick, ->-, opacity=0.2] (0, 1) -- (0, 0);
                \draw[thick, ->-, opacity=0.2] (0, 0) -- (0, -1);
                \node [fill=white,inner sep=1pt] at (0,1.1) {$\mathcal{Z}^\dagger_{\Gamma_{ij}}$};
                \draw[thick] (0, 0) -- (0.5, 0.75);
                \node [fill=white,inner sep=1pt] at (0.85, 0.55) {$\mathcal{Z}_{\Gamma_{ij}}$};
                \filldraw[black, fill=white] (0, 0) circle (4pt);
            \end{tikzpicture}
        };
        \node at (4.15+2, 0.15) {$=$};
        \node at (5+2, 0.15) {
            \begin{tikzpicture}[scale=0.75]
                \draw[thick, ->-, opacity=0.2] (0, 1) -- (0, 0);
                \draw[thick, ->-, opacity=0.2] (0, 0) -- (0, -1);
                \node [fill=white,inner sep=1pt] at (0,-1.1) {$\mathcal{Z}^\dagger_{\Gamma_{ij}}$};
                \draw[thick] (0, 0) -- (0.5, 0.75);
                \node [fill=white,inner sep=1pt] at (0.85, 0.55) {$\mathcal{Z}_{\Gamma_{ij}}$};
                \filldraw[black, fill=white] (0, 0) circle (4pt);
            \end{tikzpicture}
        };
        
    \end{tikzpicture}
    }
\label{eq:rep-g-edge-tensor-identities}
\end{equation}

\subsection{Flux ribbons from emergent 1-form symmetry}

We can generate flux ribbons from a starting seed of $R_g$ or $L_g$ on some vertex. By pushing this operation through onto the virtual degrees of freedom, and then up onto the edges, we can reproduce the structure of the ribbon operators originally proposed by Kitaev \cite{kitaevFaulttolerantQuantumComputation2003, bombinFamilyNonAbelianKitaev2008}. Consider applying $R^\dagger_g$ onto some vertex. Using the $Rep(G)$ PEPO tensor identities in Eqs. \eqref{eq:rep-g-vertex-tensor-identities} and \eqref{eq:rep-g-edge-tensor-identities}, we can re-write the action in the following way: 
\begin{equation}
    \begin{tikzpicture}
        \node at (0, 0) {
            \begin{tikzpicture}[scale=0.75]
                \draw[thick, -<-, opacity=0.2] (-1, 0) -- (0, 0);
    	    \draw[thick, -<-, opacity=0.2] (0, 0) -- (1, 0);
    	    \draw[thick, -<-, opacity=0.2] (0, 0) -- (0, 1);
    	    \draw[thick, ->-, opacity=0.2] (0, 0) -- (0, -1);
    	    \draw[thick] (0, 0) -- (-0.5, -0.75);
                \node [fill=white,inner sep=1pt] at (-0.6, -0.85) {$R^\dagger_g$};
                \filldraw[color=black, fill=black, thick] (0, 0) circle (4pt);
            \end{tikzpicture}
        };
        \node at (1.75, 0) {\footnotesize $= \sum_{h \in G}$};
        \node at (3.5, -0.25) {
            \begin{tikzpicture}[scale=0.75]
                \draw[thick, -<-, opacity=0.2] (-1, 0) -- (0, 0);
    	    \node [fill=white,inner sep=1pt] at (-1.1,0) {$T_h$};
    	    \draw[thick, -<-, opacity=0.2] (0, 0) -- (1, 0);
    	    \draw[thick, -<-, opacity=0.2] (0, 0) -- (0, 1);
    	    \draw[thick, ->-, opacity=0.2] (0, 0) -- (0, -1);
                \node [fill=white,inner sep=1pt] at (0, -1.4) {$L^\dagger_{\bar{h}gh}$};
    	    \draw[thick] (0, 0) -- (-0.5, -0.75);
                \node [fill=white,inner sep=1pt] at (-0.7, -0.75) {$L^\dagger_g R^\dagger_g$};
                \filldraw[color=black, fill=black, thick] (0, 0) circle (4pt);
            \end{tikzpicture}
        };
    \end{tikzpicture}
\end{equation}

We can now push the two virtual operators onto/through their neighboring edge degrees of freedom. The projector $T_h$ is a sum of $Z_{\Gamma}$ operators and so can be trivially pushed onto the edge. The $L^\dagger_{\bar{h} g h}$ operator, however, will leave some remnant action on the virtual degrees of freedom on the \emph{other side} of the edge:  
\begin{equation}
\begin{tikzpicture}[scale=0.75]
	\node at (3.75+3, 0) {\footnotesize $= \sum_{h \in G}$};
        \node at (5.75+4, 0) {
        \begin{tikzpicture}[scale=0.75]
            \draw[thick, -<-, opacity=0.2] (-2, 0) -- (0, 0);
    	\draw[thick, -<-, opacity=0.2] (-2.5, 0) -- (-2, 0);
    	\draw[thick, -<-, opacity=0.2] (0, 0) -- (1, 0);
    	\draw[thick, -<-, opacity=0.2] (0, 0) -- (0, 1);
    	\draw[thick, ->-, opacity=0.2] (0, 0) -- (0, -2);
            \draw[thick, ->-, opacity=0.2] (0, -2) -- (0, -2.5);
    	\draw[thick] (0, 0) -- (-0.5, -0.75);
            \draw[thick] (-1.8, 0) -- (-1.3, 0.75);
            \draw[thick] (0, -1.8) -- (0.5, -1.05);
            \filldraw[color=black, fill=black, thick] (0, 0) circle (4pt);
            \filldraw[color=black, fill=white, thick] (-1.8, 0) circle (4pt);
            \filldraw[color=black, fill=white, thick] (0, -1.8) circle (4pt);

            \node [fill=white,inner sep=1pt] at (-1.5, 0.75) {$T_h$};
            \node [fill=white,inner sep=1pt] at (-0.7, -0.75) {$L^\dagger_g R^\dagger_g$};
            \node [fill=white,inner sep=1pt] at (0.8, -1.2) {$L_{\bar{h}gh}$};
            \node [fill=white,inner sep=1pt] at (0, -2.6) {$L^\dagger_{\bar{h}gh}$};
            
        \end{tikzpicture}
        };
\end{tikzpicture}
\end{equation}

The final $L^\dagger_{\bar{h}gh}$ acting on the bottom-most virtual leg can be pushed onto the next vertex: 
\begin{equation}
\begin{tikzpicture}[scale=0.75]
	\node at (3.75+3, 0) {\footnotesize $= \sum_{h \in G}$};
        \node at (5.75+4, 0) {
        \begin{tikzpicture}[scale=0.75]
            \draw[thick, -<-, opacity=0.2] (-2, 0) -- (0, 0);
    	\draw[thick, -<-, opacity=0.2] (-2.5, 0) -- (-2, 0);
    	\draw[thick, -<-, opacity=0.2] (0, 0) -- (1, 0);
    	\draw[thick, -<-, opacity=0.2] (0, 0) -- (0, 1);
    	\draw[thick, ->-, opacity=0.2] (0, 0) -- (0, -2);
            \draw[thick, ->-, opacity=0.2] (0, -2) -- (0, -3.2);
    	\draw[thick] (0, 0) -- (-0.5, -0.75);
            \draw[thick] (-1.8, 0) -- (-1.3, 0.75);
            \draw[thick] (0, -1.8) -- (0.5, -1.05);
            \draw[thick] (0, -3.2) -- (-0.5, -0.75-3.2);
            \filldraw[color=black, fill=black, thick] (0, 0) circle (4pt);
             \filldraw[color=black, fill=black, thick] (0, -3.2) circle (4pt);
            \filldraw[color=black, fill=white, thick] (-1.8, 0) circle (4pt);
            \filldraw[color=black, fill=white, thick] (0, -1.8) circle (4pt);

            \node [fill=white,inner sep=1pt] at (-1.5, 0.75) {$T_h$};
            \node [fill=white,inner sep=1pt] at (-0.7, -0.75) {$L^\dagger_g R^\dagger_g$};
            \node [fill=white,inner sep=1pt] at (0.8, -1.2) {$L_{\bar{h}gh}$};
            \node [fill=white,inner sep=1pt] at (-0.5, -4) {$R^\dagger_{\bar{h}gh}$};
            
        \end{tikzpicture}
        };
\end{tikzpicture}
\end{equation}

We can now repeat the same steps with the $R^\dagger$ operator bottom vertex, extending the operator down another edge:
\begin{equation}
\begin{tikzpicture}[scale=0.75]
	\node at (3.75+3, 0.5) {\footnotesize $= \sum_{h_1, h_2  \in G}$};
        \node at (5.75+4, 0) {
        \begin{tikzpicture}[scale=0.75]
            \draw[thick, -<-, opacity=0.2] (-2, 0) -- (0, 0);
    	\draw[thick, -<-, opacity=0.2] (-2.5, 0) -- (-2, 0);
    	\draw[thick, -<-, opacity=0.2] (0, 0) -- (1, 0);
    	\draw[thick, -<-, opacity=0.2] (0, 0) -- (0, 1);
    	\draw[thick, ->-, opacity=0.2] (0, 0) -- (0, -2);
            \draw[thick, ->-, opacity=0.2] (0, -2) -- (0, -3.2);
            \draw[thick, ->-, opacity=0.2] (0, -3.2) -- (0, -5.2);
            \draw[thick, ->-, opacity=0.2] (0, -5.2) -- (0, -6);
            \draw[thick, ->-, opacity=0.2] (0, -3.2) -- (-2, -3.2);
            \draw[thick, ->-, opacity=0.2] (-2, -3.2) -- (-2.5, -3.2);
            \draw[thick, ->-, opacity=0.2] (1, -3.2) -- (0, -3.2);
    	\draw[thick] (0, 0) -- (-0.5, -0.75);
            \draw[thick] (-1.8, 0) -- (-1.3, 0.75);
            \draw[thick] (0, -1.8) -- (0.5, -1.05);
            \draw[thick] (-1.8, -3.2) -- (-1.3, -3.2+0.75);
            \draw[thick] (0, -3.2) -- (-0.5, -0.75-3.2);
            \draw[thick] (0, -4.8) -- (0.5, -4.8+0.75);
            \filldraw[color=black, fill=black, thick] (0, 0) circle (4pt);
            \filldraw[color=black, fill=black, thick] (0, -3.2) circle (4pt);
            \filldraw[color=black, fill=white, thick] (-1.8, 0) circle (4pt);
            \filldraw[color=black, fill=white, thick] (-1.8, -3.2) circle (4pt);
            \filldraw[color=black, fill=white, thick] (0, -1.8) circle (4pt);
            \filldraw[color=black, fill=white, thick] (0, -4.8) circle (4pt);

            \node [fill=white,inner sep=1pt] at (-1.5, 0.75) {$T_{h_1}$};
            \node [fill=white,inner sep=1pt] at (-0.7, -0.75) {$L^\dagger_g R^\dagger_g$};
            \node [fill=white,inner sep=1pt] at (0.8, -1.2) {$L_{\bar{h}_1gh_1}$};
            \node [fill=white,inner sep=1pt] at (-1.2, -4) {$L^\dagger _{\bar{h}_1 g h_1}R^\dagger_{\bar{h}_1gh_1}$};
             \node [fill=white,inner sep=1pt] at (-1.5, 0.75-3.2) {$T_{h_2}$};
             \node[black] at (-1.2, -4) {$L^\dagger _{\bar{h}_1 g h_1}R^\dagger_{\bar{h}_1gh_1}$};
             \node [fill=white,inner sep=1pt] at (1.4, -4.2) {$L_{\bar{h}_2\bar{h}_1gh_1h_2}$};
             \node [fill=white,inner sep=1pt] at (0.6, -6) {$L^\dagger_{\bar{h}_2\bar{h}_1gh_1h_2}$};

            \draw[thick, opacity=0.2] (0, -3.2) -- (0, -4.2);

        \end{tikzpicture}
        };
\end{tikzpicture}
\end{equation}

This process can be repeated along a closed loop to give a 1-form symmetry operator; we can also terminate the procedure to end up with a pair of flux excitations at each end. The resultant (open or closed) ribbon operators have an action on the edge degrees of freedom matching the flux ribbons originally proposed by Kitaev \cite{kitaevFaulttolerantQuantumComputation2003}. However, there is still some residual action on the vertices which prevents a complete correspondence. This can be resolved by noting all of the vertices are acted on by matching $L$ and $R$ operators; in other words, we only act on the vertices by conjugation. We are using a trivial $Rep(G)$ paramagnet as our input state, which is invariant under conjugation: $L_g R_g \ket{e} = \ket{g \bar{g}} = \ket{e}$. Thus, the action on the vertices is trivial.

\section{Internal states from non-symmetric gauging inputs}

We elaborate on a comment in the main text by demonstrating that the information about the internal state of a topological defect or anyon is captured by non-local operators that result from non-symmetric inputs to the gauging map. We focus on the particular example of charge conjugation defects in the $\mathbb{Z}_3$ toric code. 

\begin{figure}
    \centering
    \includegraphics[width=1\linewidth]{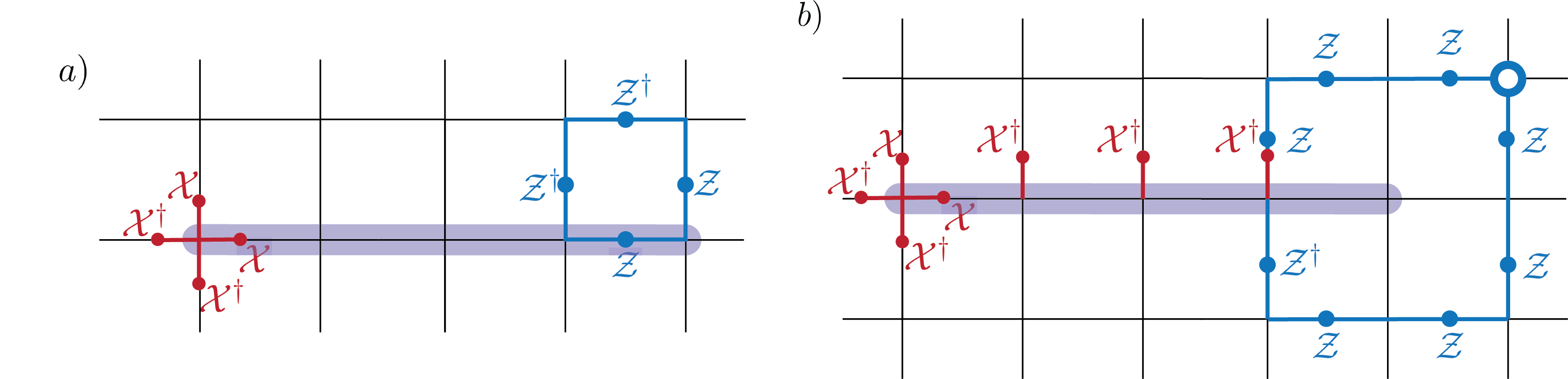}
    \caption{\textbf{Non-local stabilizers encode internal states:} a) The region where we apply the 1D $\mathbb{Z}_3$ gauging map is highlighted in purple; the indicated vertex and plaquette stabilizers will become non-local after the full gauging and un-gauging process. b) There is a nontrivial commutation relation between the nonlocal vertex stabilizer of a charge-conjugation defect and an encircling $e$ anyon. The starting point of the $e$ string is indicated by the white circle; there will be a residual $e$ anyon located there after the braiding process. The missing $\bar{e}$ needed to keep the overall state neutral is delocalized along the defect line, and the vertex stabilizer will have an corresponding eigenvalue of $\bar \omega$.}
    \label{fig:nonlocal}
\end{figure}

Recall that the charge conjugation defect ribbon circuit derived in the main text consists of (1) ungauging along a line using $\mathsf{KW}_{\mathbb{Z}_3}$, (2) applying the following finite-depth circuit between the newly recovered vertex degrees of freedom and their neighboring edges:
\begin{equation}
\raisebox{-15pt}{
\begin{tikzpicture}
    \node at (0, 0) {FDLU = $\prod\limits_{i \in \gamma} \text{C}\mathcal{X}^\dagger_{i \rightarrow e_i}~,$};
    \node at (2.25, 0) {
        \begin{tikzpicture}[scale=0.25]
            \draw[-, opacity=0.2] (0, 0) -- (2, 0);
            \draw[-, opacity=0.2] (2, 0) -- (2, 2);
            \node at (1, 0) {\footnotesize $i$};
             \node at (2, 1) {\footnotesize $e_i$};
        \end{tikzpicture}
    };
\end{tikzpicture}
}
\end{equation}
and (3) re-gauge to return to edge degrees of freedom everywhere. For this section, we will use the unitary, sequential circuit implementation of the 1D $\mathbb{Z}_3$ gauging map: \begin{equation}
    {\mathsf{KW}_{\mathbb{Z}_3}} = C\mathcal{X}_{i_n \rightarrow i_{n-1}}\cdots C\mathcal{X}_{i_1 \rightarrow i_2} C\mathcal{X}_{i_0 \rightarrow i_1}
\end{equation}
for edges $i_0, i_1, \dots, i_n$ making up an open path $\gamma$ along the lattice. In the main text, we demonstrated how the stabilizers along the bulk of this line are transformed under the charge conjugation circuit. Now we consider the stabilizers at the two endpoints, $i_0$ and $i_n$ (see Fig. \ref{fig:nonlocal}a). 
We are gauging the 1-form symmetry, which looks like a global $\prod \mathcal{Z}$ symmetry along the path $\gamma$. The vertex operator with only one edge overlapping with gamma will have a non-symmetric support $\mathcal{X}$ on $\gamma$, and so must map to something nonlocal after the first gauging step: 
\begin{equation}
\raisebox{-27pt}{
\begin{tikzpicture}
    \node at (0, 0) {
        \begin{tikzpicture}[scale=0.75]
            \draw[thick, opacity=0.2] (-1.1, 0) -- (0, 0);
            \draw[thick, opacity=0.2] (0, 0) -- (1.1, 0);
            \draw[thick, opacity=0.2] (0, 0) -- (0, 1.1);
            \draw[thick, opacity=0.2] (0, 0) -- (0, -1.1);
            \node[black] at (0.75, 0) {$\mathcal{X}^{(i_0)}$};
            \node[black] at (0, 0.5) {$\mathcal{X}$};
            \node[black] at (-0.5, 0) {$\mathcal{X}^\dagger$};
            \node[black] at (0, -0.5) {$\mathcal{X}^\dagger$};
        \end{tikzpicture}
    };
    \draw[->] (1.25,0) -- node[above,midway] {\footnotesize $\mathsf{KW}^{1D}_{\mathbb{Z}_3}$} (2.25,0);
    \node at (4, 0) {
        \begin{tikzpicture}[scale=0.75]
            \draw[thick, opacity=0.2] (-1.1, 0) -- (0, 0);
            \draw[thick, opacity=0.2] (0, 0) -- (2.1, 0);
            \draw[thick, opacity=0.2] (0, 0) -- (0, 1.1);
            \draw[thick, opacity=0.2] (0, 0) -- (0, -1.1);
            \node[black] at (0.75, 0) {$\mathcal{X}^{(i_0)}$};
            \node[black] at (1.75, 0) {$\mathcal{X}^{\dagger(i_1)}$};
            \node[black] at (0, 0.5) {$\mathcal{X}$};
            \node[black] at (-0.5, 0) {$\mathcal{X}^\dagger$};
            \node[black] at (0, -0.5) {$\mathcal{X}^\dagger$};
        \end{tikzpicture}
    };
    \node at (5.5, 0) {$\cdots$};
    \node at (7, 0) {
        \begin{tikzpicture}
            \draw[thick, opacity=0.2] (0, 0) -- (2.1, 0);
            \node[black] at (0.75, 0) {$\mathcal{X}^{(i_{n-1})}$};
            \node[black] at (1.75, 0) {$\mathcal{X}^{(i_n)}$};
        \end{tikzpicture}
    };
\end{tikzpicture}
}
\end{equation}

The plaquette operator at the end of the path $\gamma$ has support $\mathcal{Z}$ on $\gamma$, which is symmetric, and so stays local: 
\begin{equation}
\raisebox{-23pt}{
\begin{tikzpicture}
    \node at (0, 0) {
        \begin{tikzpicture}[scale=1.25]
            \draw[-,opacity=0.2] (-0.2,0.3) -- (0.9,0.3);
            \draw[-,opacity=0.2] (-0.2,-0.3) -- (0.9,-0.3);
            \draw[-,opacity=0.2] (0,-0.55) -- (0,0.55);
            \draw[-,opacity=0.2] (0.7,-0.55) -- (0.7,0.55);
            \node at (0,0) {$\mathcal Z^\dagger$};
            \node at (0.42,0.3) {$\mathcal Z^\dagger$};
            \node at (0.35,-0.3) {$\mathcal Z^{(i_n)}$};
            \node at (0.7,0) {$\mathcal Z$};
        \end{tikzpicture}
    };
    \draw[->] (1.25,0) -- node[above,midway] {\footnotesize $\mathsf{KW}^{1D}_{\mathbb{Z}_3}$} (2.25,0);
    \node at (4, 0) {
        \begin{tikzpicture}[scale=1.25]
            \draw[-,opacity=0.2, thick] (-0.2,0.3) -- (0.9,0.3);
            \draw[-,opacity=0.2, thick] (-1.1,-0.3) -- (0.9,-0.3);
            \draw[-,opacity=0.2, thick] (0,-0.55) -- (0,0.55);
            \draw[-,opacity=0.2, thick] (0.7,-0.55) -- (0.7,0.55);
            \node at (0,0) {$\mathcal Z^\dagger$};
            \node at (0.42,0.3) {$\mathcal Z^\dagger$};
            \node at (0.35,-0.3) {$\mathcal Z^{(i_n)}$};
            \node at (-0.5,-0.3) {$\mathcal Z^{\dagger (i_{n-1})}$};
            \node at (0.7,0) {$\mathcal Z$};
        \end{tikzpicture}
    };
\end{tikzpicture}
}
\end{equation}
We apply the finite-depth circuit between the newly gauged degrees of freedom and the vertical edges just above: 
\begin{equation}
    \raisebox{-50pt}{
    \begin{tikzpicture}
        \node at (0, 0) {
            \begin{tikzpicture}
                \node at (0, 0) {
                    \begin{tikzpicture}[scale=0.75]
                        \draw[thick, opacity=0.2] (-1.1, 0) -- (0, 0);
                        \draw[thick, opacity=0.2] (0, 0) -- (2.1, 0);
                        \draw[thick, opacity=0.2] (0, 0) -- (0, 1.1);
                        \draw[thick, opacity=0.2] (0, 0) -- (0, -1.1);
                        \node[black] at (0.75, 0) {$\mathcal{X}^{(i_0)}$};
                        \node[black] at (1.75, 0) {$\mathcal{X}^{(i_1)}$};
                        \node[black] at (0, 0.5) {$\mathcal{X}$};
                        \node[black] at (-0.5, 0) {$\mathcal{X}^\dagger$};
                        \node[black] at (0, -0.5) {$\mathcal{X}^\dagger$};
                    \end{tikzpicture}
                };
                \node at (1.5, 0) {$\cdots$};
                \node at (3, 0) {
                    \begin{tikzpicture}
                        \draw[thick, opacity=0.2] (0, 0) -- (2.1, 0);
                        \node[black] at (0.75, 0) {$\mathcal{X}^{(i_{n-1})}$};
                        \node[black] at (1.75, 0) {$\mathcal{X}^{(i_n)}$};
                    \end{tikzpicture}
                };
            \end{tikzpicture}
        };
        \draw[->] (3,0) -- node[above,midway] {\footnotesize FDLU} (4.25,0);
        \node at (7.5, 0) {
            \begin{tikzpicture}[scale=0.75]
                \draw[thick, opacity=0.2] (-1.1, 0) -- (0, 0);
                \draw[thick, opacity=0.2] (0, 0) -- (2.1, 0);
                \draw[thick, opacity=0.2] (0, 0) -- (0, 1.1);
                \draw[thick, opacity=0.2] (0, 0) -- (0, -1.1);
                \draw[thick, opacity=0.2] (1, 1.1) -- (1, -0.1);
                \draw[thick, opacity=0.2] (2, 1.1) -- (2, -0.1);
                
                \node[black] at (0.75, 0) {$\mathcal{X}^{(i_0)}$};
                \node[black] at (1.75, 0) {$\mathcal{X}^{(i_1)}$};
                \node[black] at (0, 0.5) {$\mathcal{X}$};
                \node[black] at (-0.5, 0) {$\mathcal{X}^\dagger$};
                \node[black] at (0, -0.5) {$\mathcal{X}^\dagger$};
                \node[black] at (1, 0.5) {$\mathcal{X}^\dagger$};
                \node[black] at (2, 0.5) {$\mathcal{X}^\dagger$};

                \node at (2.75, 0) {$\cdots$};
                \node at (4.5, 0.355) {
                    \begin{tikzpicture}[scale=0.75]
                        \draw[thick, opacity=0.2] (0, 0) -- (2.5, 0);
                        \node[black] at (0.75, 0) {$\mathcal{X}^{(i_{n-1})}$};
                        \node[black] at (2, 0) {$\mathcal{X}^{(i_n)}$};
                        \draw[thick, opacity=0.2] (1, 1.1) -- (1, -0.1);
                        \node[black] at (1, 0.5) {$\mathcal{X}^\dagger$};
                        
                    \end{tikzpicture}
                };
            \end{tikzpicture}
        };
        \node at (0, -2.2) {
            \begin{tikzpicture}
                \begin{tikzpicture}[scale=1.25]
                    \draw[-,opacity=0.2, thick] (-0.2,0.3) -- (0.9,0.3);
                    \draw[-,opacity=0.2, thick] (-1.1,-0.3) -- (0.9,-0.3);
                    \draw[-,opacity=0.2, thick] (0,-0.55) -- (0,0.55);
                    \draw[-,opacity=0.2, thick] (0.7,-0.55) -- (0.7,0.55);
                    \node at (0,0) {$\mathcal Z^\dagger$};
                    \node at (0.42,0.3) {$\mathcal Z^\dagger$};
                    \node at (0.35,-0.3) {$\mathcal Z^{(i_n)}$};
                    \node at (-0.5,-0.3) {$\mathcal Z^{\dagger (i_{n-1})}$};
                    \node at (0.7,0) {$\mathcal Z$};
                \end{tikzpicture}
            \end{tikzpicture}
        };
        \draw[->] (3,-1.7) -- node[above,midway] {\footnotesize FDLU} (4.25,-1.7);
        \node at (6, -1.5) {
            \begin{tikzpicture}[scale=1.25]
                \draw[-,opacity=0.2, thick] (-0.2,0.3) -- (0.9,0.3);
                \draw[-,opacity=0.2, thick] (-1.1,-0.3) -- (0.9,-0.3);
                \draw[-,opacity=0.2, thick] (0,-0.55) -- (0,0.55);
                \draw[-,opacity=0.2, thick] (0.7,-0.55) -- (0.7,0.55);
                \node at (0,0) {$\mathcal Z^\dagger$};
                \node at (0.42,0.3) {$\mathcal Z^\dagger$};
                \node at (0.35,-0.3) {$\mathcal Z^{(i_n)}$};
                \node at (-0.5,-0.3) {$\mathcal Z^{(i_{n-1})}$};
                \node at (0.7,0) {$\mathcal Z$};
            \end{tikzpicture}
        };
        
    \end{tikzpicture}
    }
\end{equation}
The plaquette term is no longer symmetric under the dual $\prod \mathcal{X}$ symmetry (this is the new global symmetry of the system after the first gauging step); upon undoing the gauging, it will also become nonlocal. 
\begin{equation}
    \raisebox{-50pt}{
    \begin{tikzpicture}
        \node at (0, 0) {
            \begin{tikzpicture}[scale=0.75]
                \draw[thick, opacity=0.2] (-1.1, 0) -- (0, 0);
                \draw[thick, opacity=0.2] (0, 0) -- (2.1, 0);
                \draw[thick, opacity=0.2] (0, 0) -- (0, 1.1);
                \draw[thick, opacity=0.2] (0, 0) -- (0, -1.1);
                \draw[thick, opacity=0.2] (1, 1.1) -- (1, -0.1);
                \draw[thick, opacity=0.2] (2, 1.1) -- (2, -0.1);
                
                \node[black] at (0.75, 0) {$\mathcal{X}^{(i_0)}$};
                \node[black] at (1.75, 0) {$\mathcal{X}^{(i_1)}$};
                \node[black] at (0, 0.5) {$\mathcal{X}$};
                \node[black] at (-0.5, 0) {$\mathcal{X}^\dagger$};
                \node[black] at (0, -0.5) {$\mathcal{X}^\dagger$};
                \node[black] at (1, 0.5) {$\mathcal{X}^\dagger$};
                \node[black] at (2, 0.5) {$\mathcal{X}^\dagger$};

                \node at (2.75, 0) {$\cdots$};
                \node at (4.5, 0.355) {
                    \begin{tikzpicture}[scale=0.75]
                        \draw[thick, opacity=0.2] (0, 0) -- (2.5, 0);
                        \node[black] at (0.75, 0) {$\mathcal{X}^{(i_{n-1})}$};
                        \node[black] at (2, 0) {$\mathcal{X}^{(i_n)}$};
                        \draw[thick, opacity=0.2] (1, 1.1) -- (1, -0.1);
                        \node[black] at (1, 0.5) {$\mathcal{X}^\dagger$};
                    \end{tikzpicture}
                };
            \end{tikzpicture}
        };
        \draw[->] (3,0) -- node[above,midway] {\footnotesize $\mathsf{KW}^\dagger_{\mathbb{Z}_3}$} (4.25,0);
        \node at (7.25, 0) {
            \begin{tikzpicture}[scale=0.75]
                \draw[thick, opacity=0.2] (-1.1, 0) -- (0, 0);
                \draw[thick, opacity=0.2] (0, 0) -- (2.1, 0);
                \draw[thick, opacity=0.2] (0, 0) -- (0, 1.1);
                \draw[thick, opacity=0.2] (0, 0) -- (0, -1.1);
                \draw[thick, opacity=0.2] (1, 1.1) -- (1, -0.1);
                \draw[thick, opacity=0.2] (2, 1.1) -- (2, -0.1);
                
                \node[black] at (0.75, 0) {$\mathcal{X}^{(i_0)}$};
                \node[black] at (0, 0.5) {$\mathcal{X}$};
                \node[black] at (-0.5, 0) {$\mathcal{X}^\dagger$};
                \node[black] at (0, -0.5) {$\mathcal{X}^\dagger$};
                \node[black] at (1, 0.5) {$\mathcal{X}^\dagger$};
                \node[black] at (2, 0.5) {$\mathcal{X}^\dagger$};

                \node at (2.75, 0) {$\cdots$};
                \node at (4.25, 0.5) {
                    \begin{tikzpicture}[scale=0.75]
                        \draw[thick, opacity=0.2] (0, 0) -- (2.1, 0);
                        \draw[thick, opacity=0.2] (1, 1.1) -- (1, -0.1);
                        \node[black] at (1, 0.5) {$\mathcal{X}^\dagger$};
                    \end{tikzpicture}
                };
            \end{tikzpicture}
        };

        \node at (1.25, -2.2) {
            \begin{tikzpicture}[scale=1.25]
                \draw[-,opacity=0.2, thick] (-0.2,0.3) -- (0.9,0.3);
                \draw[-,opacity=0.2, thick] (-1.1,-0.3) -- (0.9,-0.3);
                \draw[-,opacity=0.2, thick] (0,-0.55) -- (0,0.55);
                \draw[-,opacity=0.2, thick] (0.7,-0.55) -- (0.7,0.55);
                \node at (0,0) {$\mathcal Z^\dagger$};
                \node at (0.42,0.3) {$\mathcal Z^\dagger$};
                \node at (0.35,-0.3) {$\mathcal Z^{(i_n)}$};
                \node at (-0.5,-0.3) {$\mathcal Z^{(i_{n-1})}$};
                \node at (0.7,0) {$\mathcal Z$};
            \end{tikzpicture}
        };
        \draw[->] (3,-2.2) -- node[above,midway] {\footnotesize $\mathsf{KW}^\dagger_{\mathbb{Z}_3}$} (4.25,-2.2);
        \node at (7.5, -2.2) {
            \begin{tikzpicture}[scale=1.25]
                \draw[-,opacity=0.2, thick] (-0.2,0.3) -- (0.9,0.3);
                \draw[-,opacity=0.2, thick] (-2.1,-0.3) -- (0.9,-0.3);
                \draw[-,opacity=0.2, thick] (0,-0.55) -- (0,0.55);
                \draw[-,opacity=0.2, thick] (0.7,-0.55) -- (0.7,0.55);
                \node at (-2.35, -0.3) {$\cdots$};
                \node at (0,0) {$\mathcal Z^\dagger$};
                \node at (0.42,0.3) {$\mathcal Z^\dagger$};
                \node at (0.35,-0.3) {$\mathcal Z^{(i_n)}$};
                \node at (-0.5,-0.3) {$\mathcal Z^{\dagger(i_{n-1})}$};
                \node at (-1.5,-0.3) {$\mathcal Z^{\dagger(i_{n-2})}$};
                \node at (-3,-0.3) {$\mathcal Z^{\dagger(i_{0})}$};
                \node at (0.7,0) {$\mathcal Z$};

                \draw[-,opacity=0.2, thick] (-2.6,-0.3) -- (-3.6,-0.3);
            \end{tikzpicture}
        };
    \end{tikzpicture}
        
    }
\end{equation}

How do these nonlocal stabilizers capture the internal states of the defect? (1) The number of eigenvalues of these stabilizers matches the expected quantum dimension $d = 3$ of the charge conjugation defect, and (2) these nonlocal stabilizers can track the braiding history of the charge conjugation defects, which is exactly the information stored in the internal state. 

Consider a process where an $e$ anyon braids nontrivially with a charge conjugation defect. This requires that the $e$ anyon crosses the charge conjugation defect line. When the $e$ anyon string operator crosses the defect line, it will excite the nonlocal vertex stabilizer (see Fig. \ref{fig:nonlocal}b). If we fuse the $e$ anyon (which has been transmuted to an $\bar{e}$) with its partner $\bar{e}$, there will now be a residual $e$ in the system. The nonlocal defect vertex now contains the information about the missing $\bar{e}$. The nonlocal plaquette plays an identical role for any $m$ anyons that braid with the charge conjugation defect.  

\section{\emph{D(S3)} Ribbon Operators}

\begin{table}
    \centering
    \begin{tabular}{ccccc}
    \hline 
         $D(S_3)$ \textbf{anyon} & \textbf{Conjugacy class} & \textbf{Irrep} & \textbf{Dimension} &  $\mathbb{Z}_3$ \textbf{excitations}\\
    \hline 
    \hline 
         $[1]$ & $\{1\}$ & $[+]$ & $d=1$ & vaccuum \\
         $[-]$ charge & $\{1\}$ & $[-]$ & $d=1$ & c.c. charge \\
         $[2]$ charge & $\{1\}$ & $[2]$ & $d = 2$ &$\{e, \bar{e}\}$ \\
         $[C_3]$ & $\{\mu, \bar{\mu}\} $ & $[+]$ & $d=2$ & $\{m, \bar{m}\}$ \\
         $[C_3, \omega]$ & $\{\mu, \bar{\mu}\} $ & $[\omega]$ & $d=2$ & $\{em, \bar{e}\bar{m}\}$ \\
         $[C_3, \bar{\omega}]$ & $\{\mu, \bar{\mu}\} $ & $[\bar{\omega}]$ & $d=2$ & $\{\bar{e}m, e\bar{m}\}$ \\
         $[C_2]$ & $\{\sigma, \mu \sigma, \bar{\mu} \sigma\}$  & $[+]$ & $d=3$ & c.c. defect \\
         $[C_2, -]$ & $\{\sigma, \mu \sigma, \bar{\mu} \sigma\}$  & $[-]$ & $d=3$ & c.c. defect $\times$ cc. charge \\
    \hline
    \end{tabular}
    \caption{\textbf{Correspondence between $S_3$ and $\mathbb{Z}_3$ excitations}: The anyonic excitations of $D(S_3)$ along with their properties are listed to the left. The final column indicates the precursor excitations in $\mathbb{Z}_3$ for each anyon. Any $\mathbb{Z}_3$ excitations related by the charge conjugation symmetry are combined into one non-Abelian anyon in $D(S_3)$; each precursor excitation becomes a basis state in the internal Hilbert space of the non-Abelian anyon. }
    \label{tab:s3-anyon-correspondence}
\end{table}

As discussed in the main text, we can construct ribbon operators for $D(S_3)$ anyons by combining the $\mathbb{Z}_2$ gauging map and the $\mathbb{Z}_3$ toric code ribbons. This is because $S_3$ is obtained from $\mathbb{Z}_3$ by promoting the charge conjugation outer automorphism $\sigma \in Out(\mathbb{Z}_3)$ to a group element in its own right: 
\begin{equation}
    S_3 = \langle \mu^a \sigma^b \vert \mu^3 = 1, \sigma^2 = 1, \mu \sigma = \sigma \bar{\mu} \rangle
\end{equation}
Here $\mu$ is the generator of $\mathbb{Z}_3$. This choice of group presentation for $S_3$ allows us to represent each group element as a qutrit and a qubit in some state:
\begin{equation}
    \ket{\mu^a \sigma^b \in S_3} = \ket{a \in \mathbb{Z}_3} \ket{b \in \mathbb{Z}_2}
\end{equation}
Physically, we can obtain $D(S_3)$ from the $\mathbb{Z}_3$ toric code by gauging the charge conjugation symmetry \cite{verresenEfficientlyPreparingSchr2022}. Each anyon in $D(S_3)$ has a precursor anyon or defect in the $\mathbb{Z}_3$ toric code (see Table. \ref{tab:s3-anyon-correspondence}). Thus, we can create any $D(S_3)$ anyon by: (1) ungauging the $\mathbb{Z}_2$ charge conjugation along a line, (2) creating the corresponding precursor excitation in the uncovered $\mathbb{Z}_3$ toric code, (3) re-gauging charge conjugation to return to $D(S_3)$ everywhere. We can re-write the known ribbon operators for $D(S_3)$ \cite{kitaevFaulttolerantQuantumComputation2003, bombinFamilyNonAbelianKitaev2008} in such a way that this structure is apparent. For instance, the ribbon operator for the $d=2$ non-Abelian flux $[C_3] \in D(S_3)$ can be written as a circuit in the following way:
\begin{equation}
\raisebox{-70pt}{
    \begin{tikzpicture}
        \node at (0, 0) {
            \begin{tikzpicture}[scale=0.5]
                \draw[thick] (-6,0) -- node[below]{$b_1$} (-4,0) ;
                \draw[thick] (-4, 0) -- node[right]{$\alpha_1$} (-4, 2);
                \draw[thick] (-4,0) -- node[below]{$b_2$} (-2,0);
                \draw[thick] (-2, 0) -- node[right]{$\alpha_2$} (-2, 2);
                \draw[thick] (-2,0) -- node[below]{$b_3$} (0,0);
                \draw[thick] (0, 0) -- node[right]{$\alpha_3$} (0, 2);
                \draw[thick] (0,0) -- node[below]{$b_4$} (2,0);
                \draw[thick] (2, 0) -- node[right]{$\alpha_4$} (2, 2);
            \end{tikzpicture}
        };
        \node at (6, 0) {
            \begin{tikzpicture}

                \begin{yquant}[vertical]
                qubit {$\ket{\alpha_1}$} a1;
                qubit {$\ket{b_1}$} b1;
                qubit {$\ket{\alpha_2}$} a2;
                qubit {$\ket{b_2}$} b2;
                qubit {$\ket{\alpha_3}$} a3;
                qubit {$\ket{b_3}$} b3;
                qubit {$\ket{\alpha_4}$} a4;
                qubit {$\ket{b_4}$} b4;
                
                cnot b2 | b1;
                cnot b3 | b2;
                cnot b4 | b3;

                barrier (a1, a2, a3, a4, b1, b2, b3, b4);
                
                align a2, a3, a4;
                box {$\mathcal{C}$} a2 | b1;
                box {$\mathcal{C}$} a3 | b2;
                box {$\mathcal{C}$} a4 | b3;
                
                barrier (a1, a2, a3, a4, b1, b2, b3, b4);
                
                align a1, a2, a3, a4;
                box {$\mathcal{X}$} a1; 
                box {$\mathcal{X}$} a2; 
                box {$\mathcal{X}$} a3; 
                box {$\mathcal{X}$} a4; 
                
                barrier (a1, a2, a3, a4, b1, b2, b3, b4);
                
                box {$\mathcal{C}$} a4 | b3;
                box {$\mathcal{C}$} a3 | b2;
                box {$\mathcal{C}$} a2 | b1;

                barrier (a1, a2, a3, a4, b1, b2, b3, b4);
                
                cnot b4 | b3;
                cnot b3 | b2;
                cnot b2 | b1;
                
                \end{yquant}
                
            \end{tikzpicture}
        };
        
        \node at (9, 2) {\Huge $\left. \right \}$};
        \node at (10.25, 2) {\footnotesize $\mathbb{Z}_2$ ungauging};
        \node at (9, 0.74) {\huge $\left. \right \}$};
        \node[align=left] at (10.5, 0.74) {\footnotesize undoing symmetry\\ \footnotesize enrichment};
        \node at (9, -0.24) {\huge $\left. \right \}$};
        \node at (10.25, -0.24) {\footnotesize $\mathbb{Z}_3$ $m$ string};
        \node at (9, -1.24) {\huge $\left. \right \}$};
        \node[align=left] at (10.5, -1.24) {\footnotesize redoing symmetry\\ \footnotesize enrichment};
        \node at (9, -2.5) {\Huge $\left. \right \}$};
        \node at (10.25, -2.5) {\footnotesize $\mathbb{Z}_2$ gauging};
    \end{tikzpicture}
}
\end{equation}
The support of the ribbon operator is indicated on the left; $\alpha_i$ are the qutrit degrees of freedom on the vertical links, while $b_i$ are the qubit degrees of freedom on the horizontal edges. The ribbon circuit consists of several steps: the first is gauging the $\mathbb{Z}_2$ 1-form symmetry generated by the sign representation $[-]$ (in other words, un-gauging the $\mathbb{Z}_2$ part of the topological order). This leaves us with a small strip of symmetry-enriched $\mathbb{Z}_3$ toric code \cite{verresenEfficientlyPreparingSchr2022}; to return to the usual $\mathbb{Z}_3$ toric code we apply a finite-depth circuit undoing the symmetry-enrichment. We then apply the corresponding Abelian anyon string, in this case an $m$ string. Then we redo the symmetry-enrichment and undo the 1-form gauging, returning to $D(S_3)$ fully. 

In the case of the $d=3$ non-Abelian flux, $[C_2]$, its precursor excitation is the non-Abelian charge conjugation defect. Thus, we must add an extra step; we return to the trivial $\mathbb{Z}_3$ paramagnet to create the precursor excitation to the charge conjugation defect. Again, the known ribbon operators for the $[C_2]$ anyon have this structure built in: 
\begin{equation}
    \raisebox{-150pt}{
        \begin{tikzpicture}
            \node at (2, 6) {
                \begin{tikzpicture}[scale=0.5]
                    \draw[red, thick] (-6,0) -- node[below]{$a_1, b_1$} (-4,0) ;
                    \draw[blue, thick] (-4, 0) -- node[right]{\footnotesize $\alpha_1, \beta_1$} (-4, 2);
                    \draw[red, thick] (-4,0) -- node[below]{$a_2, b_2$} (-2,0);
                    \draw[blue, thick] (-2, 0) -- node[right]{\footnotesize $\alpha_2, \beta_2$} (-2, 2);
                    \draw[red, thick] (-2,0) -- node[below]{$a_3, b_3$} (0,0);
                    \draw[blue, thick] (0, 0) -- node[right]{\footnotesize $\alpha_3, \beta_3$} (0, 2);
                    \draw[red, thick] (0,0) -- node[below]{$a_4, b_4$} (2,0);
                    \draw[blue, thick] (2, 0) -- node[right]{\footnotesize $\alpha_4, \beta_4$} (2, 2);
                \end{tikzpicture}
            };
            \node at (0, 0) {
                \begin{tikzpicture}[scale=0.8]
                    \begin{yquant}[vertical]
                    [red]
                    qubit {$\ket{a_1}$} a[1];
                    [red]
                    qubit {$\ket{b_1}$} b[1];
                    [blue]
                    qubit {$\ket{\alpha_1}$} alp[1];
                    [blue]
                    qubit {$\ket{\beta_1}$} beta[1];
                    
                    [red]
                    qubit {$\ket{a_2}$} a[+1];
                    [red]
                    qubit {$\ket{b_2}$} b[+1];
                    [blue]
                    qubit {$\ket{\alpha_2}$} alp[+1];
                    [blue]
                    qubit {$\ket{\beta_2}$} beta[+1];
                    
                    [red]
                    qubit {$\ket{a_3}$} a[+1];
                    [red]
                    qubit {$\ket{b_3}$} b[+1];
                    [blue]
                    qubit {$\ket{\alpha_3}$} alp[+1];
                    [blue]
                    qubit {$\ket{\beta_3}$} beta[+1];
                    
                    [red]
                    qubit {$\ket{a_4}$} a[+1];
                    [red]
                    qubit {$\ket{b_4}$} b[+1];
                    [blue]
                    qubit {$\ket{\alpha_4}$} alp[+1];
                    [blue]
                    qubit {$\ket{\beta_4}$} beta[+1];
                    
                    cnot b[1] | b[0];
                    cnot b[2] | b[1];
                    cnot b[3] | b[2];
                    
                    align a[1], a[2], a[3];
                    box {$\mathcal{C}$} a[1] | b[0];
                    box {$\mathcal{C}$} a[2] | b[1];
                    box {$\mathcal{C}$} a[3] | b[2];
                    
                    barrier (a, b, alp, beta);
                    
                    box {$\mathcal{X}$} a[1] | a[0];
                    box {$\mathcal{X}$} a[2] | a[1];
                    box {$\mathcal{X}$} a[3] | a[2];
                    
                    barrier (a, b, alp, beta);
                    
                    align alp;
                    box {$X$} beta; 
                    box {$\mathcal{C}$} alp;
                    
                    box {$\mathcal{C}$} a[0] | b[0];
                    box {$\mathcal{C}$} a[1] | b[1];
                    box {$\mathcal{C}$} a[2] | b[2];
                    box {$\mathcal{C}$} a[3] | b[3];
                    
                    box {$\mathcal{X}$} alp[0] | a[0];
                    box {$\mathcal{X}$} alp[1] | a[1];
                    box {$\mathcal{X}$} alp[2] | a[2];
                    box {$\mathcal{X}$} alp[3] | a[3];
                    
                    box {$\mathcal{C}$} a[0] | b[0];
                    box {$\mathcal{C}$} a[1] | b[1];
                    box {$\mathcal{C}$} a[2] | b[2];
                    box {$\mathcal{C}$} a[3] | b[3];
                    
                    
                    
                    barrier (a, b, alp, beta);
                    
                    box {$\mathcal{X}^{\dagger}$} a[3] | a[2];
                    box {$\mathcal{X}^{\dagger}$} a[2] | a[1];
                    box {$\mathcal{X}^{\dagger}$} a[1] | a[0];
                    
                    barrier (a, b, alp, beta);
                    
                    align a[1], a[2], a[3];
                    box {$\mathcal{C}$} a[3] | b[2];
                    box {$\mathcal{C}$} a[2] | b[1];
                    box {$\mathcal{C}$} a[1] | b[0];
                    
                    cnot b[3] | b[2];
                    cnot b[2] | b[1];
                    cnot b[1] | b[0];
                    
                    \end{yquant}
                \end{tikzpicture}
            };
            \node at (5, 3) {\Huge $\left. \right \}$};
            \node[align=left] at (11.35-4.5, 3) {\footnotesize $\mathbb{Z}_2$ ungauging + undoing \\ \footnotesize  symmetry enrichment};
            \node at (5, 1.5) {\Huge $\left. \right \}$};
            \node[align=left] at (10.5-4.5, 1.5) {\footnotesize $\mathbb{Z}_3$ gauging};

            \node at (5, -0.25) {\Huge $\left. \right \}$};
            \node[align=left] at (11-4.5, -0.25) {\footnotesize FDLU c.c. circuit};
            \node at (5, -2) {\Huge $\left. \right \}$};
            \node[align=left] at (10.75-4.5, -2) {\footnotesize $\mathbb{Z}_3$ ungauging};
            \node at (5, -3.5) {\Huge $\left. \right \}$};
            \node[align=left] at (11.25-4.5, -3.5) {\footnotesize $\mathbb{Z}_2$ gauging + \\ \footnotesize symmetry enrichment};
        \end{tikzpicture}
    }
\end{equation}

\section{Twisted Quantum Double Ribbon Operators}

\begin{figure}
    \centering
    \includegraphics[width=0.5\linewidth]{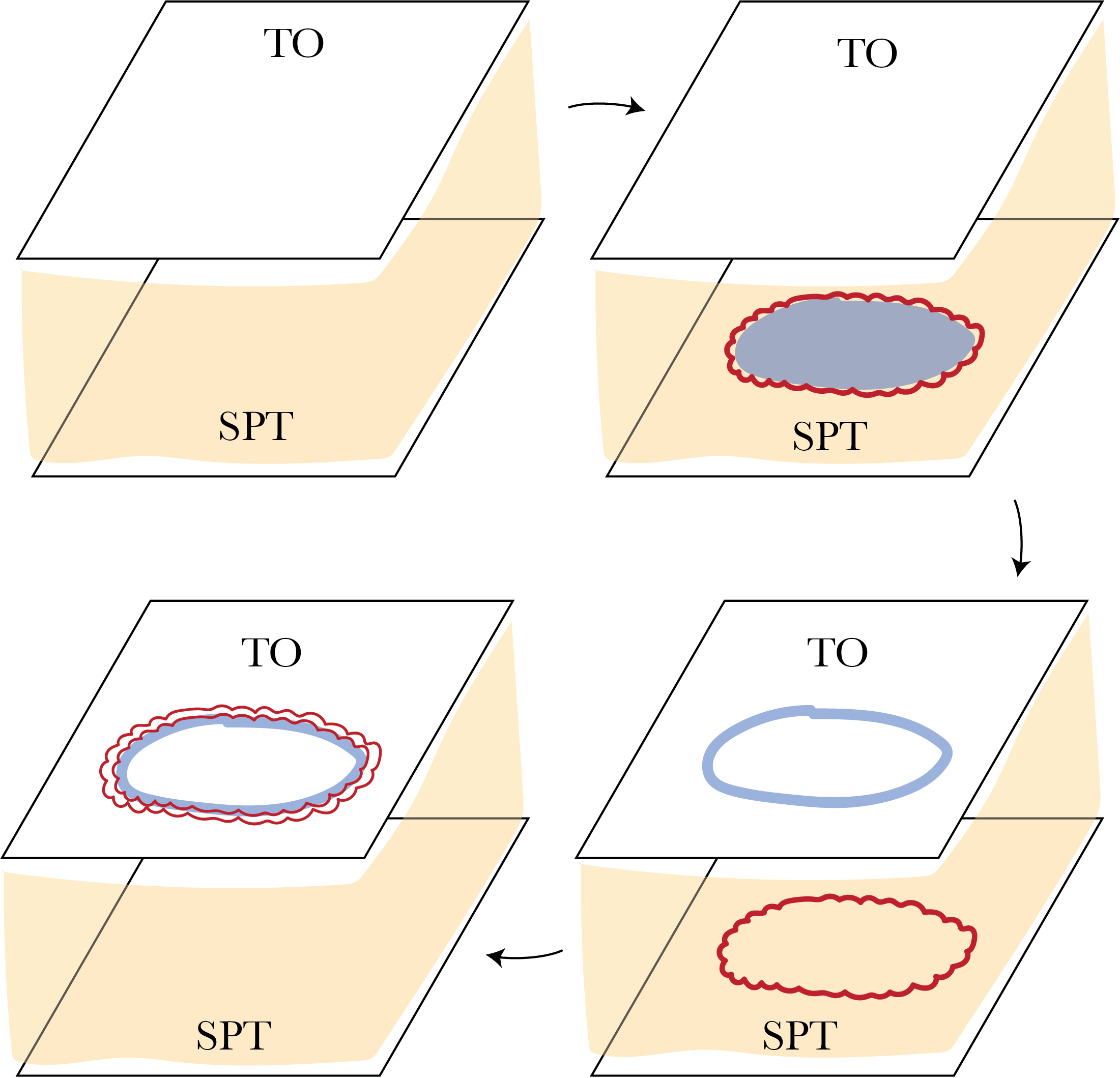}
    \caption{\textbf{Deriving Twisted Ribbons:} Proceeding clockwise from the top left corner, we demonstrate the procedure for deriving closed ribbon operators for excitations in twisted quantum doubles. a) We begin with some nontrivial SPT, which is connected to our twisted quantum double (TO) by some gauging map (the yellow bulk). b) We apply a finite symmetry membrane (blue), decorated on its boundary (red line) to ensure it acts trivially on the SPT. c) Using the properties of the gauging map, we can push the symmetry membrane through the gauging map to obtain some effective boundary action on the TO (which may span the gauging map, as with the charge-conjugation membrane in the $\mathbb{Z}_3$ case). d) If we want a ribbon purely in terms of edge variables, we can push the boundary decoration through as well--- there is no guarantee that this decoration is symmetric with respect to the gauging map, and so this may end up being a non-local, or linear-depth, circuit.}
    \label{fig:twisted-schematic}
\end{figure}

We can consider different input states to the $G$-gauging map than a trivial paramagnet. This allows us to explore properties of states beyond conventional quantum double models. For instance, gauging the $\mathbb{Z}_2$ global symmetry of the Levin-Gu SPT yields the doubled-semion topological order, while gauging the $\mathbb{Z}_2^3$ symmetry of the $\mathbb{Z}_2^3$ type III SPT yields a topological order isomorphic to $D(D_4)$.

We can straightforwardly obtain the ribbon operators for doubled semion excitations using the properties of the $\mathbb{Z}_2$ gauging map and the Levin-Gu SPT; see Fig. \ref{fig:twisted-schematic} for a schematic illustration of the procedure. The main difference when the input state is a nontrivial SPT rather than a trivial paramagnet is the need to account for the symmetry fractionalization on the SPT. We still identify the excitations on top of the twisted quantum double with excitations or symmetry defects on top of the SPT, and use the properties of the 1D gauging map to access the SPT degrees of freedom. Now, however, there might be some nontrivial action on the SPT degrees of freedom as part of the ribbon operation, whereas before they simply acted as controls. 

We derive ribbon operators for three example topological orders which can be obtained by gauging an SPT: (1) doubled semions, (2) the twisted quantum double realizing $D(D_4)$, and (3) quaternions (or $D(Q_8)$). 

\subsection{Doubled Semion}

The doubled semion topological order is a twisted version of the $\mathbb{Z}_2$ toric code. It contains anyons $\{1, s, \bar{s}, s\bar{s}\}$ with the following nontrivial fusion rules: \begin{equation}
    \begin{aligned}
    s \times s &= 1 \\
    \bar{s} \times \bar{s} &= 1\\
    s \times \bar{s} &= s\bar{s}
    \end{aligned}
\end{equation}
The anyons $s$ and $\bar{s}$ are semions and anti-semions, respectively, meaning they have exchange phases $i$ and $-i$; they have bosonic mutual statistics. Both $1$ and $s \bar{s}$ are self-bosons. 

Doubled semion topological order can be obtained by gauging the Levin-Gu SPT, the only nontrivial bosonic SPT with a $\mathbb{Z}_2$ global symmetry. This SPT is the ground state of the following Hamiltonian, defined with qubits living on the sites of the triangular lattice:
\begin{equation}
\raisebox{-23pt}{
    \begin{tikzpicture}
        \node at (0, 0) {$H_{LG} = -\sum\limits_v $};
        \node at (2, 0.1){
            \begin{tikzpicture}[scale = 0.75]
                \draw[-, opacity=0.2] (0, 1) -- (0, 0) -- (0.5*1.73, -0.5) -- (1.73, 0) -- (1.73, 1) -- (0.5*1.73, 1.5) -- (0, 1);
                \draw[-, opacity=0.2] (0.5*1.73, -0.5) -- (0.5*1.73, 1.5);
                \draw[-, opacity=0.2] (0, 0) -- (1.73, 1);
                \draw[-, opacity=0.2] (0, 1) -- (1.73, 0);

                \draw[decorate, decoration={snake, segment length = 2.5pt, amplitude=1pt}] (0, 1) -- (0, 0) -- (0.5*1.73, -0.5) -- (1.73, 0) -- (1.73, 1) -- (0.5*1.73, 1.5) -- (0, 1);
                \filldraw[black] (0, 0) circle (2pt);
                \filldraw[black] (0, 1) circle (2pt);
                \filldraw[black] (0.5*1.73, -0.5) circle (2pt);
                \filldraw[black] (1.73, 0) circle (2pt);
                \filldraw[black] (1.73, 1) circle (2pt);
                \filldraw[black] (0.5*1.73, 1.5) circle (2pt);

                \node at (0.5*1.73, 0.5) {$X_v$};
                
            \end{tikzpicture}
        };
        \node at (3, 0) {$,$};
        \node at (4.5, 0.2){
            \begin{tikzpicture}
                \node at (0, 0.3) {$i$};
                \node at (1, 0.3) {$j$};
                \filldraw[black] (0, 0) circle (2pt);
                \filldraw[black] (1, 0) circle (2pt);
                \draw[decorate, decoration={snake, segment length = 2.5pt, amplitude=1pt}] (0, 0) -- (1, 0);            
            \end{tikzpicture}
        };
        \node at (6.1, 0) {$=~ e^{i \frac{\pi}{4} Z_i Z_j}$};
    \end{tikzpicture}
}
\label{eq:levin-gu-hamiltonian}
\end{equation}

Gauging the global $\prod X$ symmetry using the $\mathbb{Z}_2$ gauging map will give us the doubled semion Hamiltonian. On the triangular lattice, the $\mathbb{Z}_2$ gauging map acts in the following way: 
\begin{equation}
    \raisebox{-23pt}{
    \begin{tikzpicture}
        \node at (0, 0) {
            \begin{tikzpicture}[scale=0.75]
                \draw[-, opacity=0.2] (0, 1) -- (0, 0) -- (0.5*1.73, -0.5) -- (1.73, 0) -- (1.73, 1) -- (0.5*1.73, 1.5) -- (0, 1);
                \draw[-, opacity=0.2] (0.5*1.73, -0.5) -- (0.5*1.73, 1.5);
                \draw[-, opacity=0.2] (0, 0) -- (1.73, 1);
                \draw[-, opacity=0.2] (0, 1) -- (1.73, 0);

                \node[black] at (0.5*1.73, 0.5) {$X$};

                \draw[->] (2.25,0.4) -- node[above,midway] {\footnotesize $\mathsf{KW}_{\mathbb{Z}_2}$} (3.5,0.4);
                
            \end{tikzpicture}
        };
        \node at (2.5, 0) {
            \begin{tikzpicture}[scale=0.75]
                \draw[-, opacity=0.2] (0, 1) -- (0, 0) -- (0.5*1.73, -0.5) -- (1.73, 0) -- (1.73, 1) -- (0.5*1.73, 1.5) -- (0, 1);
                \draw[-, opacity=0.2] (0.5*1.73, -0.5) -- (0.5*1.73, 1.5);
                \draw[-, opacity=0.2] (0, 0) -- (1.73, 1);
                \draw[-, opacity=0.2] (0, 1) -- (1.73, 0);

                \node[black] at (0.25*1.73, 0.25) {$X$};
                \node[black] at (0.75*1.73, 0.25) {$X$};
                \node[black] at (0.5*1.73, 0) {$X$};
                \node[black] at (0.5*1.73, 1) {$X$};
                \node[black] at (0.25*1.73, 0.75) {$X$};
                \node[black] at (0.75*1.73, 0.75) {$X$};
            \end{tikzpicture}
        };
        \node at (6, 0) {
            \begin{tikzpicture}[scale=0.75]
                \draw[-, opacity=0.2] (0, 1) -- (0, 0) -- (0.5*1.73, -0.5) -- (1.73, 0) -- (1.73, 1) -- (0.5*1.73, 1.5) -- (0, 1);
                \draw[-, opacity=0.2] (0.5*1.73, -0.5) -- (0.5*1.73, 1.5);
                \draw[-, opacity=0.2] (0, 0) -- (1.73, 1);
                \draw[-, opacity=0.2] (0, 1) -- (1.73, 0);

                \node[black] at (0.5*1.73, 0.5) {$Z$};
                \node[black] at (0, 0) {$Z$};

                \draw[->] (2.25,0.4) -- node[above,midway] {\footnotesize $\mathsf{KW}_{\mathbb{Z}_2}$} (3.5,0.4);
            \end{tikzpicture}
        };
        \node at (8.5, 0) {
            \begin{tikzpicture}[scale=0.75]
                \draw[-, opacity=0.2] (0, 1) -- (0, 0) -- (0.5*1.73, -0.5) -- (1.73, 0) -- (1.73, 1) -- (0.5*1.73, 1.5) -- (0, 1);
                \draw[-, opacity=0.2] (0.5*1.73, -0.5) -- (0.5*1.73, 1.5);
                \draw[-, opacity=0.2] (0, 0) -- (1.73, 1);
                \draw[-, opacity=0.2] (0, 1) -- (1.73, 0);

                \node[black] at (0.25*1.73, 0.25) {$Z$};
                
            \end{tikzpicture}
        };
    \end{tikzpicture}
    }
\label{eq:triangular-Z2-gauging}
\end{equation}
The $ZZ \rightarrow Z$ transformation law tells us how the control gates transform: 

\begin{equation}
        \begin{tikzpicture}
            \filldraw[black] (0, 0) circle (2pt);
            \filldraw[black] (1, 0) circle (2pt);
            \draw[decorate, decoration={snake, segment length = 2.5pt, amplitude=1pt}] (0, 0) -- (1, 0);

             \draw[->] (1.5,0) -- node[above,midway] {\footnotesize $\mathsf{KW}_{\mathbb{Z}_2}$} (2.75,0);

             \draw[-, opacity=0.2] (3.25, 0) -- (4.25, 0);
             \node[black] at (3.75, 0) {$R$};
        \end{tikzpicture}
\label{eq:R-transform}
\end{equation}
where $R = e^{i \frac{\pi}{4} Z}$. Going forward, we will use both pictorial equations like the ones above, and explicit mathematical formulas. To set notation: We denote the set of vertices on the triangular lattice as $V$, the set of all plaquettes $P$, and edges $E$. An edge $e \in E$ can also be labeled by the two vertices it connects; i.e. $e = {v, v'}$ for some nearest-neighor pair $v, v'$. For the dual hexagonal lattice, we denote the set of plaquettes $\tilde{P}$, the set of vertices $\tilde{V}$, and the set of edges $\tilde{E}$. We have $V \sim \tilde{P}$, $P \sim \tilde{V}$, and $E \sim \tilde{E}$. Given a set of plaquettes $S \subset P$, $\partial S\subset E$ is the set of edges forming the boundary of the region enclosed by $S$ (note that $S$ can consist of disjoint plaquettes). For some set of vertices $M \subset V$, $\partial M \subset E$ is the set of edges terminating on only one vertex in $M$. We can alternately think of $\partial M$ as $\partial \tilde{M}$, where $\tilde{M} \subset \tilde{P}$. 

Using this notation, we write the $\mathbb{Z}_2$ gauging rules for the triangular lattice explicitly:
\begin{equation}
    \begin{aligned}
        X_v &\longrightarrow \prod_{e \in \partial v} X_e \\
        Z_{v} Z_{v'} \text{, with } \langle v, v' \rangle &\longrightarrow Z_{e} \text{ s.t. }  e = \{v, v'\}
    \end{aligned}
\end{equation}

Applying these transformations gives us the doubled semion Hamiltonian\cite{Freedman04,Levin_2005}:
\begin{equation}
    \raisebox{-23pt}{
    \begin{tikzpicture}
        \node at (0, 0) {$H_{DS} =~ - \sum\limits_v $};
        \node at (2, 0.1) {
            \begin{tikzpicture}[scale = 0.75]
                \draw[-, opacity=0.2] (0, 1) -- (0, 0) -- (0.5*1.73, -0.5) -- (1.73, 0) -- (1.73, 1) -- (0.5*1.73, 1.5) -- (0, 1);
                \draw[-, opacity=0.2] (0.5*1.73, -0.5) -- (0.5*1.73, 1.5);
                \draw[-, opacity=0.2] (0, 0) -- (1.73, 1);
                \draw[-, opacity=0.2] (0, 1) -- (1.73, 0);

                \node[black] at (0.25*1.73, 0.25) {$X$};
                \node[black] at (0.75*1.73, 0.25) {$X$};
                \node[black] at (0.5*1.73, 0) {$X$};
                \node[black] at (0.5*1.73, 1) {$X$};
                \node[black] at (0.25*1.73, 0.75) {$X$};
                \node[black] at (0.75*1.73, 0.75) {$X$};

                \node[black] at (0, 0.5) {$R$};
                \node[black] at (0.25*1.73, -0.25) {$R$};
                \node[black] at (0.75*1.73, -0.25) {$R$};
                \node[black] at (1.73, 0.5) {$R$};
                \node[black] at (0.25*1.73, 1.25) {$R$};
                \node[black] at (0.75*1.73, 1.25) {$R$};
            \end{tikzpicture}
        };
        \node at (3.5, 0) {$- \sum\limits_p$};
        \node at (4.25, 0.1){
            \begin{tikzpicture}[scale=0.75]
                \draw[-, opacity=0.2] (0, 1) -- (0, 0) -- (0.5*1.73, 0.5) -- (0, 1);
                \node[black] at (0, 0.5) {$Z$};
                \node[black] at (0.25*1.73, 0.25) {$Z$};
                \node[black] at (0.25*1.73, 0.75) {$Z$};
            \end{tikzpicture}
        };
    \end{tikzpicture}
    }
\end{equation}

Where $v$ denotes a sum over vertices, and $p$ denotes a sum over plaquettes (triangles). More detailed derivations of this result can be found in \cite{Levin_2012,tantivasadakarnLongrangeEntanglementMeasuring2022a}.

We want to construct ribbons for the nontrivial anyons $s, \bar{s}, s\bar{s}$. As discussed in the main text, we should see what precursor excitations exist in the SPT, and trace how they transform under the gauging map. We note that, after gauging, nearest-neighbor vertex stabilizers no longer commute. This can be rectified by restricting to the ``zero-flux'' sector, where all plaquette stabilizers are $+1$ (in other words, taking the weight of the plaquette term in the Hamiltonian to $-\infty$). However, we will still need to be careful about the order in which we push vertex stabilizers through the gauging map.

\begin{figure}
    \centering
    \includegraphics[width=0.65\linewidth]{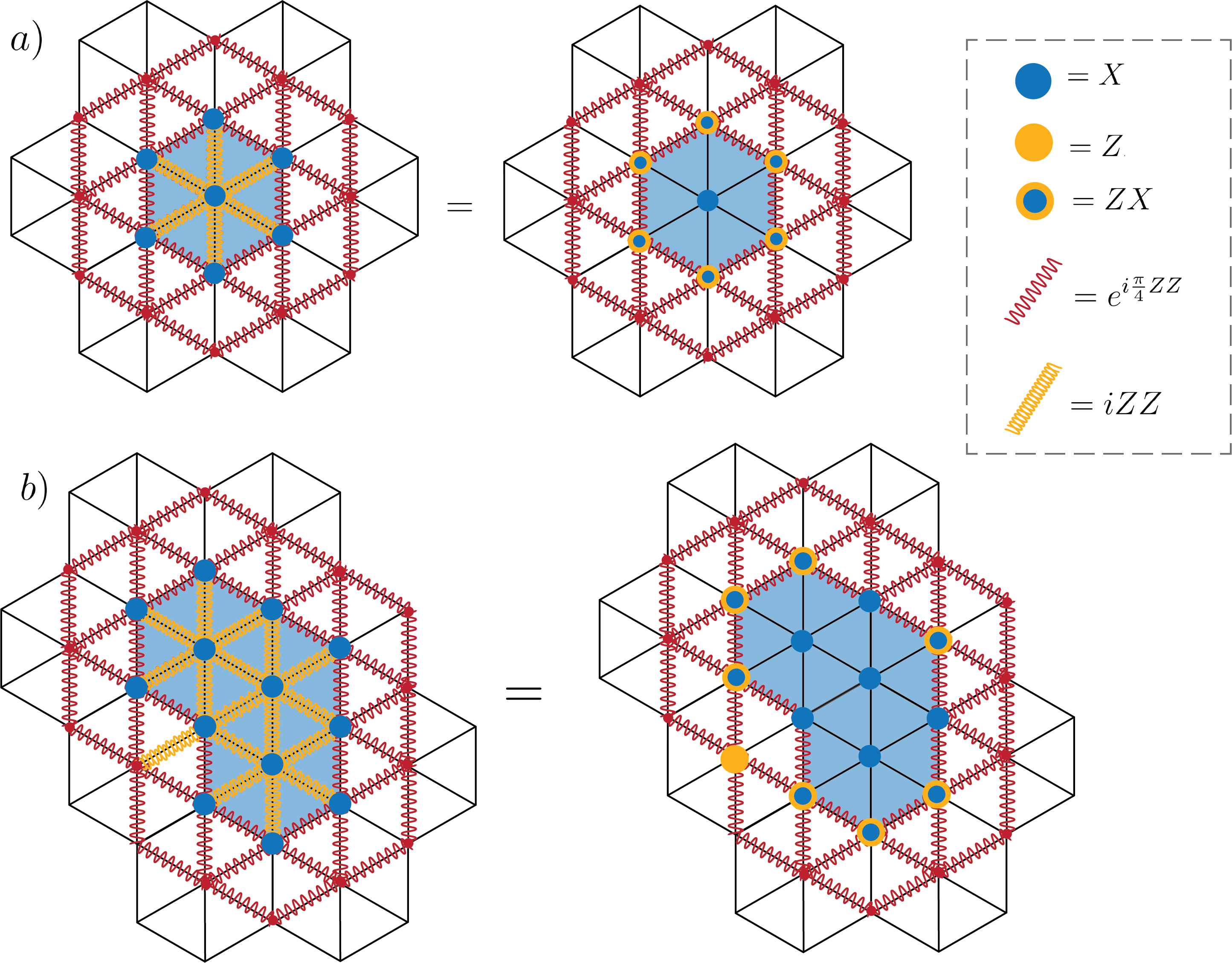}
    \caption{\textbf{Levin-Gu SPT Fractionalization:} a) The leftmost picture shows the product of the seven Levin-Gu stabilizers corresponding to the seven vertices in the blue region. The edges connected to the central vertex all have two stabilizers acting on them, indicated by the overlapping squiggly lines. Moving to the center, the overlapping $e^{i \pi/4 Z Z}$  gates square to $ZZ$. Finally, given the even number of $ZZ$ incident on the central vertex, the support there disappears, leaving solely boundary decoration to the bulk $X$ membrane (blue). b) A second example, for a region with a different shape.}
    \label{fig:levin-gu-fractionalization}
\end{figure}

The simplest excitations we can create on top of the Levin-Gu SPT are a pair of charges, using $Z_i Z_j$. This anti-commutes with the two stabilizers at sites $i, j$. When we push $Z_i Z_j$ through the gauging map, it turns into a string of $Z$ operators along a path of edges connecting sites $i, j$ (due to the emergent 1-form symmetry generated by the $\mathbb{Z}_2$ gauging map). This string has bosonic self-exchange statistics, so it must be the string operator for $s \bar{s}$: \begin{equation}
    R^\gamma_{s \bar{s}} = \prod_{e \in \gamma} Z_e
\end{equation}
where $\gamma = \{e\}$ is a collection of edges forming a 1D path on the triangular lattice.

We can also consider inserting a symmetry defect into the Levin-Gu SPT, by applying a finite membrane of $X$ operators. This defect operator anti-commutes with a single charge $Z$, and so its corresponding anyon should have nontrivial mutual statistics with the $s\bar{s}$ anyon. This could be either $s$ or $\bar{s}$; as we will see, there are two distinct ways of pushing the fractionalization pattern of a finite $X$ membrane through the gauging map, generating the $s$ and $\bar{s}$ ribbons.

First, we need to find the symmetry localization for a finite $X$ membrane on top of the Levin-Gu SPT. The form of the stabilizers already tells us what the localization looks like for a membrane consisting of a single $X$;
\begin{equation}
    A_v = X_v \prod_{j, k \in \langle v \rangle} e^{i \frac{\pi}{4} Z_j Z_k}\qquad \longrightarrow \qquad X_v \ket{\Omega_{LG}} = \prod_{j, k \in \langle v \rangle} e^{i \frac{\pi}{4} Z_j Z_k} \ket{\Omega_{LG}}
\end{equation}
Where $A_v$ denotes the stabilizer centered on vertex $v$, $\langle v \rangle$ denotes the set of all nearest-neighbors of $v$, and $\ket{\Omega_{LG}}$ is the Levin-Gu SPT (the ground state of Eq. \eqref{eq:levin-gu-hamiltonian}).
We can take products of stabilizers to find the set of operators that trivializes the action of a larger membrane. Two examples of the symmetry localization for a multi-vertex membrane are illustrated in Fig. \ref{fig:levin-gu-fractionalization}. Written out explicitly, a generic product of Levin-Gu stabilizers for some set of vertices $M$ takes the form: 
\begin{equation}
\begin{aligned}
    \prod_{v \in M}^{\#} A_v &= \prod^{\#}_{v \in M} \left (X_v \prod_{j, k \in \langle v \rangle} e^{i \frac{\pi}{4} Z_j Z_k}  \right ) \\
    &= \left (\prod_{\substack{\langle j, k \rangle \\ \langle j\rangle \cap \langle k\rangle \subseteq M}} i Z_j Z_k \right ) \left ( \prod^{\#}_{v \in M} X_v  \left (\prod_{\substack{j, k \in \langle v \rangle\\ \langle j\rangle \cap \langle k\rangle \cap M \neq \emptyset \\ \langle j \rangle \cap \langle k \rangle \nsubseteq M }} 
 e^{i \frac{\pi}{4} Z_j Z_k}  \right) \right)
\end{aligned}
\label{eq:explicit-z2-gauging}
\end{equation}
Here, $\#$ indicates an ordered product. The first product in the second line are over all nearest-neighbor pairs $\langle j, k \rangle$ in the lattice satisfying the given conditions. Intuitively, each nearest-neighbor pair $j, k$ share two of their other nearest neighbors; the set $\langle j \rangle \cap \langle k \rangle$ contains the two vertices completing the triangles sharing the edge $e = \{j, k\}$. When both of these vertices are included in $M$, there will be two stabilizers containing $e^{i \frac{\pi}{4} Z_j Z_k}$ terms, and so the overall action on $j, k$ is $i Z_j Z_k$. These terms can be pulled out of the ordered product because they necessarily act on edges disjoint from $\partial M$, and so must commute with all other operators in the product. When only one of the vertices in $\langle j \rangle \cap \langle k \rangle$ is contained in $M$, only one stabilizer in the product contains a $e^{i \frac{\pi}{4} Z_j Z_k}$ term. If neither vertex is contained in $M$, there is no support on either $j$ or $k$.

\begin{figure}
    \centering
    \includegraphics[width=0.75\linewidth]{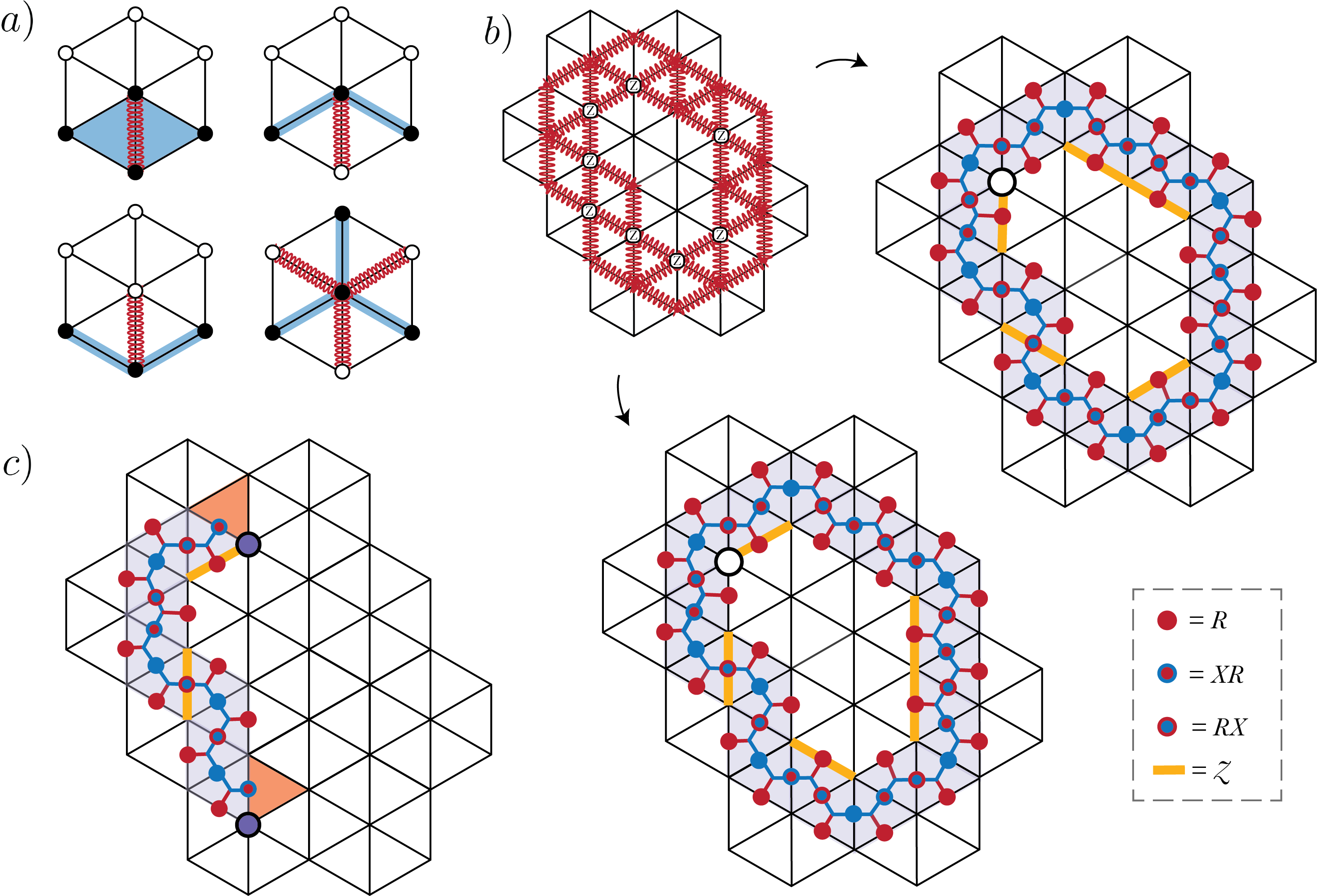}
    \caption{\textbf{Doubled Semion Ribbons:} a) The four cases (up to rotation) where a vertex $j$ (the central vertex) satisfies $\abs{\alpha}_j = 1$. The blue regions indicate the support of the membrane $M$. b) Given some boundary fractionalization pattern (upper left), there are two choices when pushing through the gauging map. Starting from the vertex indicated by the white circle, we can pair $Z$ operators clockwise (bottom) or counter-clockwise (right). These choices correspond to the distinct $s$ and $\bar{s}$ semions. c) Open semion ribbons can be created by removing part of a closed ribbon. This open ribbon commutes with all stabilizers along its bulk, but excites one vertex (indicated by a purple circle) and one plaquette (highlighted orange) at each of its ends.}
    \label{fig:semion-ribbons}
\end{figure}

We can further simplify the first product of $Z_j Z_k$ terms. A single vertex $j$ might have several neighbors $\{k_\alpha\}$ for which $\langle j \rangle \cap \langle k_\alpha \rangle \subseteq M$. If there are an even number of such $k_\alpha$, $Z_j$ will drop out of the product. Denote the parity of $\{k_\alpha\}$ for vertex $j$ as $\abs{\alpha}_j$. The only cases where $\abs{\alpha}_j = 1$ are illustrated in Fig. \ref{fig:semion-ribbons}a); they correspond to cases where $j$ lives at a certain convex turn along the support of the boundary decoration. Denote the set of vertices $j$ with $\abs{\alpha}_j = 1$ as $\alpha$. Using this fact, we have:
\begin{equation}
    \prod_{v \in M}^{\#} A_v = \left (\prod_{j \in \alpha} Z_j \right ) \left ( \prod^{\#}_{v \in M} X_v  \left (\prod_{\substack{j, k \in \langle v \rangle\\ \langle j\rangle \cap \langle k\rangle \cap M \neq \emptyset \\ \langle j \rangle \cap \langle k \rangle \nsubseteq M }} 
 e^{i \frac{\pi}{4} Z_j Z_k}  \right) \right)
\end{equation}

Since we are considering only closed loops, the number of $Z_j$ in the above product will be zero or even. Applying Eq. \eqref{eq:explicit-z2-gauging}, the mapping for $X$ and $e^{i \frac{\pi}{4} Z_j Z_k}$ is unambiguous. However, there are two distinct ways to pair the $Z_j$; we can begin at some $Z_j$ and construct pairs clockwise or counterclockwise along the boundary. These two choices differ by a $\prod Z$ ($s \bar{s}$) string along the boundary, indicating they give us $s$ and $\bar{s}$ ribbons, although it remains to determine which choice corresponds to which semion.
\begin{equation}
     \prod^{\#}_{v \in M} A_v \longrightarrow \left (\prod_{\substack{\gamma \\ \gamma = \{j, j'\} \subset \alpha}} \prod_{e \in \gamma} i Z_e \right ) \left ( \prod^{\#}_{v \in M} \left(\prod_{e \in \partial v} X_e \right)  \left (\prod_{\substack{e = \{j, k\}\\ \langle j\rangle \cap \langle k\rangle \cap M \neq \emptyset \\ \langle j \rangle \cap \langle k \rangle \nsubseteq M }} R_e \right) \right)
\end{equation}
where $\gamma$ are the paths connecting either clockwise or counterclockwise nearest neighbors $j, j' \in \alpha$ (here nearest-neighbor is with respect to the support of $\alpha$). See Fig. \ref{fig:semion-ribbons}b for an example ribbon. 

We have constructed closed ribbon operators for $s, \bar{s}$. Open ribbon operators can be obtained by simply terminating these closed ribbon expressions; the particular choice of termination doesn't matter, since the ribbons will only differ by a local operator acting at the ends. We can verify that both open and closed ribbons commute with all doubled semion stabilizers along their bulk; the open ribbon will excite a plaquette and vertex stabilizer at each of its ends (see Fig. \ref{fig:semion-ribbons}c for an example).
Immediately, we can see both closed and open ribbons commute with the plaquette stabilizers everywhere in their bulk; there are two $X$ operators acting on every triangle. At each end of the open ribbon, one triangle will only have a single $X$ acting on it, and so will be excited.
The commutation relations with the vertex stabilizers are more complicated, but using the commutation relations below, it can be verified that a closed ribbon commutes will all elementary vertex stabilizers. $R = e^{i \frac{\pi}{4} Z}$ has the following commutation relations with $X, Z$:
\begin{equation}
    \begin{aligned}
        RX &= (iZ) XR \\
        RZ &= ZR 
    \end{aligned}
\label{eq:R-commutations}
\end{equation}

Intuitively, the gauging map will preserve commutation relations, and the closed ribbon operators are simply the dual operators to products of commuting SPT stabilizers. 

\begin{figure}
    \centering
    \includegraphics[width=0.75\linewidth]{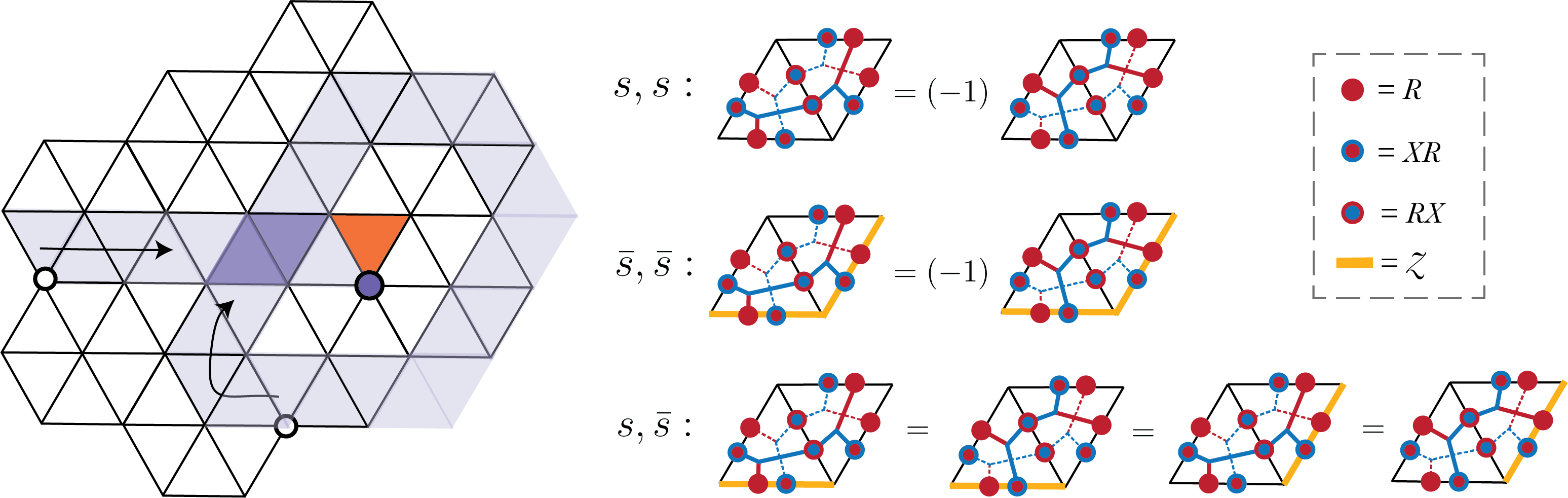}
    \caption{\textbf{Verifying semionic braiding statistics:} A full braiding process of one semion around another is depicted. Consider the overlap in support of the two ribbons highlighted in purple to the left (their direction and starting points are indicated). To the right, we picture the three possible scenarios: overlapping $s, s$, $\bar{s}, \bar{s}$ and $s, \bar{s}$. The overlapping segments of each ribbon are drawn, with the ribbon applied first indicated by dotted lines, the second ribbon indicated by solid lines. Commuting the first ribbon through the second yields the expected phases.}
    \label{fig:semion-braiding}
\end{figure}

Now we verify the self and mutual statistics of the anyons created by these ribbons. We expect that $s$ and $\bar{s}$ should have trivial mutual statistics, while they should each pick up a phase of $-1$ when braiding fully around any anyon of the same type. In other words, two overlapping $s$ or $\bar{s}$ ribbons anti-commute, while an $s$ and $\bar{s}$ ribbon commute. See Fig. \ref{fig:semion-braiding} for a visual demonstration using the commutation relations given in Eq. \eqref{eq:R-commutations}.

\subsection{\emph{D(D4)}}
    
The (untwisted) quantum double of $D_4$, the group of symmetries of a square, is equivalent to the theory obtained by gauging the global symmetries of the $\mathbb{Z}_2^3$ type III SPT \cite{propitiusTopologicalInteractionsBroken1995, Yoshida_2016}. Intriguingly, this demonstrates that gauging (twisted) Abelian symmetries can yield \emph{non-Abelian} topological orders; we will show a simple origin for these emergent non-Abelian statistics in this derivation. 

The $\mathbb{Z}_2^3$ type III SPT has a structure very similar to the Levin-Gu SPT considered in the previous section. We consider a three-colored triangular lattice, with sublattices $a = R, B, G$. The Hamiltonian for the $\mathbb{Z}_2^3$ Type III SPT is given by: 
\begin{equation}
\raisebox{-23pt}{
\begin{tikzpicture}
    \node at (0, 0) {$H_{III} = - \sum\limits_{v \in R}$};
    \node at (2, 0){
        \begin{tikzpicture}[scale=0.75]
            \draw[-, opacity=0.2] (0, 1) -- (0, 0) -- (0.5*1.73, -0.5) -- (1.73, 0) -- (1.73, 1) -- (0.5*1.73, 1.5) -- (0, 1);
            \draw[-, opacity=0.2] (0.5*1.73, -0.5) -- (0.5*1.73, 1.5);
            \draw[-, opacity=0.2] (0, 0) -- (1.73, 1);
            \draw[-, opacity=0.2] (0, 1) -- (1.73, 0);

            \draw[-, very thick, black]  (0, 1) -- (0, 0) -- (0.5*1.73, -0.5) -- (1.73, 0) -- (1.73, 1) -- (0.5*1.73, 1.5) -- (0, 1);
            \filldraw[blue] (0, 0) circle (3pt);
            \filldraw[green] (0, 1) circle (3pt);
            \filldraw[green] (0.5*1.73, -0.5) circle (3pt);
            \filldraw[blue] (1.73, 0) circle (3pt);
            \filldraw[green] (1.73, 1) circle (3pt);
            \filldraw[blue] (0.5*1.73, 1.5) circle (3pt);
            \node[red] at (0.5*1.73, 0.5) {$X_v$};
        \end{tikzpicture}
    };
    \node at (6.5, 0) {$- \sum\limits_{v \in G}$};
    \node at (8, 0){
        \begin{tikzpicture}[scale=0.75]
            \draw[-, opacity=0.2] (0, 1) -- (0, 0) -- (0.5*1.73, -0.5) -- (1.73, 0) -- (1.73, 1) -- (0.5*1.73, 1.5) -- (0, 1);
            \draw[-, opacity=0.2] (0.5*1.73, -0.5) -- (0.5*1.73, 1.5);
            \draw[-, opacity=0.2] (0, 0) -- (1.73, 1);
            \draw[-, opacity=0.2] (0, 1) -- (1.73, 0);

            \draw[-, very thick, black]  (0, 1) -- (0, 0) -- (0.5*1.73, -0.5) -- (1.73, 0) -- (1.73, 1) -- (0.5*1.73, 1.5) -- (0, 1);
            \filldraw[red] (0, 0) circle (3pt);
            \filldraw[blue] (0, 1) circle (3pt);
            \filldraw[blue] (0.5*1.73, -0.5) circle (3pt);
            \filldraw[red] (1.73, 0) circle (3pt);
            \filldraw[blue] (1.73, 1) circle (3pt);
            \filldraw[red] (0.5*1.73, 1.5) circle (3pt);
            \node[green] at (0.5*1.73, 0.5) {$X_v$};
        \end{tikzpicture}
    };
    \node at (3.5, 0) {$-\sum\limits_{v \in B}$};
    \node at (5, 0){
        \begin{tikzpicture}[scale=0.75]
            \draw[-, opacity=0.2] (0, 1) -- (0, 0) -- (0.5*1.73, -0.5) -- (1.73, 0) -- (1.73, 1) -- (0.5*1.73, 1.5) -- (0, 1);
            \draw[-, opacity=0.2] (0.5*1.73, -0.5) -- (0.5*1.73, 1.5);
            \draw[-, opacity=0.2] (0, 0) -- (1.73, 1);
            \draw[-, opacity=0.2] (0, 1) -- (1.73, 0);

            \draw[-, very thick, black]  (0, 1) -- (0, 0) -- (0.5*1.73, -0.5) -- (1.73, 0) -- (1.73, 1) -- (0.5*1.73, 1.5) -- (0, 1);
            \filldraw[green] (0, 0) circle (3pt);
            \filldraw[red] (0, 1) circle (3pt);
            \filldraw[red] (0.5*1.73, -0.5) circle (3pt);
            \filldraw[green] (1.73, 0) circle (3pt);
            \filldraw[red] (1.73, 1) circle (3pt);
            \filldraw[green] (0.5*1.73, 1.5) circle (3pt);
            \node[blue] at (0.5*1.73, 0.5) {$X_v$};
        \end{tikzpicture}
    };
    \node at (10.5, 0){
    \begin{tikzpicture}
        \node at (0, 0.3) {$i$};
        \node at (1, 0.3) {$j$};
        \filldraw[black] (0, 0) circle (2pt);
        \filldraw[black] (1, 0) circle (2pt);
        \draw[-, very thick, black] (0, 0) -- (1, 0);  
    \end{tikzpicture}
    };
    \node at (11.9, -0.3) {$ = CZ_{ij}$};
\end{tikzpicture}
}
\label{eq:typeiii-spt-hamiltonian}
\end{equation}

Gauging the $\mathbb{Z}_2$ global symmetries for each sublattice separately will yield $D(D_4)$ topological order; details of this gauging procedure can be found in \cite{iqbalCreationNonAbelianTopological2023}. The mapping is the same as that discussed in the doubled semion section, applied to a single triangular sublattice rather than the full lattice. The resultant $D(D_4)$ Hamiltonian is:
\begin{equation}
\raisebox{-23pt}{
\begin{tikzpicture}
    \node at (0, 0) {$H_{D_{4}} = - \sum\limits_{v \in R}$};
    \node at (2.1, 0){
        \begin{tikzpicture}[scale=0.9]
            \draw[-, opacity=0.2] (0, 1) -- (0, 0) -- (0.5*1.73, -0.5) -- (1.73, 0) -- (1.73, 1) -- (0.5*1.73, 1.5) -- (0, 1);
            \draw[-, opacity=0.2] (0.5*1.73, -0.5) -- (0.5*1.73, 1.5);
            \draw[-, opacity=0.2] (0, 0) -- (1.73, 1);
            \draw[-, opacity=0.2] (0, 1) -- (1.73, 0);

            \node[red] at (0, 0.5) {\footnotesize $X$};
            \node[red] at (0.25*1.73, 1.25) {\footnotesize $X$};
            \node[red] at (0.75*1.73, 1.25) {\footnotesize $X$};
            \node[red] at (1.73, 0.5) {\footnotesize $X$};
            \node[red] at (0.75*1.73, -0.25) {\footnotesize $X$};
            \node[red] at (0.25*1.73, -0.25) {\footnotesize $X$};

            \draw[-, very thick, black] (0.25*1.73, 0.25) --  (0.25*1.73, 0.75) -- (0.5*1.73, 1) -- (0.75*1.73, 0.75) -- (0.75*1.73, 0.25) -- (0.5*1.73, 0) -- (0.25*1.73, 0.25);
            \filldraw[blue, thick, fill=white] (0.5*1.73, 0) circle (3pt);
            \filldraw[blue, thick, fill = white] (0.75*1.73, 0.75) circle (3pt);
            \filldraw[blue, thick, fill=white] (0.25*1.73, 0.75) circle (3pt);
            
            \filldraw[green, thick, fill=white] (0.25*1.73, 0.25) circle (3pt);
            \filldraw[green, thick, fill=white] (0.75*1.73, 0.25) circle (3pt);
            \filldraw[green, thick, fill=white] (0.5*1.73, 1) circle (3pt);

        \end{tikzpicture}
    };
    \node at (3.5, 0) {$+$};
    \node at (4.75, 0) {
        \begin{tikzpicture}[scale=0.9]
            \draw[-, opacity=0.2] (0, 1) -- (0, 0) -- (0.5*1.73, -0.5) -- (1.73, 0) -- (1.73, 1) -- (0.5*1.73, 1.5) -- (0, 1);
            \draw[-, opacity=0.2] (0.5*1.73, -0.5) -- (0.5*1.73, 1.5);
            \draw[-, opacity=0.2] (0, 0) -- (1.73, 1);
            \draw[-, opacity=0.2] (0, 1) -- (1.73, 0);

            \node[red] at (0.25*1.73, 0.25) {\footnotesize $Z$};
            \node[red] at (0.75*1.73, 0.25) {\footnotesize $Z$};
            \node[red] at (0.5*1.73, 1) {\footnotesize $Z$};
        \end{tikzpicture}
    };
    \node at (7.5, 0.25) {\small $+~ (R \rightarrow B \rightarrow G \rightarrow R)$};
    \node at (7.5, -0.25) {\small $+~ (R \rightarrow G \rightarrow B \rightarrow R)$};

\end{tikzpicture}
}
\label{eq:d4-hamiltonian}
\end{equation}

The anyons in $D(D_4)$ will correspond to charges and flux membranes on top of the type III SPT. Charges on top of the SPT are created by $Z_v Z_{v'}$ for some pair of vertices $v, v'$, as with the Levin-Gu SPT. Crucially, in this case, $v, v'$ must belong to the same sublattice $a$, since we are applying gauging maps to each sublattice separately. If $v, v'$ belong to different sublattices, we cannot form a string connecting them using the emergent 1-form symmetries. 

\begin{figure}
    \centering
    \includegraphics[width=0.75\linewidth]{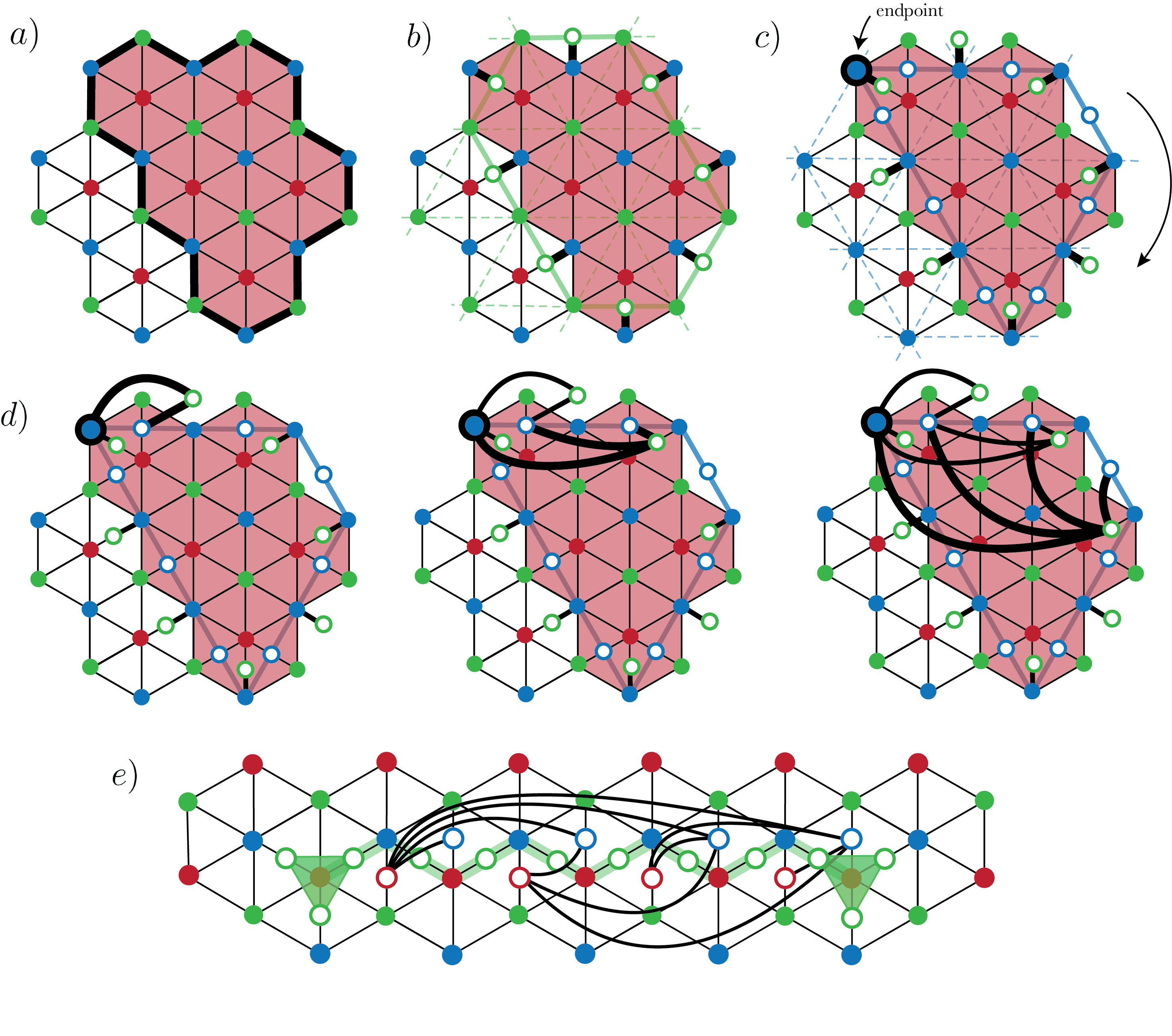}
    \caption{\textbf{Deriving} $D(D_4)$ \textbf{ribbon operators:} a) The boundary fractionalization pattern for the product of $R$ vertex stabilizers, indicated by the highlighted region. Here, bolded lines indicate $CZ$ gates. b) Each blue qubit on the boundary is connected to two green qubits with a $CZ$ gate; we can apply the $\mathbb{Z}_2$ gauging map to the green sublattice and collapse these $CZ$ pairs to single $CZ$ gates between a blue site and a green edge. c) To push the non-symmetric operators on the blue sites through the gauging map, we pick an endpoint to terminate all nonlocal strings on, and we construct the nonlocal strings clockwise. d) Proceeding clockwise from the endpoint, we construct the nonlocal $CZ$ strings resulting from pushing through operators acting on successive blue sites. In each case, the green edge connected to the original blue site becomes connected to all blue edges between it and the endpoint site. The new CZ gates at each step are indicated by thicker lines. e) An open green flux string, constructed using the same method as above, but on an open line. The highlighted green lines indicates an $X$ string, while the bolded black lines still indicate $CZ$. We can see that a plaquette operator on either side of the string is excited.}
    \label{fig:d4-ribbons}
\end{figure}

Closed flux ribbons correspond to finite $\prod X$ symmetry membranes on each sublattice. Again, we need to use the symmetry fractionalization properties of the type III SPT to fully understand how these finite membranes push through the $\mathbb{Z}_2^3$ gauging map. Consider a product of vertex stabilizers living on the $a=R$ sublattice (see Fig. \ref{fig:d4-ribbons}a)---the $CZ$ gates in the bulk cancel, leaving only those gates on the boundary. The structure of the type III SPT ensures that these boundary CZ gates only act on the $B, G$ sublattices; we only need consider the $B, G$ $\mathbb{Z}_2$ gauging maps when pushing the boundary decoration through to $D(D_4)$. We will proceed in stages; first, we apply the gauging map to the $G$ sublattice. This turns each pair of $CZ$ gates between a blue site and its neighboring green sites along the boundary to a single $CZ$ between this blue site and the corresponding green edge qubit (see Fig. \ref{fig:d4-ribbons}b). This is a consequence of the $Z_{v}Z_{v'} \rightarrow Z_{e}$ mapping: 
\begin{equation}
\begin{aligned}
    CZ_{b g_1} CZ_{b g_2} &= (\ket{0}\bra{0}_b + Z_{g_1} \ket{1}\bra{1}_b) (\ket{0}\bra{0}_b + Z_{g_2} \ket{1}\bra{1}_b) \\
    &= \ket{0}\bra{0}_b + Z_{g_1} Z_{g_2} \ket{1}\bra{1}_b \\
    &\rightarrow \ket{0}\bra{0}_b + Z_{e_{12}} \ket{1}\bra{1}_b\\
    &= \ket{0}\bra{0}_{e_{12}} + Z_b \ket{1}\bra{1}_{e_{12}}
\end{aligned}
\end{equation}
where $e_{12}$ is the edge connecting vertices $g_1$, $g_2$. Now we want to push the remaining action on blue sites through to blue edges. However, we have only a single $CZ$ acting on each blue site, containing a single $Z_b$ which is not symmetric under the $\mathbb{Z}_2$ gauging map on the $a = B$ sublattice. This is the origin of the non-Abelian nature of the $D(D_4)$ fluxes; the non-symmetric input prevents the final circuit on the edge degrees of freedom from being finite-depth. We pick a blue vertex $b_f$ to be the endpoint along the ribbon (see Fig. \ref{fig:d4-ribbons}c)---we will terminate all the non-local strings we get by pushing the $CZ$ through $KW_{\mathbb{Z}_2}^B$ on this site. For example, consider some green edge $e_g$ connected to a site $b$: 
\begin{equation}
\begin{aligned}
    CZ_{e_g, b} &= P_{e_g}^0 + P_{e_g}^1 Z_b \\
    &\rightarrow P_{e_g}^0 + P_{e_g}^1 \left (\prod_{e_b \in \gamma = \{b, b_f\}} Z_{e_b} \right) Z_{b_f} \\
    &=CZ_{b_f, e_g} \left ( \prod_{e_b \in \gamma} CZ_{e_g, e_b} \right)  
\end{aligned}
\end{equation}
where $\gamma = \{b, b_f\}$ is the collection of edges from vertex $b$ to the endpoint vertex $b_f$. If we repeat this process for every pair of $e_g$, $b$, we obtain a dense staircase of $CZ$ (see Fig. \ref{fig:d4-ribbons}d where every $e_g$ along the ribbon is connected to every $e_b$ between it and the ending vertex. 
\begin{equation}
\begin{aligned}
    U_{m_r} = \prod_{e_g} CZ_{b_f, e_g} \left ( \prod_{e_b \in \gamma} CZ_{e_g, e_b} \right) 
\end{aligned}
\end{equation}
where we are denoting the red symmetry flux anyon as $m_r$. There is still action on the endpoint vertex, but given a closed ribbon operator, this disappears due to the emergent 1-form symmetry on the green sublattice (the $e_g$ form a closed loop). Given an open ribbon, this residual action is keeping track of the overall charge of the $m_r$ flux pair being created.
We note that the $D(D_4)$ ribbons we have derived reproduce those in \cite{iqbalCreationNonAbelianTopological2023}.

We see that the extensive-depth of the flux ribbons results from the fact that a single $CZ$ spanning two different sublattices is not symmetric under either sublattice $\mathbb{Z}_2$ symmetry. Thus, when we push this $CZ$ through the gauging map, it must become non-local. 

\subsection{Quaternions (\emph{D(Q8)})}

Similarly to $D(D_4)$, quaternion topological order (or $D(Q_8)$) is isomorphic to the twisted quantum double arising from gauging the global symmetries of a type I+III $\mathbb{Z}_2$ SPT \cite{propitiusTopologicalInteractionsBroken1995}. Such an SPT can be constructed by starting with the type III SPT considered in the previous example, and then applying the Levin-Gu entangler to each of the three sublattices. 
\begin{equation}
\raisebox{-40pt}{
\begin{tikzpicture}
    \node at (-0.75, 0) {$H_{III+I} = - \sum\limits_{v \in R}$};
    \node at (1.75, 0){
        \begin{tikzpicture}[scale=0.65]
            \draw[-, opacity=0.2] (0, 1) -- (0, 0) -- (0.5*1.73, -0.5) -- (1.73, 0) -- (1.73, 1) -- (0.5*1.73, 1.5) -- (0, 1);
            \draw[-, opacity=0.2] (0.5*1.73, -1.5) -- (0.5*1.73, 2.5);
            \draw[-, opacity=0.2] (1.73, -1) -- (1.73, 2);
            \draw[-, opacity=0.2] (0, -1) -- (0, 2);
            \draw[-, opacity=0.2] (-0.5*1.73, -0.5) -- (1.5*1.73, 1.5);
            \draw[-, opacity=0.2] (0, -1) -- (1.5*1.73, 0.5);
            \draw[-, opacity=0.2] (-0.5*1.73, 0.5) -- (1.73, 2);
            \draw[-, opacity=0.2] (-0.5*1.73, 1.5) -- (1.5*1.73, -0.5);
            \draw[-, opacity=0.2] (0, 2) -- (1.5*1.73, 0.5);
            \draw[-, opacity=0.2] (-0.5*1.73, 0.5) -- (1.73, -1);

            \draw[-, very thick, black] (0, 1) -- (0, 0) -- (0.5*1.73, -0.5) -- (1.73, 0) -- (1.73, 1) -- (0.5*1.73, 1.5) -- (0, 1);
            \filldraw[blue] (0, 0) circle (3pt);
            \filldraw[green] (0, 1) circle (3pt);
            \filldraw[green] (0.5*1.73, -0.5) circle (3pt);
            \filldraw[blue] (1.73, 0) circle (3pt);
            \filldraw[green] (1.73, 1) circle (3pt);
            \filldraw[blue] (0.5*1.73, 1.5) circle (3pt);

            \draw[red, decorate, decoration={snake, segment length = 2.5pt, amplitude=1pt}] (-0.5*1.73, 0.5) -- (0, -1) -- (1.73, -1) -- (1.5*1.73, 0.5) -- (1.73, 2) -- (0, 2) -- (-0.5*1.73, 0.5);
            \filldraw[red] (-0.5*1.73, 0.5) circle (3pt);
            \filldraw[red] (0, -1) circle (3pt);
            \filldraw[red] (0, 2) circle (3pt);
            \filldraw[red] (1.73, -1) circle (3pt);
            \filldraw[red] (1.73, 2) circle (3pt);
            \filldraw[red] (1.5*1.73, 0.5) circle (3pt);
            
            \node[red] at (0.5*1.73, 0.5) {$X_v$};
        \end{tikzpicture}
    };
    \node at (3.5, 0) {$-\sum\limits_{v \in B}$};
    \node at (5.25, 0){
        \begin{tikzpicture}[scale=0.65]
            \draw[-, opacity=0.2] (0, 1) -- (0, 0) -- (0.5*1.73, -0.5) -- (1.73, 0) -- (1.73, 1) -- (0.5*1.73, 1.5) -- (0, 1);
            \draw[-, opacity=0.2] (0.5*1.73, -1.5) -- (0.5*1.73, 2.5);
            \draw[-, opacity=0.2] (1.73, -1) -- (1.73, 2);
            \draw[-, opacity=0.2] (0, -1) -- (0, 2);
            \draw[-, opacity=0.2] (-0.5*1.73, -0.5) -- (1.5*1.73, 1.5);
            \draw[-, opacity=0.2] (0, -1) -- (1.5*1.73, 0.5);
            \draw[-, opacity=0.2] (-0.5*1.73, 0.5) -- (1.73, 2);
            \draw[-, opacity=0.2] (-0.5*1.73, 1.5) -- (1.5*1.73, -0.5);
            \draw[-, opacity=0.2] (0, 2) -- (1.5*1.73, 0.5);
            \draw[-, opacity=0.2] (-0.5*1.73, 0.5) -- (1.73, -1);

            \draw[-, very thick, black] (0, 1) -- (0, 0) -- (0.5*1.73, -0.5) -- (1.73, 0) -- (1.73, 1) -- (0.5*1.73, 1.5) -- (0, 1);
            \filldraw[green] (0, 0) circle (3pt);
            \filldraw[red] (0, 1) circle (3pt);
            \filldraw[red] (0.5*1.73, -0.5) circle (3pt);
            \filldraw[green] (1.73, 0) circle (3pt);
            \filldraw[red] (1.73, 1) circle (3pt);
            \filldraw[green] (0.5*1.73, 1.5) circle (3pt);
            \node[blue] at (0.5*1.73, 0.5) {$X_v$};

            \draw[blue, decorate, decoration={snake, segment length = 2.5pt, amplitude=1pt}] (-0.5*1.73, 0.5) -- (0, -1) -- (1.73, -1) -- (1.5*1.73, 0.5) -- (1.73, 2) -- (0, 2) -- (-0.5*1.73, 0.5);
            \filldraw[blue] (-0.5*1.73, 0.5) circle (3pt);
            \filldraw[blue] (0, -1) circle (3pt);
            \filldraw[blue] (0, 2) circle (3pt);
            \filldraw[blue] (1.73, -1) circle (3pt);
            \filldraw[blue] (1.73, 2) circle (3pt);
            \filldraw[blue] (1.5*1.73, 0.5) circle (3pt);
        \end{tikzpicture}
    };
    \node at (7, 0) {$- \sum\limits_{v \in G}$};
    \node at (8.75, 0){
        \begin{tikzpicture}[scale=0.65]
            \draw[-, opacity=0.2] (0, 1) -- (0, 0) -- (0.5*1.73, -0.5) -- (1.73, 0) -- (1.73, 1) -- (0.5*1.73, 1.5) -- (0, 1);
            \draw[-, opacity=0.2] (0.5*1.73, -1.5) -- (0.5*1.73, 2.5);
            \draw[-, opacity=0.2] (1.73, -1) -- (1.73, 2);
            \draw[-, opacity=0.2] (0, -1) -- (0, 2);
            \draw[-, opacity=0.2] (-0.5*1.73, -0.5) -- (1.5*1.73, 1.5);
            \draw[-, opacity=0.2] (0, -1) -- (1.5*1.73, 0.5);
            \draw[-, opacity=0.2] (-0.5*1.73, 0.5) -- (1.73, 2);
            \draw[-, opacity=0.2] (-0.5*1.73, 1.5) -- (1.5*1.73, -0.5);
            \draw[-, opacity=0.2] (0, 2) -- (1.5*1.73, 0.5);
            \draw[-, opacity=0.2] (-0.5*1.73, 0.5) -- (1.73, -1);

            \draw[-, very thick, black] (0, 1) -- (0, 0) -- (0.5*1.73, -0.5) -- (1.73, 0) -- (1.73, 1) -- (0.5*1.73, 1.5) -- (0, 1);
            \filldraw[red] (0, 0) circle (3pt);
            \filldraw[blue] (0, 1) circle (3pt);
            \filldraw[blue] (0.5*1.73, -0.5) circle (3pt);
            \filldraw[red] (1.73, 0) circle (3pt);
            \filldraw[blue] (1.73, 1) circle (3pt);
            \filldraw[red] (0.5*1.73, 1.5) circle (3pt);
            \node[green] at (0.5*1.73, 0.5) {$X_v$};

            \draw[green, decorate, decoration={snake, segment length = 2.5pt, amplitude=1pt}] (-0.5*1.73, 0.5) -- (0, -1) -- (1.73, -1) -- (1.5*1.73, 0.5) -- (1.73, 2) -- (0, 2) -- (-0.5*1.73, 0.5);
            \filldraw[green] (-0.5*1.73, 0.5) circle (3pt);
            \filldraw[green] (0, -1) circle (3pt);
            \filldraw[green] (0, 2) circle (3pt);
            \filldraw[green] (1.73, -1) circle (3pt);
            \filldraw[green] (1.73, 2) circle (3pt);
            \filldraw[green] (1.5*1.73, 0.5) circle (3pt);
        \end{tikzpicture}
    };
    \node at (12, 0.1){
    \begin{tikzpicture}
        \node at (0, 1){
            \begin{tikzpicture}[scale=0.75]
                \node at (0, 0.3) {$i$};
                \node at (1, 0.3) {$j$};
                \filldraw[black] (0, 0) circle (2pt);
                \filldraw[black] (1, 0) circle (2pt);
                \draw[-, very thick, black] (0, 0) -- (1, 0);  
            \end{tikzpicture}
        };
    \node at (1, 0.75) {\footnotesize$ = CZ_{ij}$};
    \node at (0, 0.25){
        \begin{tikzpicture}[scale=0.75]
            \node at (0, 0.3) {$i$};
            \node at (1, 0.3) {$j$};
            \filldraw[black] (0, 0) circle (2pt);
            \filldraw[black] (1, 0) circle (2pt);
            \draw[decorate, decoration={snake, segment length = 2.5pt, amplitude=1pt}] (0, 0) -- (1, 0);  
        \end{tikzpicture}
    };
    \node at (1.25, 0.1) {\footnotesize $ = e^{i \frac{\pi}{4} Z_i Z_j}$};
    
    \end{tikzpicture}
    };
\end{tikzpicture}
}
\label{eq:quaternion-spt-hamiltonian}
\end{equation}

The structure of the resultant ribbon operators are a combination of the $D(D_4)$ ribbons and the semion strings derived in the previous sections; the fractionalization of the $\prod X$ symmetry is the tensor product of the Levin-Gu and type III patterns. Since we are gauging each sublattice separately, the final $a$ sublattice flux ribbon operators will look like semion strings living purely on the $a$ sublattice decorated by a dense $CZ$ staircase between qubits on the other two sublattices.

\end{document}